\documentclass[aps,prx,twocolumn,longbibliography,superscriptaddress,showpacs]{revtex4-1}
\usepackage{graphicx}
\usepackage{latexsym}
\usepackage{amssymb}
\usepackage{amsmath}
\usepackage{amsfonts}
\usepackage{mathtools}
\usepackage{dsfont}
\usepackage{bbm}
\usepackage{bm}
\usepackage{multirow}
\usepackage{color}
\usepackage{tikz}
\usepackage{comment}
\usepackage{footmisc} 
\usepackage{xcolor}
\newcommand{\ket}[1]{\vert#1\rangle}
\newcommand{\bra}[1]{\langle#1\vert}
\usepackage[colorlinks=true, citecolor={blue!80!black}, urlcolor={blue!50!black}, linkcolor = {blue!80!black}]{hyperref}
\allowdisplaybreaks[1]
\usepackage[percent]{overpic}
 \usepackage[utf8]{inputenc}

\renewcommand{\Im}{\mathop{\mathrm{Im}}}

\DeclareMathAlphabet{\mathbbold}{U}{bbold}{m}{n}

\definecolor{forestgreen}{rgb}{0.13, 0.55, 0.13}
\definecolor{gr}{rgb}{0,0.82,0.18}
\definecolor{DarkerRed}{rgb}{0.9,0.05,0.25}

\begin{document}

\title{
Shannon entropy of the measurement record at measurement-dominated criticality and RG flow: A $c$-theorem for effective central charge and a $g$-theorem for effective boundary entropy}
\author{Rushikesh A. Patil}
\author{Andreas W. W. Ludwig}
\affiliation{Department of Physics, University of California, Santa Barbara, CA 93106, USA}

\date{\today}

\begin{abstract}
{We present two theorems demonstrating non-perturbatively the decrease under {relevant}
renormalization group (RG) flow of two quantities, $c_{\rm eff}$
and $g_{\rm eff}$ characterizing, respectively, the universal information 
content of the Shannon entropy of the measurement record for two different types of 
measurement-dominated criticality.}
First, we demonstrate
the decrease of the ``effective central charge" $c_{\text{eff}}$ of $2D$ replica field theories in the $R\rightarrow1$ replica limit  that govern the long-distance physics of weakly monitored $2D$ classical critical systems (Baysian inference problems) studied recently in {the literature} [arXiv:2504.01264; arXiv:2504.12385; arXiv:2504.08888].
In particular, we show that $c_{\text{eff}}$ is \textit{less} than the central charge $c$ of
the unmeasured critical 
{system. We refer to this result as the ``$c$-effective theorem''.}
In addition, we present an analogous ``$g$-effective theorem" demonstrating
the decrease
{under}
RG flow of the effective ``Affleck-Ludwig'' boundary entropy 
$\ln g_\text{eff}${, quantifying a corresponding contribution to the Shannon entropy} 
   for analogous $2D$ \textit{defect} replica field theories in the $R\rightarrow1$ replica limit,
   which govern the long-distance physics in the problem of performing {weak}
   {\it quantum} measurements on one-dimensional quantum critical ground states.
   {Lastly, we discuss a {possible} consequence of our 
  {theorems}
  for classical systems with
generic uncorrelated impurity-type  quenched disorder, according to which, under a certain assumption, and
as opposed to problems with measurement-induced randomness, {the}
corresponding universal quantities $c_{\text{eff}}^{(R\rightarrow0)}$ 
and $g_{\text{eff}}^{(R\rightarrow0)}$
in the $R\rightarrow0$ replica limit would  \textit{increase} under RG flow.}
\end{abstract}

\maketitle

\tableofcontents

\section{Introduction\label{Sec:Intro}}
Understanding universal critical behavior of systems 
{subjected to various forms of quenched} disorder or randomness is an important 
{problem in}
condensed matter and statistical physics. 
Recently, there has been a surge of interest in studying measurement-induced phenomena
{where measurements form a source
 of randomness that couples to the degrees of freedom in the system 
{and lead}
to novel forms of criticality  dominated by measurements}
{(including, e.g., Refs.
\cite{PhysRevB.98.205136,PhysRevB.100.134306,skinner2019measurement,chan2019unitary,potter2022entanglement,fisher2023random,choi2020quantum,PhysRevX.10.041020,PhysRevLett.125.070606,BaoChoiAltman2019,JianYouVasseurLudwig2019,ZabaloGullansPixleyEtAlPRB2020,NahumSkinnerDiffusionAnnihilationPRR2020,li2023entanglement,LiChenLudwigFisher,lavasani2021measurement,LiVasseurFisherLudwig,VasseurPotterYouLudwig,cao2019entanglement,ZhouNahum,jian2022criticalityentanglementnonunitaryquantum,NahumRoySkinnerRuhman,IppolitiEtAlMeasurementOnlyPRX2021,ZabaloGullansWilsonVasseurLudwigGopalakrishnanHusePixley,KumarKemalChakrabortyLudwigGopalakrishnanPixleyVasseur,PhysRevX.12.041002,PhysRevLett.129.200602,PhysRevLett.129.120604,li2023cross,MajidyAgrawalGopalakrishnanPotterVasseurHalpern,JianShapourianBauerLudwig,FavaPiroliSwannBernardNahum,MirlinFreeFermionsPRX2023,NahumWiese,PiroliLiVasseurNahumTrivialQuantumTrajectoriesPRB2023,FavaPiroliBernardNahum,ChahineBuchhold,PoboikoGornyiMirlin,AgrawalEtAlLearnabilityPRX2024,GarrattAltmanPRXQ2023,GuoFosterJianLudwig,PopoikoPopperlGornyiMirlin,KhannaKumarVasseurLudwig,wang2025decoherenceselfdual,pütz2025flownishimori,GarrattWeinsteinAltman2022,WeinsteinSajithAltmanGaratt,YangMaoJian,BawejaLuitzGarratt,PatilLudwig2024})}.
{{Notably,} our 
 understanding
 of criticality in such monitored systems 
 has
 benefited 
 immensely from 
 the extensive body of work over several decades
 on
 quenched disordered 
 systems, classical or quantum,
 subjected to generic, uncorrelated quenched ``impurity-type'' disorder.}
 To study such systems 
 {whose randomness arises either from generic 
 {(uncorrelated)} quenched ``impurity-type'' disorder, or from quantum measurements in monitored 
 {quantum}
 systems governed by the Born-rule,}
 one usually relies on the replica trick to average over the randomness \footnote{There are other approaches to average
 {over} randomness, such as e.g. supersymmetry {(in the absence of interactions)} 
 \cite{EfetovBook}
 and
 {the Keldysh-Schwinger} formalism {(see e.g. \cite{ChamonNayakLudwig,MSFoster2008,GuoFosterJianLudwig,PoboikoGornyiMirlin})}}
 and that naturally yields field theories involving 
 {a variable 
 {number $R$}
 of copies}
 {(``replicas'')}
 of the system, where the interaction between different replica copies is governed by the form of randomness in the original problem.
One crucial difference between the replica field theories derived for systems with 
{generic {(uncorrelated) quenched} ``impurity-disorder''} 
and 
quantum systems where the randomness originates from the intrinsic indeterministic nature of quantum mechanical measurement outcomes,
is that the former replica field theories require us to take the replica limit $R\rightarrow0$ at the end of our calculations, while the latter {replica} field theories are required to be studied in the limit $R\rightarrow1$.
(The need to take the $R \to 1$ limit for monitored  quantum systems was first established in 
\cite{JianYouVasseurLudwig2019,BaoChoiAltman2019},
{and} the crucial difference between these limits {is} highlighted
{in many subsequent works, including, e.g., 
\cite{JianShapourianBauerLudwig,NahumWiese,PhysRevLett.129.120604,ZabaloGullansWilsonVasseurLudwigGopalakrishnanHusePixley,PatilLudwig2024}.)}
This is because in a
monitored quantum system, we are required to average quantities with the Born-rule probability, and that 
turns out to naturally 
{introduce~\cite{JianYouVasseurLudwig2019,BaoChoiAltman2019}}
an additional replica copy of the system, leading to a
replica theory
{in the limit  $R\rightarrow1$}
instead of a 
replica theory {in the limit  $R\rightarrow0$}.
Interestingly,
replica theories in the $R\rightarrow1$ limit
also arise naturally in the case of monitored \textit{classical} systems, as discussed
{very recently} in Ref. \cite{NahumJacobsen}, where one is interested in the properties of the classical 
{statistical}
ensemble conditioned on measurement outcomes as given by Bayes' theorem.
{({See also}
Ref. \cite{KimKeyserlingkLamacraft,PutzGarrattNishimoriTrebstZhu,gopalakrishnan2025monitoredfluctuatinghydrodynamics} for related recent works, and for Bayesian inference problems in statistical physics see e.g. \cite{NishimoriBook,ZdeborovaKrzakala,Sourlas_1994,Iba_1999}.)}\par
{In 
{the present}
paper we focus on $(1+1)$D and $(2+0)$D systems, where infinite-dimensional conformal
{symmetry~\cite{BelavinPolyakovZamolodchikov}
leads}
to additional control over the system.}
One general replica action that has been extensively studied in the literature of $2D$ classical \textit{critical} systems with 
{generic (uncorrelated)} quenched {impurity-type} disorder {[$R\to 0$ replica limit]} is {given by
(see, e.g., 
\cite{Ludwig-Potts-OPE-RG-NPB285-1987-97,LUDWIGCardy,Cardy_book})
\begin{equation}\label{EqDisorderReplicaAction}
    -\mathbb{S}=-\sum_{a=1}^{R}S_{*}^{(a)}+\Delta\int d^2 x \sum_{\substack{a,b=1\\ a\neq b}}^{R}\varphi^{(a)}(x)\varphi^{(b)}(x),
\end{equation}
where $S_{*}$ is the action 
{of a}
$(1+1)D$ or $2D$
{conformal field theory (CFT)}
that describes 
the system without 
{disorder, $\varphi$ is a
{scalar} field from the CFT $S_{*}$ that couples to the 
disorder,  and $\Delta$ is relevant in the {renormalization group (RG)}
sense.~\footnote{For classical critical systems with
{generic uncorrelated quenched (``impurity-type'')} 
disorder {(replica limit $R \to 0$)}, a more general form of replica action is given by including
{in addition} equal replica terms of form $\sum_{a}\varphi^{(a)}\varphi^{(a)}$
in the replica action. 
However, for the \textit{monitored} classical critical systems 
\cite{NahumJacobsen}, which are described by the limit $R\rightarrow1$ of the action in Eq. \eqref{EqDisorderReplicaAction}, such equal replica terms are 
{required to disappear from
the replica action  in the replica limit $R\rightarrow1$.
This is because {in the monitored problem  (replica limit $R \to 1$)}
the measurement-averaged first moment of a correlation function
must be equal to the correlation function in the unmeasured ensemble.
However, it
is possible to have the terms of form 
$(R-1)\sum_{a}\varphi^{(a)}\varphi^{(a)}$
in the replica action, since the coupling constant for
{such a} term vanishes
in the limit $R\rightarrow1$. 
In fact, such terms are seen to be
present in replica actions for monitored classical systems
{discussed} 
in Ref. \cite{NahumJacobsen} for certain measurement protocols, and 
in App. \ref{App:VanishingSingleReplicaPerturbation}
we show that the presence of such terms does \textit{not} affect the
effective central charge
{discussed in our}
Eq. \eqref{EqDefinitionOfEffC}.}}}
{We are primarily interested in 
{replica theories}
whose long-distance physics is controlled by a {\it critical} 
fixed point {(at which the correlation length is infinite)}.}
The properties of the critical system with 
{generic (uncorrelated)} quenched {impurity-disorder} are obtained by working with the above replica action with
{a variable number $R$}
of copies, and finally taking the limit $R\rightarrow0$ in all results.
In 
{{\it clean, disorder-free} ({i.e.} 
\it{translationally invariant}) critical systems,}
the central charge is an important measure of the universal critical behavior of the system \cite{BelavinPolyakovZamolodchikov,BloteCardyNightingale,Affleck1986,CalabreseCardy_2004}.
{On the other hand, in}
critical systems with {quenched} randomness, obtaining the analogous notion of central charge is 
{much less straightforward,} since one has to rely on either the replica trick 
{or on} {other}
methods of averaging over the randomness to
{first} obtain an equivalent {(``weak'')} translationally invariant theory {to which concepts of conformal symmetry can be applied}.
Since the replica theory in Eq. \eqref{EqDisorderReplicaAction} involves $R$ copies, we can associate {with it} a $R$-dependent central charge 
{$c(R)$,}
{{which}
is the central charge of} 
{the}
corresponding 
fixed point 
{that}
governs the long-distance physics of the
critical replica
{theory}
for sufficiently small values of $R$. 
However,
in the replica limit $R\rightarrow0$, which is of relevance to systems with {generic} {(uncorrelated)} quenched {impurity-disorder}, 
{$c(R)$}
is
{always} trivially \textit{zero} and is \textit{not} a useful measure of criticality in a disordered system. 
{Instead,
in systems with {generic} (uncorrelated) quenched {impurity-type} disorder,
the so-called effective central charge defined as
\begin{equation}\label{EqDisorderReplicaZeroEffC}
    c^{(R\rightarrow0)}_{\text{eff}}:=
\frac{dc(R)}{dR}\bigg|_{R=0},
\end{equation}
represents~\cite{LUDWIGCardy}
a useful universal 
characteristic of the 
criticality.}~\footnote{{See also
Gurarie's 
$b$-parameter~\cite{GURARIE1999765,GurarieLudwig_2002,GurarieLudwig05c0log}
which is another related useful universal characterization of random critical points.}}\par
On the other hand, for
{general critical}
$(1+1)D$  {[or: $(2+0)D$]}
replica field theories
{in the $R\rightarrow 1$ limit}
{(which need \textit{not} be of the form in Eq. \eqref{EqDisorderReplicaAction}),}
{the effective central charge} 
{now defined}
{(analogous to the 
$R\rightarrow0$ limit
discussed above)}
as the derivative
{with respect to the number $R$ of replicas at $R=1$}
of the central charge 
$c(R)$
of the 
corresponding
fixed point of interest~\footnote{{I.e. {of} the fixed point that governs the long-distance 
physics of the given replica} {theory}},
turns out to be an important
universal characterization of the underlying 
criticality,
\begin{equation}\label{EqDefinitionOfEffC}
    c_{\text{eff}}\equiv c_{\text{eff}}^{(R\rightarrow1)}:=\frac{dc(R)}{dR}\bigg|_{R=1}. 
\end{equation}
This quantity was recently introduced 
in Ref.~\onlinecite{ZabaloGullansWilsonVasseurLudwigGopalakrishnanHusePixley}
in the context of measurement-induced criticality in 
$(1+1)$D monitored quantum circuits, 
{where 
{$c(R)$}
in the above equation 
then denotes the (replica-dependent) central charge 
of the 
{fixed point describing the entanglement transition.}}
{In that work,
the above $R\rightarrow1$ effective central charge $c_{\text{eff}}$}
was shown~\cite{ZabaloGullansWilsonVasseurLudwigGopalakrishnanHusePixley} to characterize the universal finite-size scaling behavior (``{\it Casimir Effect}'') of the Shannon entropy of the measurement 
record at the entanglement 
transition~\footnote{{We note that in the context of entanglement transitions in monitored quantum circuits this quantity has sometimes been confused with a different universal
quantity of these circuits (and  is at times referred to with the same name), namely the coefficient of the logarithm of a subsystem size of the entanglement entropy, which corresponds to the scaling dimension of the boundary condition changing ``twist operator'' \cite{VasseurPotterYouLudwig,JianYouVasseurLudwig2019}
used to formulate the entanglement entropy. The confusion arises because in a standard, translationally invariant CFT (without any randomness, from measurements or otherwise), these two quantities coincide 
\cite{CalabreseCardy_2004,Affleck1986,BloteCardyNightingale}.
They no longer coincide for critical CFTs which lack translational symmetry. 
(See 
Ref.~\cite{ZabaloGullansWilsonVasseurLudwigGopalakrishnanHusePixley},
and
{comments on this distinction},
e.g., 
in
Refs.~\cite{KumarKemalChakrabortyLudwigGopalakrishnanPixleyVasseur,
PutzGarrattNishimoriTrebstZhu,wang2025decoherenceselfdual}.)
As explained in \cite{wang2025decoherenceselfdual}, these two
different universal quantities appear {\it simultaneously} in the entanglement entropy of a decoherence-induced  critical mixed state that can be associated with the entanglement transition in any monitored (deep) quantum circuit in $(1+1)D$.}},
{and it}
was 
also employed in several
subsequent works, 
including~\cite{KumarKemalChakrabortyLudwigGopalakrishnanPixleyVasseur,KhannaKumarVasseurLudwig,wang2025decoherenceselfdual,pütz2025flownishimori}.
{{An analogous quantity which applies}
 to 
 $(1+1)D$ [or $2D$] defect replica field theories {in the $R\rightarrow1$ replica limit}, 
 {which arise}
 in the problem of performing quantum measurements on one-dimensional critical ground states, was later introduced in
Ref.~\cite{PatilLudwig2024}.
The
 {universal} finite-size scaling behavior of the Shannon entropy of the measurement
 record in the corresponding measurement problem 
 then
 {turns out~\cite{PatilLudwig2024} to be}
 characterized by the effective Affleck-Ludwig \textit{boundary entropy}~\cite{AffleckLudwig1991PRL} of the defect replica field theory in the $R\rightarrow1$ replica limit, in contrast to
  {$c_{\text{eff}}$ {arising from}
  the ``Casimir Effect'',
 {which is} a ``bulk'' quantity.
 [We will also discuss below
 {in more detail}
 the effective Affleck-Ludwig {boundary entropy} in the latter measurement problems.]
 }}
\par
The precise form of replica field theories for monitored {\textit{quantum}}
systems, although
{all being}
defined in the replica limit $R\rightarrow1$, varies greatly {depending on the 
type of dynamics},
the specific protocol for measurements and possibly other factors \cite{PhysRevB.98.205136,PhysRevB.100.134306,skinner2019measurement,chan2019unitary,potter2022entanglement,fisher2023random,BaoChoiAltman2019,JianYouVasseurLudwig2019,li2023entanglement,LiVasseurFisherLudwig,VasseurPotterYouLudwig,ZhouNahum,NahumWiese,NahumRoySkinnerRuhman,ZabaloGullansWilsonVasseurLudwigGopalakrishnanHusePixley,KumarKemalChakrabortyLudwigGopalakrishnanPixleyVasseur,wang2025decoherenceselfdual,pütz2025flownishimori,PhysRevX.12.041002,PhysRevLett.129.200602,PhysRevLett.129.120604,MajidyAgrawalGopalakrishnanPotterVasseurHalpern,cao2019entanglement,JianShapourianBauerLudwig,FavaPiroliSwannBernardNahum,FavaPiroliBernardNahum,PoboikoGornyiMirlin,ChahineBuchhold,GuoFosterJianLudwig,PopoikoPopperlGornyiMirlin,KhannaKumarVasseurLudwig}.
{(In particular, the replica actions for
{monitored}
\textit{quantum} 
{systems are generally not,}
at least on the face of it, 
of the form 
{of}
Eq. \eqref{EqDisorderReplicaAction}{, which assumes the existence of an ultraviolet fixed point consisting of decoupled replicas of a given CFT
$S_*$.})}
{As mentioned above, 
Nahum and Jacobsen \cite{NahumJacobsen} have recently considered the problem of
statistical mechanics
{systems}
with imperfect/weak \textit{classical} measurements.}
{{In contrast to the `Born-rule' for 
monitored 
quantum
systems, in classical monitored systems (classical) measurements simply update our knowledge of the probability distribution for the unmeasured classical 
{statistical}
ensemble according to Bayes' theorem, and
{of interest are here}
the properties of the resulting ``conditioned ensemble".}}
{{In the case 
{where}
the unmeasured classical system is a {critical} system,
the replica field theories for such $2D$ weakly monitored classical critical systems
{are, as shown in Ref.~\cite{{NahumJacobsen}},}
precisely 
given by the $R\rightarrow1$ replica limit of the general replica actions of the form in Eq. \eqref{EqDisorderReplicaAction}.~\footnote{{See, e.g., Sec.  III of Ref. \cite{NahumJacobsen}.}}
{Moreover, 
in analogy with 
{the case of entanglement transitions in $(1+1)D$ monitored {\it quantum} circuits}
mentioned above, it has been shown
{in Ref. \cite{NahumJacobsen}}
that the effective central charge $c_{\text{eff}}$ {of} Eq. \eqref{EqDefinitionOfEffC}, together with the central charge of the unmeasured  classical critical system,
characterizes the finite-size scaling behavior of the Shannon entropy of the measurement record in $2$D monitored 
{\it classical} critical systems.}
{We note that,
{as emphasized in} Ref. \cite{PutzGarrattNishimoriTrebstZhu}, and as we discuss {in a general formulation} in App. \ref{AppRKWavefunctionsAndMeasurements}, 
{such}
classical monitored systems are equivalent to the corresponding problem of performing suitable \textit{quantum} Born-rule measurements on the Rokhsar-Kivelson (RK) wavefunctions~\cite{CLHenley_2004,ArdonneFedleyFradkin,IsakovFendleyLudwigTrebstTroyer}
corresponding to the unmeasured classical stat-mech system.}
\par
In this work,
{we present two theorems: the {``$c_{\text{eff}}$ theorem''} and the {``$g_{\text{eff}}$ theorem''}.}
{First,
in Sec. \ref{Sec:Proof},}
we present
a
{general}
theorem 
{stating} that the effective central charge in Eq. \eqref{EqDefinitionOfEffC} for a $(1+1)D$ 
{[or: $(2+0)D$]}
{replica field theory}
of the form shown in Eq. \eqref{EqDisorderReplicaAction} 
is,
{in the replica limit $R\to 1$},
less than the central charge $c$ of the unperturbed CFT $S_{*}$ when the coupling 
{constant} $\Delta$ is
relevant 
in RG sense 
\footnote{The case of an irrelevant perturbation is trivial since the perturbed theory flows back to the unperturbed fixed point under RG, and $c_{\text{eff}}$ is equal to the central
charge $c$ of the unperturbed theory. We postpone the discussion of 
marginal perturbations to a future work.}, 
i.e.
\begin{equation}\label{EqCeffectivetheoremINtro}
c_{\text{eff}}<c.    
\end{equation}
We will refer to this result as the ``$c$-effective {($c_{\text{eff}}$)} theorem", since it
{represents}
an analogue of the celebrated $c$-theorem \cite{Zamolodchikovctheorem} of A.B. Zamolodchikov for $(1+1)D$ unitary CFTs. 
{It is important to note that the 
replica theory of Eq. \eqref{EqDisorderReplicaAction} {in the limit of $R\to1 $ replicas}
is a non-unitary theory.}
{Its non-unitary nature}
arises because we are interested in the \textit{analytic continuation} of physical quantities evaluated in the theory with a variable number of replica copies $R>1$ to the replica limit $R\rightarrow1$, and 
{this is physically} related to the fact that a relevant perturbation
{of strength}
$\Delta$ in Eq. \eqref{EqDisorderReplicaAction} explicitly breaks the ``strong" translational invariance of the unperturbed theory $\sum_{a}S^{(a)}_*$, 
which is invariant under independent translations in individual replica copies,
to the ``weak" translational invariance in which the action is invariant only upon
{simultaneously} 
translating \textit{every} replica copy by the same amount.
Since the $R\rightarrow1$ replica action in Eq. \eqref{EqDisorderReplicaAction} would be usually obtained by averaging over a form of randomness, the latter symmetry discussion is 
equivalent
to the {\it lack} of translational invariance in the system 
within
a given realization of randomness, 
and to the {\it presence} of
``average" (``weak" or ``statistical") translational symmetry 
{in}
the ensemble of the system with all possible realizations of 
randomness.~\footnote{{See also, e.g., already  Giorgio  Parisi's early paper 
Ref.~\onlinecite{PARISIEnergyMomentumConservation}.}}
}
{Due to this non-unitary nature of the $R\rightarrow1$ replica field theory, we cannot use reflection positivity of correlation functions, which was used by Zamolodchikov in his proof of the c-theorem for $2D$ unitary theories,
{to derive}
our 
$c$-effective theorem in Eq.~\eqref{EqCeffectivetheoremINtro}.
However, as we will discuss below in Sec.~\ref{Sec:Proof}, the correlation function that gives the difference between the central charge $c$ of the unperturbed theory and the effective central charge $c_{\text{eff}}$ of the replica field theory Eq.~\eqref{EqDisorderReplicaAction} in the $R\rightarrow1$ replica limit can be written as 
a randomness-averaged square of a correlation function (see Eq.~\eqref{EQFinalSumRuleIntegrated}), which gives us the required positivity and we obtain the $c$-effective theorem in Eq.~\eqref{EqCeffectivetheoremINtro}.
}
We note that {the theorem in Eq. \eqref{EqCeffectivetheoremINtro}} is a
{general}
statement about the replica field theory in Eq. \eqref{EqDisorderReplicaAction} in the replica limit $R\rightarrow1$, and it is applicable to any problem where such
{a}
replica theory arises as {an} effective long-wavelength description of the problem. 
For example, in addition to the 
{above-mentioned}
monitored classical critical systems, {or equivalently the corresponding monitored critical RK wavefunctions,} that {we discuss} {in some more detail in} Sec.~\ref{Sec:Proof}
{below}, it would be interesting to see if this theorem can find applications
{to}
systems with monitored 
{\it quantum} dynamics or problems
{of}
quantum error correction.
However, it should be noted that in addition to the replica limit $R\rightarrow1$, our theorem
{relies}
explicitly on the specific form of the replica perturbation in Eq. \eqref{EqDisorderReplicaAction}. 
Therefore, the main challenge in applying our theorem {directly}
to any of the above problems is to check if the replica action for a given problem can be shown to be equivalent 
in the infrared limit
{(by a choice of a suitable ultraviolet limit)}
to 
{a} replica action in
the form of 
Eq.~\eqref{EqDisorderReplicaAction} for some unperturbed 
{(ultraviolet fixed point)}
CFT $S_{*}$ 
and a {scalar}  scaling field $\varphi$ in {that} CFT.
More generally,
{it might be possible to obtain the desired theorem also in other
situations in which, analogous to our Eq.~\eqref{EQFinalSumRuleIntegrated}, the
difference between $c_{\text{eff}}$ 
in the
ultraviolet (UV) and the infrared (IR) limit
can be written in terms of randomness-averaged square of a correlation function.}
The
{ability to do this}
depends crucially on the replica limit $R\rightarrow1$ and the replica perturbation of form $\sum_{a\neq b}\varphi^{(a)}\varphi^{(b)}$ as in Eq.~\eqref{EqDisorderReplicaAction}.~\footnote{{A UV fixed point with decoupled replicas might not be necessary if a given problem with randomness allows us to write the required replica correlation functions (see Sec.~\ref{Sec:Proof}) as
{a randomness averaged square} of a correlation function.}}
{We also note the `empirical' observation that the numerically computed values for 
the quantity $c_{\rm eff}$, 
Eq.~\eqref{EqDefinitionOfEffC},
are known to decrease upon 
RG flow 
{in certain} $(1+1)D$ cases that have been investigated 
{numerically, but which are not (at least on the face of it) of the form of our
Eq.~\eqref{EqDisorderReplicaAction}.}
These include the RG flow
between the percolation and the Nishimori critical 
point~\cite{pütz2025flownishimori}, as well as 
the RG flow of the $(1+1)D$ Haar-random entanglement transition between infinite onsite Hilbert space dimension (described by percolation)
and finite onsite Hilbert space 
dimension~\cite{ZabaloGullansWilsonVasseurLudwigGopalakrishnanHusePixley,JianYouVasseurLudwig2019}.
This would be consistent with a $c_{\rm eff}$-theorem of more general validity, beyond $R\to 1$ replica theories of the specific form of our Eq.~\eqref{EqDisorderReplicaAction}
for which we have  provided a proof in the present paper.}
\par
{We note that {the long-distance properties of} $2D$ classical critical systems 
and $1d$ quantum critical
{ground} states
{(i.e. in  one spatial dimension),
are both 
{governed}
by
$2D=(1+1)D$ CFTs.}
However, it is important to distinguish 
the problem of a $2D$ monitored classical critical system studied in Ref. \cite{NahumJacobsen},
{which 
as noted above is
equivalent to the problem of performing 
suitable 
Born-rule 
{\it quantum}
measurements on the corresponding $2D$ RK wavefunction,}
from the problem of a $1d$ quantum critical ground state subjected to quantum measurements 
that was {initiated} in Ref. \cite{GarrattWeinsteinAltman2022}, and 
{pursued in} subsequent
works~\cite{WeinsteinSajithAltmanGaratt,LeeJianXu,YangMaoJian,MurcianoSalaYueMongAlicea,sun2023newcriticalstatesinduced} {as well as {in}
\cite{PatilLudwig2024}.}
This is because the $1d$ quantum critical ground state exists only in one spatial dimension (the other dimension in the CFT description is that of
{imaginary 
time $\tau$}~\footnote{{whose purpose  in this context is solely to generate the unmeasured quantum ground state in a path-integral formulation}}), 
{and}
a single round of quantum measurements are performed only on a one-dimensional arrangement of degrees of freedom {located at the $\tau=0$ time-slice in space-time}, while in the $2D$ {monitored} classical problem, 
{or in the  corresponding monitored RK wavefunction problem,} (a single round of) 
measurements are performed on degrees of freedom in the two-dimensional
{system.}}
As a consequence of this, unlike the classical problem, the replica field theories for $1d$ monitored quantum critical
states
are $R\rightarrow1$ 
{\textit{defect/boundary}}
replica field theories, i.e. instead of the
bulk perturbation $\Delta\int d^2x\sum_{\substack{a\neq b}}\varphi^{(a)}(x)\varphi^{(b)}(x)$ in Eq. \eqref{EqDisorderReplicaAction}
{of}
the unmeasured $2D$ CFT $S_{*}$, the replica perturbation for the latter problem is given by the defect perturbation $\Delta\int dx\sum_{a\neq b}\varphi^{(a)}(x,0)\varphi^{(b)}(x,0)$, where the field $\varphi^{(a)}(x,0)$ is supported on the one-dimensional imaginary time $\tau=0$ ``defect" line 
{in space-time}
[\onlinecite{GarrattWeinsteinAltman2022,WeinsteinSajithAltmanGaratt,YangMaoJian,PatilLudwig2024}].}
{Analogous to the $c_{\text{eff}}$ theorem for the {\it bulk} replica field theories in Eq. \eqref{EqDisorderReplicaAction}, in Sec. \ref{SecGEffThm}, we present a {``$g_{\text{eff}}$ theorem''} for the 
$(1+1)D$ [or $2D$] 
{\it defect} replica field theories in the $R\rightarrow1$ limit given by
\begin{equation}
 -\mathbb{S}=-\sum_{a=1}^{R}S_{*}^{(a)}+\Delta\int dx \sum_{\substack{a,b=1\\ a\neq b}}^{R}\varphi^{(a)}(x,0)\varphi^{(b)}(x,0),\label{EqDefectReplicaFieldTheoryIntro}
 \end{equation}
where $\varphi$ is a scaling field from the CFT $S_{*}$, and {the coupling constant} $\Delta$ is relevant in the RG sense. 
We 
demonstrate non-perturbatively the
decrease under
RG flow of the effective Affleck-Ludwig 
boundary entropy $s_{\text{eff}}$ defined as 
\begin{equation}
    s_{\text{eff}}=\frac{ds(R)}{dR}\Big|_{R=1},\label{EqSeffIntro}
\end{equation}
where $s(R)=\ln g(R)$ is the 
{corresponding}
boundary entropy~\cite{AffleckLudwig1991PRL} for the $R$-copy defect replica field theory in Eq. \eqref{EqDefectReplicaFieldTheoryIntro}. An equivalent quantity is the effective 
`ground state degeneracy' $g_{\text{eff}}$ defined as $g_{\text{eff}}:=e^{s_{\text{eff}}}$, which features in the name of the theorem. 
Since the effective 
{boundary}
entropy in the ultraviolet (UV) {limit},
i.e. in the absence of a defect, is zero \footnote{$s(R)=\ln g(R)=0$ for any number of replicas in the absence of a defect}, the $g_{\text{eff}}$ theorem implies that the effective 
boundary entropy $s_{\text{eff}}$ ($=\ln g_{\text{eff}}$) for the
defect replica action in Eq. \eqref{EqDefectReplicaFieldTheoryIntro} must be less than zero.
As discussed above, the 
defect replica field theories in Eq. \eqref{EqDefectReplicaFieldTheoryIntro} in the $R\rightarrow1$ replica limit govern the long-distance physics of the problem of performing weak measurements on one-dimensional quantum critical ground states. (See e.g. \cite{GarrattWeinsteinAltman2022,WeinsteinSajithAltmanGaratt,YangMaoJian, PatilLudwig2024}.) \
Moreover,
as shown in Ref.~\onlinecite{PatilLudwig2024}, the effective
boundary entropy
$s_{\text{eff}}$, Eq.~\eqref{EqSeffIntro}, of the defect replica field theory 
Eq.~\eqref{EqDefectReplicaFieldTheoryIntro} in the $R\rightarrow1$ replica limit
characterizes the universal finite-size scaling behavior of the Shannon entropy of the measurement record on the one-dimensional quantum critical ground state.}
{As we demonstrate in Sec. \ref{SecGEffThm}, an equivalent 
restatement
of 
our
$g_{\text{eff}}$ theorem
{says that} the Shannon entropy of the measurement record $S_{Shannon}(L)$ for the corresponding measurement problem on the one dimensional critical ground state of length $L$ (periodic boundary conditions) is a concave function of length $L$, for large values of $L$ where the defect replica field theory description in Eq. \eqref{EqDefectReplicaFieldTheoryIntro} of the measurement problem is valid.}
{Analogous to our $c_{\text{eff}}$ theorem, the proof of the $g_{\text{eff}}$ theorem circumvents
{the need to use} the reflection positivity required for the proof of the $g$-theorem for 
$2D$ unitary {boundary} CFTs~\cite{FriedanKonechny}
{(which is not available in the presence of measurements, of interest to us)}
by noting that the correlation function characterizing the decrease under the RG flow of $s_\text{eff}=e^{g_{\text{eff}}}$ can be written as a randomness-averaged square of a correlation function. (See App.~\ref{AppGEffTheorem} for the proof.)}
{
We 
emphasize that both the $c_{\text{eff}}$ and the $g_{\text{eff}}$ theorem 
{demonstrate the
decrease
under the RG flow}
of
{quantities characterizing}
{the}
universal information contained in the Shannon entropy of the measurement record in the respective  measurement problems corresponding to the replica field theories Eq. \eqref{EqDisorderReplicaAction} and \eqref{EqDefectReplicaFieldTheoryIntro}, respectively.
}
\par
Lastly, we
note that it is an open question to-date whether it is possible to formulate some
version
of a
``$c_{\rm eff}$-theorem'' 
for {classical}
systems with generic (uncorrelated) quenched impurity disorder, i.e. in limit of $R\to 0$ replicas, which would express some monotonicity condition on the effective central charge $c_{\text{eff}}^{(R\rightarrow0)}$ (Eq. \eqref{EqDisorderReplicaZeroEffC}) in the $R\rightarrow0$ replica limit. (See, e.g., \cite{GURARIE1999765,cardy2001stresstensorquenchedrandom,FujitaHikidaRyuTakayanagi,CabraHoneckerMussardoPujol}.)
The result we report in our present paper represents an analogue of 
such a monotonicity condition in the limit of $R \to 1$ replicas, 
which is
{of relevance}
for classical systems subject to (classical) measurements.
We discuss 
in Sec. \ref{SecReplicaLimitZeroCEff} 
a possible
{consequence}
of our
$c_{\text{eff}}$-theorem
(Eq. \eqref{EqCeffectivetheoremINtro}) for the replica field theory in Eq. \eqref{EqDisorderReplicaAction} in the $R\rightarrow1$ replica 
limit, 
{for}
the effective central charge $c_{\text{eff}}^{(R\rightarrow0)}$
{of}
the same replica field theory but in the (different) replica limit $R\rightarrow0$,
{which is 
{of relevance}
for classical systems with generic (uncorrelated) ``impurity-type" quenched disorder.} 
\footnote{{As noted in footnote \cite{Note2},}
replica actions for classical systems with generic (uncorrelated) quenched disorder often contain the additional equal replica term of form $\Delta \sum_{a}\varphi^{(a)}(x)\varphi^{(a)}(x)$ in the replica action of Eq. \eqref{EqDisorderReplicaAction}. The latter equal replica term should be thought of as using point-splitting and the operator product expansion (OPE) of the field $\varphi$ with itself. Therefore, in the case when the fields present in the operator product expansion of the field $\varphi$ with itself are irrelevant in the RG sense, we can drop such equal replica terms from the replica action and the replica action is given by Eq. \eqref{EqDisorderReplicaAction} in the limit $R\rightarrow0$. (An example is the critical $Q$-state Potts model with bond impurity disorder studied in Ref. \cite{Ludwig-Potts-OPE-RG-NPB285-1987-97,LUDWIGCardy}, and that we discuss in Sec. \ref{SecReplicaLimitZeroCEff}.)}}
{In particular, we note that the function 
{$f(R)$ defined by}
$f(R)=c(R)-R.c$,
representing the difference between the 
{central charge $c(R)$  of the replica action in Eq. \eqref{EqDisorderReplicaAction} and the central charge $R c$ of the unperturbed action $\sum_aS_*
^{(a)}$,}
has 
two `natural'
zeroes, supported by concrete physical reasoning: one at $R=0$ and one at $R=1$.
Now if we assume that these are the only zeroes 
{of the function}
$f(R)$ in the interval $R\in[0,1]$~\footnote{{The assumption about zeroes of $f(R)$ is also supported by epsilon expansion results for systems discussed in Ref. \cite{LUDWIGCardy,SHIMADA2009707}.}}, {and that $f(R)$ is a 
differentiable
function of $R$~\footnote{{The differentiability/continuity of this function is intimately connected to the validity of assumptions stated at the end of Sec. \ref{Sec:Intro} in the entire interval of values of $R\in[0,1]$. In particular, as we discuss in Sec. \ref{SecReplicaLimitZeroCEff}, if the equal replica term $\sum_{a}\varphi^{(a)}\varphi^{(a)}$ is relevant, additive operator renormalizations 
{might be}
required in the replica action of Eq. \eqref{EqDisorderReplicaAction} away from the $R\rightarrow1$ replica limit, and one 
{might need}
to consider a fine-tuned replica action to have a 
{smooth}
function $f(R)=c(R)-Rc$ in the interval $R\in[0,1]$.}}}, our $c_{\text{eff}}$ theorem for the replica field theory in Eq. \eqref{EqDisorderReplicaAction} in the $R\rightarrow1$ replica limit then implies that
{the}
$R\rightarrow0$ effective central charge $c_{\text{eff}}^{(R\rightarrow0)}$ for the same replica field theory is \textit{greater} than the central charge of the unperturbed CFT $S_*$.
{Moreover, under}
the analogous assumptions, we {can} also obtain
{the result}
that the effective Affleck-Ludwig boundary entropy $s_{\text{eff}}^{(R\rightarrow0)}=(ds(R)/dR)|_{R=0}$ for the defect replica action Eq. \eqref{EqDefectReplicaFieldTheoryIntro} in the $R\rightarrow0$ replica limit is \textit{greater} than zero, i.e. it 
increases
in going {{from} the ultraviolet (UV) to
 the infrared (IR).}
We discuss
{details of}
{our arguments for both $c_{\text{eff}}^{(R\rightarrow0)}$ and $s_{\text{eff}}^{(R\rightarrow0)}$}
in Sec. \ref{SecReplicaLimitZeroCEff}.
\par
In the next section, {Sec. \ref{Sec:Proof}}, we present the proof of our 
{$c_{\text{eff}}$-theorem (Eq. \eqref{EqCeffectivetheoremINtro})
for the effective central charge in the $R\rightarrow1$ replica limit (Eq. \eqref{EqDefinitionOfEffC}).}
Apart from usual assumptions about translational and rotational (Lorentz) invariance {of the replica theory}
\footnote{As discussed above, given a replica field theory for a problem with randomness, the translational or rotational invariance of the replica field theory, which correspond to {simultaneous} translations or rotations by the same amount in every replica copy, are related to ``average'' (``weak'' or ``statistical'') translational or rotational symmetry in the problem with randomness.
{(In particular, in a fixed realization of randomness translational and rotational invariance are absent, and are \textit{not} required for our proof. What is required is {\it statistical} translational and rotational invariance, i.e. the corresponding invariance properties of the probability distribution.)}},
we will make the following 
{three fairly} {standard} assumptions in our proof for the theorem:
\\
\\*
\centerline{\bf Assumptions}
\begin{enumerate}
\item  {The renormalized operator $\sum_{a\neq b}\varphi^{a}\varphi^{b}$ in the perturbed theory of Eq. \eqref{EqDisorderReplicaAction} is the same as the operator $\sum_{a\neq b}\varphi^{a}\varphi^{b}$ in the unperturbed theory $\sum_{a}S_{*}^{(a)}$.}
Or equivalently, the replica field theory in Eq. \eqref{EqDisorderReplicaAction} is well-defined in the ultraviolet (UV) \textit{without} the need for ``counter terms", i.e. additive
{operator} renormalization. 
    \item There is \textit{no} replica symmetry breaking, i.e. the symmetry under permutation of replicas is \textit{not} spontaneously broken in the perturbed theory in Eq. \eqref{EqDisorderReplicaAction}. 
    \item The correlation functions of replicated copies of field $\varphi$ in the replica theory (Eq. \eqref{EqDisorderReplicaAction}) and the partition function $Z_R$ of the replica theory
    {[see Eq.~\eqref{EqReversingToRamdomnessFirstStep} of
    App.~\ref{AppReIntroOfRandom}]}
    are analytic functions of the number of replicas $R$,
    {at least for sufficiently small values of $R$ comprising the interval $0 \leq R \leq 1$.}
    (This assumption is also intimately connected to the validity of replica trick and applicability of replica action in Eq. \eqref{EqDisorderReplicaAction} to a given problem with randomness.)
\end{enumerate}
{The first assumption is required in our proof because we obtain a relation between the central charge of the UV fixed point and the (effective) central charge of the IR fixed point for the $R\rightarrow1$ replica field theory {Eq.~\eqref{EqDisorderReplicaAction}}, and thus having a well defined UV fixed point is necessary in our proof of the theorem. In App. \ref{App:VanishingSingleReplicaPerturbation}, we show that the first assumption is automatically true in the replica limit $R\rightarrow1$ whenever $\varphi$ is the most relevant field in a closed subalgebra of operator product expansions (OPEs) in the 
CFT $S_*$.~\footnote{We also show in App. B that
{even if}
additive operator renormalizations (“counter terms”)
{are}
required to have a well-defined UV fixed point away from the 
$R\rightarrow1$ limit {(i.e. for $R \not = 1$), they} do not affect the 
conclusion of our $c$-effective theorem in 
Eq. \eqref{EqCeffectivetheoremINtro}{, which requires {\it taking} the $R\to 1$ limit.}}}~\footnote{{We also note that we have shown in App. \ref{App:VanishingSingleReplicaPerturbation}}
that the 
{$c_{\text{eff}}$}
theorem is unaffected by the presence of {``counter terms"} of the form $(R-1)\sum_{a}\chi^{(a)}$ in the replica action of Eq. \eqref{EqDisorderReplicaAction}{,
where $\chi$ is} 
{a field in the OPE of the field $\varphi$ with itself.}
The latter result 
is 
of physical importance,
since for certain measurement protocols of monitored classical critical systems in Ref. \cite{NahumJacobsen}, the replica action in Eq. \eqref{EqDisorderReplicaAction} is known to contain the {additional} term $(R-1)\sum_{a}\varphi^{(a)}(x)\varphi^{(a)}(x)$, which is of the form $(R-1)\sum_{a}\chi^{(a)}(x)$ after using point splitting and the OPE of the field $\varphi$ with itself.
{(Compare also footnote \cite{Note2}.)}}
For example, in the case
{where}
$S_{*}$ is the
{2D} critical $Q-$state ($Q>2$) Potts model \cite{BelavinPolyakovZamolodchikov,FriedanQiuShenker,DotsenkoFateev,di1997conformal}, this assumption is therefore valid
{in the $R\rightarrow1$ replica limit,}
{when $\varphi$ is}
the leading spin field, 
and also 
{when $\varphi$ is}
the leading energy field of the $Q$-state Potts CFT.
{[The replica action  
in
the latter case was studied, in the $R\rightarrow1$ replica limit, in detail in Ref. \cite{NahumJacobsen}, and we also discuss it below in Sec. 
\ref{Sec:Proof}
]}
We note that if the replica action in Eq. \eqref{EqDisorderReplicaAction} is obtained as an effective long-wavelength description for a lattice problem, one would generically expect $\varphi$ to be the most relevant field in 
{the}
closed OPE subalgebra 
{generated by itself}.

{
The remaining {parts of the paper are}
structured in the following manner: 
As discussed above, in the next section, Sec.~\ref{Sec:Proof}, we discuss the proof of our $c_{\text{eff}}$ theorem.
In Section \ref{SecGEffThm}, we discuss our $g_{\text{eff}}$ theorem and the proof 
{of}
it is deferred to App. \ref{AppGEffTheorem}.
In Section \ref{SecReplicaLimitZeroCEff}, we discuss the consequence of the $c_{\text{eff}}$ and the $g_{\text{eff}}$ theorem in the $R\rightarrow1$ replica limit 
{for}
the effective central charge 
$c_{\text{eff}}^{(R\rightarrow0)}$ and the effective boundary entropy $s_{\text{eff}}^{(R\rightarrow0)}$ in the $R\rightarrow0$ replica limit, respectively.}
\vskip .8cm
\section{The \texorpdfstring{$c$}{Lg}-effective theorem\label{Sec:Proof}}
{We consider the replica action in Eq. \eqref{EqDisorderReplicaAction} formed by taking 
a variable number $R$ 
{of}
copies of a $(1+1)$D CFT $S_{*}$ which are interacting with each other with the replica interaction $\sum_{a\neq b}\varphi^{(a)}\varphi^{(b)}$, where $\varphi$ is a \textit{scaling} field from the unperturbed CFT $S_*$.
{Ultimately, we will be interested in studying the properties of 
this replica action
in the replica limit $R\rightarrow 1$ which is relevant for weakly monitored $2D$ classical critical systems.}
We will be interested in the case 
{where}
the perturbation 
in Eq. \eqref{EqDisorderReplicaAction} is relevant, i.e. 
{where}
the RG eigenvalue of the coupling constant $\Delta$ is 
{positive,}
{
\begin{equation}\label{EqPerturbationIsRelevant}
\text{RG Eigenvalue of }\Delta=2-2X_{\varphi}>0 \implies X_{\varphi} < 1,
\end{equation}}
where $X_{\varphi}$ is the scaling dimension of the field $\varphi$.}
\par
{We will now consider the}
 trace of the stress-energy tensor, denoted by  $\Theta(\vec{x})$, for the theory in 
 Eq. \eqref{EqDisorderReplicaAction},
 {which is known to be} 
 proportional to the perturbation 
 {of}
 the CFT and is
given~\cite{Zamolodchikovctheorem,Cardy1988}\footnote{Given the first assumption 
in the list at the end of Sec. \ref{Sec:Intro},
there are \textit{no} higher-order corrections in $\Delta$ to the expression for the trace of the stress-energy tensor in Eq. \eqref{EqTraceOfTheStressEnergyTensor}. 
Also, see App. \ref{App:VanishingSingleReplicaPerturbation}.}
by
\begin{equation}\label{EqTraceOfTheStressEnergyTensor}
\Theta(\vec{x})=-2\pi\Delta(2-2X_{\varphi})\sum_{\substack{a,b=1\\ a\neq b}}^{R}\varphi^{(a)}(\vec{x})\varphi^{(b)}(\vec{x}).
\end{equation}
{The well-known differential equation
of Zamolodchikov, which is used in his proof for the
{$c$-theorem}, relates the derivative of the 
{so-called}
$C$-function to the connected correlation of the trace of the stress-energy tensor $\Theta(\vec{x})$~\cite{Zamolodchikovctheorem,Cardy1988},
\begin{equation}\label{EqCtheorem}
   \frac{dC(r)}{dr}=-\frac{3}{2}r^3\,\big(\langle \Theta(r)\Theta(0)\rangle-\langle\Theta(0)\rangle^2\big).
\end{equation}}
{Here}
{$\langle\Theta(r)\Theta(0)\rangle$} {denotes the correlation function (evaluated in the perturbed theory) of fields $\Theta(\vec{x})$ and $\Theta(0)$,
where $r=|\vec{x}|$ is the 
distance of point $\vec{x}$  from the origin} (rotational invariance is assumed).
{The $C$-function $C(r)$
{is}
{in turn}
defined in terms of correlation functions
{of various components}
of the stress-energy tensor \footnote{{Explicitly, $C=2F-G+\frac{3}{8}H$, where $F=z^4\langle T(z,\bar{z})T(0,0)\rangle$, $G=z^2r^2\langle T(z,\bar{z}) \Theta(0,0)\rangle$ and $H=r^4\big(\langle\Theta(z,\bar{z})\Theta(0,0)\rangle-\langle\Theta(0,0)^2\rangle\big)$ (see e.g. Ref. \cite{Cardy1988}) and $T, \bar{T}$ and $\Theta$ are components of the stress-energy tensor in complex coordinates (see our Eqs. \ref{LabelEqStressTensorComponentszz}, \ref{LabelEqStressTensorComponentsz*z*}, \ref{LabelEqStressTensorComponentszz*}).}}, and importantly it interpolates between the UV and the IR central charge for the given $2D$ field theory, where $C(r=0)=c_{UV}$ is the UV central charge and $C(r=\infty)=c_{IR}$ is the IR central charge of the field theory.}
{For example, for the replica field theory in Eq. \eqref{EqDisorderReplicaAction} 
the $C$-function 
$C_R(r)$,
where the extra argument $R$ on the $C$-function denotes the number of replica copies in the theory, interpolates between 
\begin{equation}\label{EqLimitsOfCfunction}
    C_R(r=0)=R\times c, \;\;\;\text{and}\;\;\; 
     C_R(r=\infty)=c(R),
\end{equation}
where $R\times c$ is the central charge of the unperturbed action $\sum_{a}S_*^{(a)}$ and $c(R)$ is the central charge of the replica field theory in Eq. \eqref{EqDisorderReplicaAction} {in the infrared limit}.}
{We {now} apply}
{Zamolodchikov's differential equation}
in Eq. \eqref{EqCtheorem} to the replica action in Eq. \eqref{EqDisorderReplicaAction}. 
\footnote{
{The sum-rule that is obtained by integrating Zamolodchikov's differential equation}
has been {previously} applied to replica field theories 
in the literature, see e.g. \cite{cardy2001stresstensorquenchedrandom}, \cite{CabraHoneckerMussardoPujol}.}
{Then from Eq. \eqref{EqTraceOfTheStressEnergyTensor}, we obtain that}
{
\begin{align}
   &
   \frac{dC_R(r)}{dr}
   =-\frac{3}{2}(2\pi\Delta(2-2X_{\varphi}))^2r^3\times\nonumber\\&\qquad\;\;\;\;\;\;\;\;\;\times
  \bigg(\langle \sum_{\substack{a,b=1\\ a\neq b}}^{R}\varphi^{(a)}(r)\varphi^{(b)}(r)\sum_{\substack{c,d=1\\ c\neq d}}^{R}\varphi^{(c)}(0)\varphi^{(d)}(0)\rangle_{R}\nonumber\\
   &\;\;\;\;\;\;\;\;\;\;\;\;\;\;\;\;\;\;\;\;\;\;\;\;\;\;\;\;\;\;\;\;\;\;\;\;\;\;\;\;-\big(\langle \sum_{\substack{a,b=1\\ a\neq b}}^{R}\varphi^{(a)}(0)\varphi^{(b)}(0)\rangle_R\big)^2
   \bigg),
\label{EqSumRuleAppliedToReplicaTheory} 
\end{align}}
where the subscript $R$ on the correlation functions $\langle \dots\rangle_R$ 
in the above {equation} indicates that the correlation functions 
are calculated in the $R-$replica 
theory in Eq. \eqref{EqDisorderReplicaAction}.
{Let us first simplify the second term in the {parentheses of the above equation.}
{Given the replica symmetry, i.e. absence of replica symmetry breaking 
(``Assumption 2.'' above)}, we can 
{use permutation symmetry of}
the replica indices on the fields 
{
(see 
e.g.~\cite{Ludwig-Potts-OPE-RG-NPB285-1987-97,LUDWIGCardy,
LUDWIG1990infinitehierarchy,
CardyLog2013,VasseurJacobsenSaleur})}
{to}
simplify the second term in the 
{parentheses}
of Eq. \eqref{EqSumRuleAppliedToReplicaTheory}
{to}}
{
\begin{align}
   \big(\langle \sum_{\substack{a,b=1\\ a\neq b}}^{R}\varphi^{(a)}(0)\varphi^{(b)}(0)\rangle_R\big)^2&=((R(R-1)\langle\varphi^{(1)}(0)\varphi^{(2)}(0)\rangle_{R})^2\nonumber\\=&\,R^2(R-1)^2(\langle\varphi^{(1)}(0)\varphi^{(2)}(0)\rangle_{R})^2\label{EqTraceSquare}.
\end{align}
It is important for our 
{following} discussion to note that the contribution from Eq. \eqref{EqTraceSquare} to Eq. \eqref{EqSumRuleAppliedToReplicaTheory} is quadratic in $(R-1)$, i.e. 
{of order}
$\mathcal{O}((R-1)^2)$ in the replica limit
{$R \to 1$.}}
{Analogously, we can simplify 
{in the same way}
the first correlation function 
in the 
{parentheses}
of Eq. \eqref{EqSumRuleAppliedToReplicaTheory} by 
{using again permutation symmetry of}
the replica indices,}
\begin{align}
    &\langle \sum_{\substack{a,b=1\\ a\neq b}}^{R}\varphi^{(a)}(r)\varphi^{(b)}(r)\sum_{\substack{c,d=1\\ c\neq d}}^{R}\varphi^{(c)}(0)\varphi^{(d)}(0)\rangle_{R}=2R(R-1)\times\nonumber\\
    &\times\bigg(\langle \varphi^{(1)}(r)\varphi^{(1)}(0)\varphi^{(2)}(r)\varphi^{(2)}(0)\rangle_R+\nonumber
    \\&\;\;\;\;\;\;\;\;+\frac{(R-2)(R-3)}{2}\langle\varphi^{(1)}(r)\varphi^{(2)}(r)\varphi^{(3)}(0)\varphi^{(4)}(0)\rangle_R\nonumber\\&\;\;\;\;\;\;\;\;+2(R-2)\langle \varphi^{(1)}(r)\varphi^{(1)}(0)\varphi^{(2)}(r)\varphi^{(3)}(0)\rangle_R\bigg).\label{EqCorrelatorSimplified1}
\end{align}
We are interested in the behavior of the above correlation function in the limit $R\rightarrow1$,
{where it behaves as}
\begin{align}
     &\langle \sum_{\substack{a,b=1\\ a\neq b}}^{R}\varphi^{(a)}(r)\varphi^{(b)}(r)\sum_{\substack{c,d=1\\ c\neq d}}^{R}\varphi^{(c)}(0)\varphi^{(d)}(0)\rangle_{R}=\nonumber\\
     &=2(R-1)\bigg(\langle \varphi^{(1)}(r)\varphi^{(1)}(0)\varphi^{(2)}(r)\varphi^{(2)}(0)\rangle_{R=1}\nonumber\\&\;\;\;\;\;\;\;\;\;\;\;\;\;\;\;\;+\langle\varphi^{(1)}(r)\varphi^{(2)}(r)\varphi^{(3)}(0)\varphi^{(4)}(0)\rangle_{R=1}\nonumber\\&\;\;\;\;\;\;\;\;\;\;\;\;\;\;\;\;-2\langle \varphi^{1}(r)\varphi^{(1)}(0)\varphi^{(2)}(r)\varphi^{(3)}(0)\rangle_{R=1}\bigg)+\nonumber\\&\;\;\;\;\;\;\;\;\;\;\;\;\;\;\;\;\;\;\;\;\;\;\;\;\;\;\;\;\;\;\;\;\;\;\;\;\;\;\;\;\;\;\;\;\;\;\;\;\;\;\;\;\;+ \mathcal{O}((R-1)^2).\label{EqCorrelatorSimplified2}
\end{align}
{Here,} the correlation functions 
{denoted by}
$\langle \dots\rangle_{R=1}$ on the right hand side of the above equation
{are}
evaluated in the replica theory {of} Eq. \eqref{EqDisorderReplicaAction} {in the replica limit $R\rightarrow1$}.
{Then using Eq. \eqref{EqSumRuleAppliedToReplicaTheory}, \eqref{EqTraceSquare}, and \eqref{EqCorrelatorSimplified2}, we obtain
\begin{align}
   &
   \frac{dC_R(r)}{dr}
   =-3(R-1)(2\pi\Delta(2-2X_{\varphi}))^2r^3\times\nonumber\\&\qquad\;\;\;\;\;\;\;\;\;\;\;\;\;\;\bigg(\langle \varphi^{(1)}(r)\varphi^{(1)}(0)\varphi^{(2)}(r)\varphi^{(2)}(0)\rangle_{R=1}\nonumber\\&\;\;\;\;\;\;\;\;\;\;\;\;\;\;\;\;\;\;\;\;\;\;\;\;\;+\langle\varphi^{(1)}(r)\varphi^{(2)}(r)\varphi^{(3)}(0)\varphi^{(4)}(0)\rangle_{R=1}\nonumber\\&\;\;\;\;\;\;\;\;\;\;\;\;\;\;\;\;\;\;\;\;\;\;\;\;\;-2\langle \varphi^{1}(r)\varphi^{(1)}(0)\varphi^{(2)}(r)\varphi^{(3)}(0)\rangle_{R=1}\bigg)\nonumber\\&\;\;\;\;\;\;\;\;\;\;\;\;\;\;\;\;\;\;\;\;\;\;\;\;\;\;\;\;\;\;\;\;\;\;\;\;\;\;\;\;\;\;\;\;\;\;\;\;\;\;\;\;\;+ \mathcal{O}((R-1)^2),
   \label{EqReplicaSumRuleSimplified}
\end{align}
where the contribution from Eq. \eqref{EqTraceSquare} is absorbed
{into the}
corrections of order $\mathcal{O}((R-1)^2)$ in the above equation.}
\par
Note that 
replica field theories 
{such as those given}
in Eq. \eqref{EqDisorderReplicaAction}, are usually obtained by using the replica trick and 
{by}
averaging over a form 
{of}
randomness that is present in the original problem.
{In fact it turns out, as we will show, that}
it is
{always}
possible to introduce {a form of} ``randomness" back
{into the replicated action}, and write the {limit $R\rightarrow1$} replica theory in Eq. \eqref{EqDisorderReplicaAction} as a theory of
{an}
interaction between a single copy of the 
CFT $S_{*}$  
{and 
a
{(real)} ``randomness field'' $h(x)$} {coupled to the 
{scalar} field $\varphi$:}
{{We}
will denote the action of such a theory by $\mathcal{S}[\{h(x)\}]$ {and an explicit form for it is given in Eq. \eqref{EqInteractingTheoryWithRandomness} of App. \ref{AppReIntroOfRandom}}.
{Any} 
correlation functions,
{or}
{products}
of correlation functions, that are calculated in this theory
{with action}
$\mathcal{S}[\{h(x)\}]$ 
{for}
a fixed realization of randomness field $h(x)$ 
{can}
then be averaged over 
{the}
randomness field $h(x)$.}
In particular, since we are interested in obtaining a replica theory in the limit $R\rightarrow1$ {(and not
{in the limit} $R\rightarrow0$)}, 
{it turns out that} we 
have to introduce an additional copy of the 
{action}
$\mathcal{S}[\{h(x)\}]$ 
{in the averaging weight,}
while performing this average, and the details of this procedure are presented in App. \ref{AppReIntroOfRandom}.
This 
allows us to write the correlation functions of replicated fields, which are evaluated in the replica limit $R\rightarrow1${,
e.g. those appearing in Eq. \eqref{EqReplicaSumRuleSimplified}}, as 
randomness-averaged correlation functions 
{or}
{{as randomness-averaged} products of correlation functions} in the theory 
{with action}
$\mathcal{S}[\{h(x)\}]$. 
{In particular, it turns out that
{(using Eq. \eqref{EqReplicaTrickCorrelationFunction} of App. \ref{AppReIntroOfRandom})}
the correlation functions in Eq. \eqref{EqReplicaSumRuleSimplified} can be written as}
\begin{subequations}\label{EqReplicaCorrelationFunctionsToRandomnessOnes}
\begin{eqnarray}
      &&\langle \varphi^{(1)}(r)\varphi^{(1)}(0)\varphi^{(2)}(r)\varphi^{(2)}(0)\rangle_{R=1}=\overline{\langle\varphi(r)\varphi(0)\rangle_{\{h\}}^2}\nonumber\\&&\\
    &&\langle\varphi^{(1)}(r)\varphi^{(2)}(r)\varphi^{(3)}(0)\varphi^{(4)}(0)\rangle_{R=1}=\overline{\langle\varphi(r)\rangle_{\{h\}}^2\langle\varphi(0)\rangle_{\{h\}}^2}\nonumber\\&&\\
    &&\langle \varphi^{1}(r)\varphi^{(1)}(0)\varphi^{(2)}(r)\varphi^{(3)}(0)\rangle_{R=1}\nonumber\\&&\;\;\;\;\;\;\;\;\;\;\;\;\;\;\;\;\;\;\;\;\;\;\;\;\;\;\;\;\;\;\;=\overline{\langle \varphi(r)\varphi(0)\rangle_{\{h\}}\langle\varphi(r)\rangle_{\{h\}}\langle\varphi(0)\rangle_{\{h\}}}.\nonumber\\
\end{eqnarray}
\end{subequations}
The correlation functions $\langle\dots\rangle_{R=1}$ on the left hand side of the above equation are evaluated using the replica action in Eq. \eqref{EqDisorderReplicaAction} in the replica limit $R\rightarrow1$. 
On the other hand, the correlation functions $\langle\dots\rangle_{\{h\}}$ on the right hand side of the above equation are evaluated {in the single copy theory $\mathcal{S}[\{h(x)\}]$ interacting with a \textit{fixed} realization of randomness $h(x)$.}
{The average denoted by 
{an overbar
{$\overline{(\dots)}$}}
(see Eq. \eqref{EqWhatDoesanOverlineMean} for its explicit form) denotes an average performed over the randomness 
{field} 
$h(x)$  \textit{with an additional copy} of the 
{theory
$\mathcal{S}[\{h(x)\}]$ in the averaging}
{weight}, 
 which is important to obtain the limit $R\rightarrow1$ (instead of a  $R\rightarrow0$) replica theory.
Qualitatively similar averaging procedures with ``an additional copy of the
theory" naturally arise  in 
{quantum} systems
{monitored by measurements} (see
\cite{JianYouVasseurLudwig2019,BaoChoiAltman2019}),
 and in classical disordered systems with analogues of the Nishimori line \cite{Nishimori_1980,GRL2001}
 which are also relevant for quantum error correction (see e.g. \cite{DennisKitaevLandahlPreskill}).}\par
{We remark as an aside that, if}
we start out with a problem with quenched randomness which corresponds to 
{the $R\rightarrow1$ limit of a} replica field theory of the form
{of}
Eq. \eqref{EqDisorderReplicaAction}, then the procedure we describe in App. \ref{AppReIntroOfRandom} to write the correlation functions in the replica theory as randomness-averaged (products of) correlation functions in the single copy theory $\mathcal{S}[\{h(x)\}]$ with
{an}
``artificial" randomness field $h(x)$ might not be necessary.~\footnote{{I.e. in such a case, the correlation functions on the left-hand side of Eq. \eqref{EqReplicaCorrelationFunctionsToRandomnessOnes} could then be written in terms of randomness-averaged (products of) correlation functions in the original problem with quenched randomness, rather than introducing the ``artificial" randomness field $h(x)$.}}
For example, in the case of 
{weakly} monitored classical critical systems in Ref. \cite{NahumJacobsen} the (replica) correlation functions on the left-hand side of Eq. \eqref{EqReplicaCorrelationFunctionsToRandomnessOnes} correspond
{directly} to 
{suitable}
measurement-averaged products of the correlation functions in the monitored problem.
The precise form for the latter is the same as that of the corresponding correlation functions on the right-hand side of Eq. \eqref{EqReplicaCorrelationFunctionsToRandomnessOnes}, except
{that}
$\langle \dots\rangle_{\{h\}}$ now represents the correlation functions evaluated in a given (classical) ``measurement trajectory $\{h(x)\}$" and the overbar $\overline{(\dots)}$ denotes the average over measurement outcomes.
It should be noted, however, that our theorem is general, and {it} does not rely on the specific form of randomness that has been averaged over 
to obtain the replica theory in Eq. \eqref{EqDisorderReplicaAction}. In particular, it can be applied to 
any given replica field theory of the form
{of}
Eq. \eqref{EqDisorderReplicaAction} in the $R\rightarrow1$ replica limit, and it need not a priori correspond to a
{physical} problem with randomness. \par
{We now continue with our proof.}
{By}
using Eq. \eqref{EqReplicaCorrelationFunctionsToRandomnessOnes},
{we can write Eq. \eqref{EqReplicaSumRuleSimplified} as}
\begin{align}
   &
   \frac{dC_R(r)}{dr}
   =-3(R-1)(2\pi\Delta(2-2X_{\varphi}))^2r^3\times\nonumber\\&\qquad\qquad\;\;\;\times \bigg(\overline{\langle\varphi(r)\varphi(0)\rangle_{\{h\}}^2}+\overline{\langle\varphi(r)\rangle_{\{h\}}^2\langle\varphi(0)\rangle_{\{h\}}^2}+\nonumber\\
   &\;\;\;\;\;\;\;\;\;\;\;\;\;\;\;\;\;\;\;\;\;\;\;\;-2\overline{\langle\varphi(r)\varphi(0)\rangle_{\{h\}}\langle\varphi(r)\rangle_{\{h\}}\langle\varphi(0)\rangle_{\{h\}}}\bigg)+\nonumber\\
   &\;\;\;\;\;\;\;\;\;\;\;\;\;\;\;\;\;\;\;\;\;\;\;\;\;\;\;\;\;\;\;\;\;\;\;\;\;\;\;\;\;\;\;\;\;\;\;\;\;\;\;\;\;+\mathcal{O}((R-1)^2)\nonumber\\
   =&-3(R-1)(2\pi\Delta(2-2X_{\varphi}))^2r^3\times\nonumber\\&\qquad \times
   \bigg(\,\overline{(\langle\varphi(r)\varphi(0)\rangle_{\{h\}}-\langle\varphi(r)\rangle_{\{h\}}\langle\varphi(0)\rangle_{\{h\}})^2}\bigg)+\nonumber\\
   &\;\;\;\;\;\;\;\;\;\;\;\;\;\;\;\;\;\;\;\;\;\;\;\;\;\;\;\;\;\;\;\;\;\;\;\;\;\;\;\;\;\;\;\;\;\;\;\;\;\;\;\;\;+\mathcal{O}((R-1)^2),\label{EqSumRuleSimplified}
\end{align}
To zeroth order in $(R-1)$,
the {above 
{differential equation}
Eq.~\eqref{EqSumRuleSimplified}
implies} 
{that 
${dC_{R=1}(r)}/{dr}=0$ 
i.e. the $C$-function for the replica field theory with $R=1$ number of replicas does not change with distance $r$.}
{In particular, using Eq. \eqref{EqLimitsOfCfunction} it implies that the central charge of the replica field theory 
$c(R=1)=C_{R=1}(r=\infty)$
and the central charge of the unperturbed CFT 
$C_{R=1}(r=0)=1\times c=c$
are equal to each other, i.e.}
\begin{equation}\label{EqCR=1isc}
    c(R=1)=c, 
\end{equation}
which intuitively makes sense since exactly at $R=1$, we only have a single copy of the unperturbed CFT with central charge $c$. The more interesting result comes after differentiating Eq. \eqref{EqSumRuleSimplified} with respect to the number of replica copies $R$ at $R=1$,
\begin{align}
    &\frac{d}{dr}\bigg(
    \frac{dC_R(r)}{dR}\bigg|_{R=1}\bigg)=-48\pi^2\Delta^2(1-X_{\varphi})^2r^3\times\nonumber\\&\qquad\qquad\;\;\;\;\;\;\;\;\times\ \overline{\left(\langle\varphi(r)\varphi(0)\rangle_{\{h\}}-\langle\varphi(r)\rangle_{\{h\}}\langle\varphi(0)\rangle_{\{h\}}\right)^2}.\nonumber\\
    &\;\;\;\;\;\;\;\;\;\;\;\;\;\;\;\;\;\;\;\;\;\;\;\;\;\;\;\;\;\;\;\;\;\;\;\;\;\;\;\;\;\;\;\;\;\;\;\;\;\;\;\;\;\;\;\;\;\;\;\;\;\;\;\;\;\;\;\;\;\;\;\;\;\;\;\;\label{EQFinalSumRule}
\end{align}
{For any unitary unperturbed CFT $S_{*}$ (and also for 
certain~\footnote{{{An example is the $Q-$state Potts model CFT
used in Ref. \cite{NahumJacobsen} in an expansion about $Q=2$, which is nonunitary 
for generic real values {of} $Q\in (2,4)$, however the unperturbed} action $S_*$ 
and the correlation functions of the energy field (or of the spin field) are real-valued 
for arbitrary values of $2<Q<4$ (see e.g. Ref. \cite{DotsenkoFateev})}}
non-unitary unperturbed CFTs,
having real action $S_*$), 
the correlation function 
$\left (\langle \varphi(r)\varphi(0)\rangle_{\{h\}}-\langle \varphi(r)\rangle_{\{h\}}\langle\varphi(0)\rangle_{\{h\}}\right )$
{appearing 
on the right hand side of Eq.~\eqref{EQFinalSumRule}}, evaluated in the theory
{with action} $\mathcal{S}[\{h(x)\}]$ (Eq. \eqref{EqInteractingTheoryWithRandomness}), 
{is}
real-valued}~\footnote{This is because 
{any} correlation functions 
for the
{scalar} field 
$\varphi$ in the theory $\mathcal{S}[\{h(x)\}]$ (Eq. \eqref{EqInteractingTheoryWithRandomness}), {in which the real randomness field $h(x)$ is
coupled to $\varphi(x)$}, can be expanded (in a functional Taylor expansion {in $h(x)$}) in terms of 
multipoint correlation functions of the field $\varphi$ in the unperturbed CFT $S_{*}$, {which are \textit{real} e.g. in a unitary CFT $S_*$,} and the \textit{real} randomness field $h(x)$.}.
{Then the expression in the parentheses on the right hand side of 
Eq. \eqref{EQFinalSumRule} is a manifestly positive 
{number, {and} we obtain, due to the overall negative sign in Eq. \eqref{EQFinalSumRule}, that the `effective' $C$-function 
$\frac{dC_R(r)}{dR}|_{R=1}$
in the $R\rightarrow1$ replica limit \textit{decreases monotonically} with increasing distance $r$,
\begin{equation}\label{EqMonotonicDecOfCfn}
    \frac{d}{dr}\bigg(\frac{dC_R(r)}{dR}\bigg|_{R=1}\bigg)
    \leq0.
\end{equation}
Now from Eq. \eqref{EqLimitsOfCfunction} and Eq. \eqref{EqDefinitionOfEffC},
\begin{equation}\label{EqLimitsOfCEfffunction}
\frac{dC_R(r=0)}{dR}\bigg|_{R=1}=c
\;\;\;\text{and}\;\;\;
\frac{dC_R(r=\infty)}{dR}\bigg|_{R=1}
=c_{\text{eff}},
\end{equation}
} 
and
from 
the monotonic decrease of the `effective' $C$-function in Eq. \eqref{EqMonotonicDecOfCfn}, we 
obtain the result}
\begin{equation}\label{EqCEffectiveTheorem}
    \boxed{c_{\text{eff}}<c}.
\end{equation}
{The above boxed inequality is the {first}
key result of our paper. It}
implies that the effective central charge that characterizes the universal {critical behavior of
replica field theories}
{with an action of the form of}
Eq. \eqref{EqDisorderReplicaAction}
 in the limit $R\rightarrow1$
 {and a relevant coupling constant $\Delta$}, is always less than the central charge of the unperturbed CFT.~\cite{Note17}
{
Lastly, we can 
{also}
integrate
the differential equation in Eq. \eqref{EQFinalSumRule} from $r=0$ to $r=\infty$ and using Eq. \eqref{EqLimitsOfCEfffunction}, we obtain the following `sum-rule'
\begin{align}
    &
    c_{\text{eff}}
    -c=-48\pi^2\Delta^2(1-X_{\varphi})^2\times\nonumber\\&\;\;\;\;\;\;\;\;\times\int_{0}^{\infty} dr\,r^3\,\overline{\left(\langle\varphi(r)\varphi(0)\rangle_{\{h\}}-\langle\varphi(r)\rangle_{\{h\}}\langle\varphi(0)\rangle_{\{h\}}\right)^2}.\nonumber\\
    &\;\;\;\;\;\;\;\;\;\;\;\;\;\;\;\;\;\;\;\;\;\;\;\;\;\;\;\;\;\;\;\;\;\;\;\;\;\;\;\;\;\;\;\;\;\;\;\;\;\;\;\;\;\;\;\;\;\;\;\;\;\;\;\;\;\;\;\;\;\;\;\;\;\;\;\;\label{EQFinalSumRuleIntegrated}
\end{align}}
As we noted earlier
{(in the paragraph preceding
Eq.~\eqref{EqSumRuleSimplified}),}
if the correlation functions $\langle \cdots\rangle_{\{h\}}$ in Eq. \eqref{EqReplicaCorrelationFunctionsToRandomnessOnes} have a
{direct} natural interpretation on their own in a given problem with randomness, 
{the above sum-rule
in Eq. \eqref{EQFinalSumRuleIntegrated}}
serves as a useful relation between the
{difference of the} 
effective central charge $c_{\text{eff}}$ in the  $R\rightarrow1$ limit
{and}
the central charge $c$ of the unperturbed theory, and the integral of the randomness-averaged square of the connected correlation function of the field $\varphi$, {which couples to the randomness characterized by $h(x)$}.
{For example, in a $2D$ CFT $S_{*}$ under weak classical monitoring of the field $\varphi$ (\'a la Ref.~\cite{NahumJacobsen}), the sum-rule in Eq.~\eqref{EQFinalSumRuleIntegrated} relates the difference $c_{\text{eff}}-c$ to the integral of the measurement-averaged second moment of the connected correlation function of the field $\varphi$.}
\par
{We now}
discuss an application of the theorem in Eq. \eqref{EqCEffectiveTheorem} to a 
{particular example of a} \textit{classical} monitored system 
{discussed}
in Ref. \cite{NahumJacobsen},
{corresponding to}
weak monitoring of bond energies in the classical critical $Q-$state Potts model, {which is equivalent to the problem of performing weak Born-rule measurements of bond energies on the RK wavefunction corresponding to the critical $Q-$state Potts model (see App. \ref{AppRKWavefunctionsAndMeasurements}).}
As shown in Ref. \cite{NahumJacobsen}, the replica field theory for this problem
{consists of}
the 
replica action
in Eq. \eqref{EqDisorderReplicaAction}
{in the replica limit $R\rightarrow1$,}
with $S_{*}$ and $\varphi$ given by the action and the
leading
energy field of the critical $Q-$state Potts model, respectively.
{As mentioned, the}
effective central charge $c_{\text{eff}}$ defined in Eq. \eqref{EqDefinitionOfEffC}, together with the central charge $c$ of the unmeasured critical system, has been shown~\cite{NahumJacobsen} to characterize the universal finite-size scaling of the 
{Shannon entropy}
of the measurement record in these systems.
{By using} the result of
a controlled epsilon expansion in 
$\epsilon=Q-2$ {about $Q=2$} developed in  earlier works 
\cite{LUDWIGCardy,Ludwig-Potts-OPE-RG-NPB285-1987-97} (generalized to higher loop order in \cite{DotsenkoPiccoPujol,DotsenkoJacobsenLewisPicco})
and {by} applying it to the $R\to 1$ replica limit,
Ref.~\cite{NahumJacobsen} has demonstrated that the infrared behavior of this system in the case of weak {classical monitoring} 
is governed by a new RG fixed point
and
{has}
obtained the following expression for the effective central charge
{as an expansion in $\epsilon=Q-2$}
{\begin{equation}
\begin{split}
     &c_{\text{eff}}-c=-\frac{y^3}{8}-\frac{3y^4}{16}+\mathcal{O}(y^5),
\end{split}
\label{LabelEqQPottsR0Increase}
\end{equation}
where $y=\frac{4\epsilon}{3\pi}-\frac{4\epsilon^2}{9\pi^2}+\mathcal{O}(\epsilon^3)>0$ is the RG eigenvalue {(Eq. \eqref{EqPerturbationIsRelevant})} of the coupling constant $\Delta$ of the replica action (Eq. \eqref{EqDisorderReplicaAction}) for this problem.~\footnote{{The mentioned earlier works applied this
$\epsilon=Q-2$ {expansion} to study the 
{replica actions}
in the $R\rightarrow 0$
replica limit relevant for 
{systems with}
generic (uncorrelated) quenched randomness. 
See also the discussion 
{in Sec. 
\ref{SecReplicaLimitZeroCEff}
}
}}}
From this epsilon expansion result, it is clear that the effective central charge in this problem, at least for small values of $\epsilon=Q-2$, is \textit{less} than the central charge $c$ of the unmeasured critical $Q-$state Potts model. 
{This is consistent with the general statement of our 
``$c$-effective theorem"{, Eq.~\eqref{EqCEffectiveTheorem}.}
\vskip .8cm
\section{The \texorpdfstring{$g$}{Lg}-effective theorem\label{SecGEffThm}}
In this section, we 
{present} the 
{defect} 
analogue of our  $R\rightarrow1$ replica limit $c_{\text{eff}}$ theorem : The {``$g_{\text{eff}}$ theorem''}. 
{For this,} we 
consider the 
\textit{defect} replica action in Eq. \eqref{EqDefectReplicaFieldTheoryIntro} that, {as discussed in Sec.~\ref{Sec:Intro}, in the $R\rightarrow1$ replica limit governs the long-distance physics of the problem of performing weak Born-rule measurements with a local operator, which corresponds to the field $\varphi$, on the one-dimensional quantum critical ground state corresponding to the $(1+1)D$ CFT $S_*$ {(see e.g. Refs. \onlinecite{GarrattWeinsteinAltman2022,WeinsteinSajithAltmanGaratt,YangMaoJian,PatilLudwig2024})}. 
The \textit{defect} replica action in Eq. \eqref{EqDefectReplicaFieldTheoryIntro} consists of}
a variable number $R$ of copies of a $(1+1)D$ CFT $S_*$ 
interacting with each other with the replica
interaction $\sum_{\substack{a,b=1\\a\neq b}}^R\varphi^{(a)}(x,0)\varphi^{(b)}(x,0)$ \textit{only} on the \textit{one dimensional} $\tau=0$ {time-slice, a} ``defect" line in space-time, where $\varphi$ is a scaling field from the unperturbed CFT $S_{*}$.
{We will be interested in the case where the 
replica perturbation
{on the defect} in the 
{defect replica
action of Eq. \eqref{EqDefectReplicaFieldTheoryIntro}} is relevant, i.e. where the RG eigenvalue of the coupling constant $\Delta$ is positive
\footnote{{Note the difference in the RG eigenvalue
{of the coupling constant of the above defect replica action and of the corresponding coupling constant in the bulk replica action in}
Eq.~\eqref{EqPerturbationIsRelevant}.}},
\begin{equation}\label{EqDefectPerturbationIsRelevant}
\text{RG Eigenvalue of }\Delta=1-2X_{\varphi}>0 \implies X_{\varphi} < 1/2.
\end{equation}
}\par
{In} the case
we consider, where the unmeasured critical ground state is that of a CFT Hamiltonian on a space of length $L$ with periodic boundary conditions,
the corresponding {defect} replica field theory in Eq. \eqref{EqDefectReplicaFieldTheoryIntro} is 
defined on an infinite cylinder 
{of circumference $L$,}
{with a defect line}
situated at the 
{`axial'} coordinate ``$\tau=0$" along the circumference.
(See e.g. Ref. \onlinecite{PatilLudwig2024}.)
As demonstrated in Ref. \cite{PatilLudwig2024}, the defect 
{free energy}
of the defect replica field theory in Eq. \eqref{EqDefectReplicaFieldTheoryIntro},
which is equivalent to the boundary 
{free energy of the boundary theory}
obtained by folding the defect theory along the defect~\cite{LeclairLudwig} (see Sec. \ref{SubSecDefectFolding}),
{is related} in the $R\rightarrow1$ replica limit 
to the Shannon entropy of the measurement record as follows: Given the defect
{free energy}
(or equivalently, the boundary 
{free energy}
obtained upon folding) of the $R$-copy defect replica field theory in Eq. \eqref{EqDefectReplicaFieldTheoryIntro} on an infinite cylinder of circumference $L$, 
\begin{eqnarray}
    \text{Defect Free Energy}:=-\ln z_R(L).
\end{eqnarray}
{Then, in the measurement problem,} the Shannon entropy of the measurement record on the CFT ground state defined with periodic boundary conditions on space of length $L$ is given by~\cite{PatilLudwig2024}
\begin{equation}\label{EqShannonInTermsOfBoundaryPartFun}
     S_{Shannon}(L)=-\frac{d\ln z_R(L)}{dR}\Big|_{R=1}.
\end{equation}
The defect free energy
of the defect replica field theory 
in Eq. \eqref{EqDefectReplicaFieldTheoryIntro} in the $L\rightarrow\infty$ limit is given by 
\begin{equation}
    -\ln z_{R}(L\rightarrow\infty)=\alpha(R)L-s(R),\label{EqDefectFreeEnergyatLlarge}
\end{equation}
where $\alpha(R)$ is a non-universal constant and $s(R)=\ln g(R)$ is the Affleck-Ludwig boundary entropy~\cite{AffleckLudwig1991PRL} of the $R$-copy defect replica field theory in Eq. \eqref{EqDefectReplicaFieldTheoryIntro}.
Then from Eq. \eqref{EqShannonInTermsOfBoundaryPartFun},
the Shannon entropy  $S_{Shannon}(L)$ of the measurement record on the critical ground state
{is 
given in the  $L\rightarrow\infty$ limit by}
\begin{equation}
    S_{Shannon}{(L\rightarrow\infty)}=\alpha_{\text{eff}} L-s_{\text{eff}},
    \label{EqShannonEBoundary}
\end{equation}
where $\alpha_{\text{eff}}=\frac{d\alpha}{dR}|_{R=1}$
is a non-universal constant 
{while the effective Affleck-Ludwig boundary entropy $s_{\text{eff}}$ defined in Eq. \eqref{EqSeffIntro} is  universal.}
{
As 
mentioned in Sec. \ref{Sec:Intro}, 
we can define 
the equivalent quantity the effective
`ground state degeneracy' $g_{\text{eff}}$ by exponentiating $s_{\text{eff}}$
\begin{equation}
g_{\text{eff}}:=e^{s_{\text{eff}}}\label{EqDefOFGeffInTermsOfSeff},
\end{equation}
which features in the name of the theorem we present below.
}
\par 
{Now for a 
CFT
ground state on a space of finite length $L$  with periodic boundary conditions, the Shannon entropy $S_{Shannon}(L)$ of the measurement record is not exactly of the form in Eq. \eqref{EqShannonEBoundary}, {since it receives corrections of order $\mathcal{O}(\frac{1}{L})$ in the length.}
{Even in the limit $L\rightarrow\infty$,}
the universal piece $s_{\text{eff}}$ {of the Shannon entropy} is masked by the non-universal, extensive contribution $\alpha_{\text{eff}} L$. To focus on the variation in the subleading part of the Shannon entropy with $L$, let us define (analogous to Ref.~\cite{FriedanKonechny})
\begin{equation}\label{EqUniversalShannon}
    \mathfrak{s}_{\text{eff}}(L):=-\Big(1-L\frac{\partial}{\partial L}\Big)S_{Shannon}(L),
\end{equation} 
where the derivative $\big(1-L\frac{\partial}{\partial L}\big)$ precisely gets rid of the non-universal, extensive 
piece present at large {$L\rightarrow\infty$} in Eq. \eqref{EqShannonEBoundary}. 
In particular, using Eq. \eqref{EqShannonInTermsOfBoundaryPartFun},
\begin{align}
    \mathfrak{s}_{\text{eff}}(L)&=\Big(1-L\frac{\partial}{\partial L}\Big)\bigg[\frac{d\ln z_R(L)}{dR}\Big|_{R=1}\bigg]\nonumber\\
    &\implies\mathfrak{s}_{\text{eff}}(L)=\frac{d\mathfrak{s}_R(L)}{dR}\Big|_{R=1}\label{EqSeffDefInTermsOfS(L))},
\end{align} 
where {we have defined}
\begin{equation}
    \mathfrak{s}_R(L):=\Big(1-L\frac{\partial}{\partial L}\Big)\ln z_R(L),
\end{equation}
and 
{in the $L\rightarrow\infty$ limit, from Eq. \eqref{EqDefectFreeEnergyatLlarge}, 
{$\mathfrak{s}_R(L)$}
is 
{equal to $s(R)=\ln g(R)$, the Affleck-Ludwig boundary entropy of the $R$-copy defect replica field theory Eq. \eqref{EqDefectReplicaFieldTheoryIntro},}
}
\begin{equation}
    \mathfrak{s}_R(L\rightarrow\infty)=s(R)=\ln g(R).
\end{equation}
(Also see App. \ref{AppGEffTheorem}.)
The 
$\mathfrak{s}_{\text{eff}}(L)$ in Eq.~\eqref{EqSeffDefInTermsOfS(L))} then interpolates between,
\begin{equation}
    \mathfrak{s}_{\text{eff}}(L\rightarrow0)=0,
\end{equation}
i.e. the effective 
boundary entropy in the absence of a defect~\cite{Note11}}, 
and 
the effective 
boundary entropy $s_{\text{eff}}=(ds(R)/dR)|_{R=1}$ of the defect replica field theory in Eq. \eqref{EqDefectReplicaFieldTheoryIntro},
\begin{equation}\label{EqSeffatLargeLisseff}
    \mathfrak{s}_{\text{eff}}(L\rightarrow\infty)=s_{\text{eff}}=\ln g_{\text{eff}}.
\end{equation}}
Under the analogous assumptions as those listed at the end of Sec. \ref{Sec:Intro}, we show in App. \ref{AppGEffTheorem} that 
{$\mathfrak{s}_{\text{eff}}(L)$ in Eq. \eqref{EqUniversalShannon}}
decreases monotonically 
{with increase in length $L$, i.e.}
\begin{equation}
   \boxed{{\frac{d\mathfrak{s}_{\text{eff}}(L)}{d\ln L}\leq 0}}.\label{EqGeffTheorem}
\end{equation}
{
In particular, since $\mathfrak{s}_{\text{eff}}(L)$ interpolates between zero, i.e. the effective 
boundary entropy in the absence of a defect, and 
the effective 
boundary entropy $s_{\text{eff}}$ for the defect replica action in Eq. \eqref{EqDefectReplicaFieldTheoryIntro}, the above equation implies that the latter must be less than zero.  
That is, the effective 
boundary entropy of the defect replica field theory in Eq. \eqref{EqDefectReplicaFieldTheoryIntro} with a relevant coupling constant $\Delta$ has decreased in going from UV to IR --- it has decreased under the RG flow.
}
{Equivalently, in the limit $L\rightarrow\infty$ the universal, constant contribution to the Shannon entropy of the measurement record, which is equal to $(-1)\times s_{\text{eff}}$ from Eq. \eqref{EqShannonEBoundary}, must be positive.}
This is, for example, consistent with the epsilon-expansion result {for $s_{\text{eff}}=\ln g_{\text{eff}}$} in Ref. \cite{PatilLudwig2024} that is obtained for the defect replica field theory in Eq. \eqref{EqDefectReplicaFieldTheoryIntro} for the problem of performing measurements with the energy operator on the tricritical Ising ground state, where $S_*$ is the tricritical Ising CFT and $\varphi(x,0)$ is the energy field from the tricritical Ising CFT.\par
{An interesting 
{restatement}
of our 
$g_{\text{eff}}$ theorem follows simply from Eq. \eqref{EqGeffTheorem} and the definition of $\mathfrak{s}_{\text{eff}}(L)$ in Eq. \eqref{EqUniversalShannon},
which 
gives
us
the following inequality for the Shannon entropy of the measurement record $S_{Shannon}(L)$
\begin{equation}
    \frac{d^2S_{Shannon}(L)}{dL^2}\leq0.\label{EqInequalityOnShannon}
\end{equation}
That is, the Shannon entropy of the measurement record
in the corresponding measurement problem on the one-dimensional quantum critical ground state of length $L$ (periodic boundary conditions) is a concave function
of length $L$,
for large values of $L$ where the defect
replica field theory description in Eq. \eqref{EqDefectReplicaFieldTheoryIntro} of the
measurement problem is valid.
It would be interesting to see if an
information-theoretic approach 
can be used
to 
derive
the inequality in Eq. \eqref{EqInequalityOnShannon}.
}\par
{Lastly, as a side remark, we note that the universal critical behavior in a problem of performing 
weak 
quantum measurements on a one-dimensional quantum critical ground state
can be also obtained by performing suitable classical `Baysian' measurements \`a la Ref. \cite{NahumJacobsen} on a one-dimensional (spatial) defect line in a $2D$ classical critical stat-mech ensemble, where the unmeasured $1D$ quantum critical ground state and the unmeasured $2D$ classical critical stat-mech ensemble are both governed by the same $(1+1)D$ [$2D$] CFT at long distances.
{Clearly,
this correspondence between the universal properties also holds for Born-rule measurements on any $d$-dimensional quantum ground state, critical or not, and the classical `Baysian' measurements on a $d$-dimensional spatial defect in the corresponding $(d+1)$-dimensional classical stat-mech system.}
Moreover, following the discussion in App.~\ref{AppRKWavefunctionsAndMeasurements},
the 
{latter}
classical measurement problem (or ``Baysian inference problem") can 
also 
be viewed as a problem of performing {suitable} Born-rule measurements 
on a 
{codimension one}
(spatial)
defect 
in the $(d+1)$-dimensional RK wavefunction corresponding to the classical stat-mech system.
}
\vskip .8cm
\section{The 
\texorpdfstring{$R\rightarrow0$ replica limit: $c^{(R\rightarrow0)}_{\text{eff}}$ and $g^{(R\rightarrow0)}_{\text{eff}}$}{Lg}\label{SecReplicaLimitZeroCEff}}
{
{In this section, we discuss}
a possible 
{consequence}
of our $c_{\text{eff}}$ theorem, {which we proved in Sec. \ref{Sec:Proof}} for the replica action in Eq. \eqref{EqDisorderReplicaAction} in the replica limit $R\rightarrow1$, 
{for}
the effective central charge $c_{\text{eff}}^{(R\rightarrow0)}$ (Eq. \eqref{EqDisorderReplicaZeroEffC}) in the $R\rightarrow0$ replica limit 
{of}
the same replica action. 
{The latter replica limit, as mentioned already, is of relevance to {classical} systems with generic, uncorrelated quenched impurity-type disorder.}
 To this end}
 let us consider the function $f(R)$
{
\begin{equation}\label{EqDefF(R)}
    f(R)=c(R)- R \cdot c,
\end{equation}
}
{which represents the difference between 
{the central charge $c(R)$ of the}
replica action
in 
Eq.~\eqref{EqDisorderReplicaAction} {and the central charge of the (unperturbed) action $\sum_{a}S_*^{(a)}$}.
The function $f(R)$ has two ``natural" zeroes -- one at $R=0$ and one at $R=1$. The latter zero follows from our Eq. \eqref{EqCR=1isc}, which intuitively makes sense since exactly at $R=1$, we only have a single copy of the unperturbed CFT with central charge $c$.
The former zero 
{results trivially}
since 
{$f(R=0)=c(R=0)-0\times c=c(R=0)$},
{which vanishes because the partition function is unity in this limit by the definition of the replica 
trick~\footnote{{or equivalently, 
since there are \textit{no} replicas at $R=0$.}}.}
Our $c$-effective theorem 
for
the $R\rightarrow1$ replica limit in Eq. \eqref{EqCEffectiveTheorem} implies that the slope of $f(R)$ at $R=1$ is
{negative
{i.e.}
$(df/dR)|_{R=1}<0$;}
therefore, when $R<1$ and near $R=1$,
{the function $f(R)$ is positive, i.e.} $f(R)>0$. 
{{Let us now}
assume that 
{$f(R)=c(R)-Rc$}
has zeroes only at $R=0$ and $R=1$ in the interval $0\leq R\leq 1$, since these are the only zeroes
 {which are forced to appear due to  the very robust and concrete physical 
 reasoning mentioned above.}
This assumption is further substantiated by the 
perturbative $\epsilon$-expansion result {mentioned at the end of Sec. \ref{Sec:Proof}} for the central change 
{$c(R)$}
of the replica field theory in Eq. \eqref{EqDisorderReplicaAction} in the case where $S_*$ is the $Q$-state Potts model and $\varphi$ is the energy scaling field~\cite{Ludwig-Potts-OPE-RG-NPB285-1987-97,LUDWIGCardy,DotsenkoJacobsenLewisPicco}, where at least for small values of $\epsilon=Q-2>0$ 
{the function  
{$f(R)=c(R)-Rc$}
is seen to have zeroes only at $R=0$ and $R=1$ in the interval $R\in [0,1]$. (This function is reproduced in 
footnote~\footnote{The epsilon expansion 
result
for the function  
{$f(R)=c(R)-Rc$}
in this 
system~\cite{Ludwig-Potts-OPE-RG-NPB285-1987-97,LUDWIGCardy,DotsenkoJacobsenLewisPicco} 
reads
{$$  f(R)=c(R)-Rc=$$}
$$=-\frac{R(R-1)}{8(R-2)^2}\bigg(y^3+\frac{3}{2(2-R)}y^4 
+\mathcal{O}(y^5)\bigg),
$$
where $y=\frac{4\epsilon}{3\pi}-\frac{4\epsilon^2}{9\pi^2}+\mathcal{O}(\epsilon^3)>0$ is the RG eigenvalue of the coupling $\Delta$ (Eq. \eqref{EqDisorderReplicaAction}).
 At least for small values of $\epsilon=Q-2>0$, it is clear that $f(R)=c_{IR}(R)-Rc$ in this particular problem does not have a zero in {the interval}
 $R\in[0,1]$ except for the ``natural'' zeroes at $R=0$ and $R=1$}.})
Another example where this assumption 
{is easily verified}
is the replica field theory for the disordered $O(n)$ 
model {($n<1$)}
{fine-tuned to a crossover trajectory,}
which was studied
in a different {but analogous} epsilon expansion in Ref. 
\cite{SHIMADA2009707}. {(Also see the discussion below.)}}
{If we assume that $R=0$ and $R=1$ are the only zeroes 
{of}
$f(R)$ in the interval $0\leq R\leq1$ and {that $f(R)$ is a 
differentiable
function of $R$ in this interval (see footnote \cite{Note14}),}
our $c$-effective theorem in the $R\rightarrow1$ replica limit in Eq. \eqref{EqCEffectiveTheorem}, which is equivalent to $(df(R)/dR)|_{R=1}<0$, implies that 
\footnote{{Since $f(R)$ is zero at $R=0$ and $R=1$, and \textit{if} it 
{does not}
have any other zeroes in $R\in (0,1)$, then $f(R)$ does not
{change sign}
in the interval $R\in (0,1)$. Since the slope of $f(R)$ defined $(df(R)/dR)|_{R=1}=\lim_{\delta\rightarrow0^{+}}\frac{(f(1)-f(1-\delta))}{\delta}=-\lim_{\delta\rightarrow0^{+}}\frac{(f(1-\delta))}{\delta}$ is negative at $R=1$ (See Eq. \eqref{EqCEffectiveTheorem} and \eqref{EqDefF(R)}), the slope 
$(df(R)/dR)|_{R=0}=\lim_{\delta\rightarrow0^{+}}\frac{(f(0+\delta)-f(0))}{\delta}=\lim_{\delta\rightarrow0^{+}}\frac{(f(0+\delta))}{\delta}$ at $R=0$ must be non-negative. {This is because $f(1-\delta)$ and $f(0+\delta)$ should 
be both positive
for $\delta>0$}}}, 
}
\begin{equation}
    \frac{df(R)}{dR}\bigg|_{R=0}\geq0 
     \ \Rightarrow \ 
    \frac{dc(R)}{dR}\bigg|_{R=0}\geq c 
    \ \Rightarrow \ 
    c_{{\rm eff}}^{(R\rightarrow0)}\geq c.
    \quad\label{EqReverseCeffThm}
\end{equation}
{Therefore, given the assumption that 
{$f(R)=c(R)-Rc$}
does not have zeroes in the 
{open interval
$0 < R < 1$}
{and that the assumptions listed at the end of Sec. \ref{Sec:Intro} are satisfied in this interval
our $c$-effective theorem in the $R\rightarrow1$ replica limit, that we proved in Sec. \ref{Sec:Proof}, implies that the
effective central charge
{in the $R\to 0$ replica limit}
$c_{\text{eff}}^{(R\rightarrow0)}$ (Eq. \eqref{EqDisorderReplicaZeroEffC}) for the same replica action in Eq. \eqref{EqDisorderReplicaAction} is \textit{greater} than the central charge of the unperturbed CFT $S_*$
{(i.e. {\it greater} than the $R\rightarrow0$ effective central charge at the ultraviolet fixed point)}.}}\par
{We think, despite 
being based on the assumption of 
{$f(R)=c(R)-R.c$}
not having zeroes in {the interval} $0<R<1$, this is a rather surprising result: The result implies that a 
system with generic (uncorrelated) quenched impurity-type disorder that is described by the replica action in Eq. \eqref{EqDisorderReplicaAction} with a relevant {replica interaction
in the $R\rightarrow0$ replica limit,} is guaranteed to be critical
{as it implies that the 
value of the effective central charge is greater than the (already non-vanishing) central charge of the}
clean, disorder-free critical system~\footnote{{Recall that a non-critical system, that is one which has a finite correlation length, will have  vanishing effective central charge $c_{\rm eff}$.}}.
However, caution must be exercised here since the replica action for a classical critical system described by a CFT
{action} $S_*$ and quenched disorder coupled to a field $\varphi$, contains 
{the} `\textit{equal}' replica term of form $\sum_a\varphi^{(a)}\varphi^{(a)}$ 
in addition to the `\textit{unequal}' replica interaction in Eq. \eqref{EqDisorderReplicaAction}. 
{[Note that the 
equal replica term $\sum_a\varphi^{(a)}\varphi^{(a)}$ 
is
required to disappear in the replica limit $R\rightarrow1$ from the replica actions for the corresponding classical monitored systems, see footnote \cite{Note2} and App. \ref{App:VanishingSingleReplicaPerturbation}.]}
Such equal replica 
{terms}
should be thought of as 
{arising from} point-splitting 
{when employing} the operator product expansion
(OPE) of the field $\varphi$ with itself {{and}
correspond to an operator {\it different} from the one of the 
{unequal}
replica term}. 
As noted in footnote \cite{Note12}, the equal replica term can be dropped from the replica actions {for generic, uncorrelated quenched disorder systems} if the fields appearing in the OPE of the field $\varphi$ with itself are irrelevant.}
{A case in point of  {\it criticality} enforced by our 
{above argument, based on} the
``only-two-zeroes' assumption for $f(R)$ {in the interval $R\in[0,1]$,} is the $Q$-state Potts model fixed point with
{generic, uncorrelated quenched}
(``statistical'', or: ``weak'') Kramers-Wannier self-dual 
{bond} impurity 
disorder
(i.e. replica limit $R \to 0$) of Ref. \cite{Ludwig-Potts-OPE-RG-NPB285-1987-97,LUDWIGCardy,LUDWIG1990infinitehierarchy,DotsenkoPiccoPujol,DotsenkoJacobsenLewisPicco,JengLudwig}{.}
{In particular, the replica limit $R\rightarrow0$ version of the perturbative epsilon-expansion result~\cite{LUDWIGCardy,Ludwig-Potts-OPE-RG-NPB285-1987-97}
discussed
at the end of Sec. \ref{Sec:Proof}, implies that the  
{$R\rightarrow0$}
effective central charge $c^{(R\rightarrow 0)}_{\text{eff}}$ (Eq. \eqref{EqDisorderReplicaZeroEffC}) for this system, at least for small values of $\epsilon=Q-2>0$, is 
\textit{greater} than the central charge $c$ of the clean, disorder-free critical $Q-$state Potts model~\cite{LUDWIGCardy}.}
In this system any equal-replica terms of the kind mentioned above are irrelevant and the criticality of the random system {for $2 < Q \leq 4$}
discovered in the $(Q-2)$-epsilon expansion~\cite{LUDWIGCardy,Ludwig-Potts-OPE-RG-NPB285-1987-97,DotsenkoPiccoPujol,DotsenkoJacobsenLewisPicco}
follows from our 
{above argument
combining the $c_{\text{eff}}$ theorem in the $R\rightarrow1$ replica limit (Sec. \ref{Sec:Proof}) and the assumption about zeroes of $f(R)$ in $R\in [0,1]$.}
{This}
criticality was also confirmed beyond the small $(Q-2)$-regime by direct numerical simulation in {Ref. \cite{JacobsenCardy,JacobsenEnergyPRE2000,JengLudwig,ChatelainNPB2000}}.\footnote{{The fact that the 
system 
{with generic uncorrelated disorder}
is also critical for $Q>4$ (while the clean
system is not), requires a different argument \cite{AizenmanWehr}}.}}
\par
{However,
if the equal replica term is not irrelevant, 
we cannot apply the 
{argument leading to}
Eq. \eqref{EqReverseCeffThm} to such 
systems with generic (uncorrelated)
quenched impurity disorder, i.e., the $R\rightarrow0$ replica limit. 
Note that our argument leading to Eq.  \eqref{EqReverseCeffThm} relies on two primary results: (i) our $c_{\text{eff}}$ theorem for the replica action in the $R\rightarrow1$ limit, which we proved in Sec. \ref{Sec:Proof} for the replica action in Eq. \eqref{EqDisorderReplicaAction}, and 
(ii) the observation that $f(R)=c(R)-Rc$ for the replica action in Eq. \eqref{EqDisorderReplicaAction} has two `natural' zeroes at $R=0$ and $R=1$, respectively.
Assuming $f(R)$ does not have any
zeroes in $R\in(0,1)$  and that it is 
a 
differentiable
function of $R$ in $R\in [0,1]$, {where the latter is intimately connected to the validity of assumptions listed at the end of Sec. \ref{Sec:Intro} (including Assumption 1.) in the \textit{entire} interval $R\in[0,1]$ (see footnote \cite{Note14}),} Eq. \eqref{EqReverseCeffThm}
then
follows. 
Firstly, 
{in the presence of an equal replica term that does not vanish in the $R\rightarrow1$ replica limit,}
our proof for the $c_{\text{eff}}$ theorem in the $R\rightarrow1$ replica limit
does not go through 
because
the expression for the trace of the stress-energy tensor $\Theta(x)$ for 
{such a}
replica action with a non-vanishing and relevant equal replica term gets non-trivially altered from that in Eq. \eqref{EqTraceOfTheStressEnergyTensor}. 
Secondly, the function $f(R)=c(R)-Rc$ no longer vanishes at $R=1${,} because a relevant equal replica term $\sum_{a}\varphi^{(a)}\varphi^{(a)}$ will 
{create}
a difference between the central charge $c(R)$ of the replica field theory in $R\rightarrow1$ limit and the central charge $c$ of the CFT $S_*$, just like in the ordinary unitary $c$-theorem. 
{[}This can be seen from Eq. \eqref{EqSumRuleSimplified}, where the RHS no longer vanishes at $R=1$ if $\Theta(x)$ in Eq. \eqref{EqTraceOfTheStressEnergyTensor} is non-trivially altered by the presence of the equal replica term in the replica action.{]}
However, both of the above failures can be circumvented in the fine-tuned scenario where {in the replica limit $R\rightarrow1$} we set the coupling constant 
of
the equal replica term to zero. 
Then provided no other additive renormalizations are required (also see App. \ref{App:VanishingSingleReplicaPerturbation}), the expression for $\Theta(x)$ in Eq. \eqref{EqTraceOfTheStressEnergyTensor} holds true {in the $R\rightarrow1$ replica limit}, 
and 
our proof of the $c_{\text{eff}}$ theorem in the $R\rightarrow1$ replica limit goes through.
Now for 
$R<1$, we tune the coupling constant for the equal replica term continuously with $R$, such that the replica field theory for {each replica index} $R$ sits on the RG trajectory (i.e. the critical manifold) of the infrared fixed point that deforms in the replica limit $R\rightarrow1$ into the infrared fixed point we get in the $R\rightarrow1$ replica limit by tuning {the strength of} the equal replica term to zero~\footnote{{This is of course provided that the fixed point exists for all $0<R<1$, which is contingent on the validity of replica trick.}}. 
Then the central charge $c(R)$ of the  infrared fixed point of the fine-tuned replica action is a 
{smooth}
function of $R$, because by construction the infrared fixed point of the fine tuned replica field theory {for any $0<R<1$} is smoothly connected to the fixed point of the replica field theory in the $R\rightarrow1$ replica limit, 
{where the}
coupling constant for the equal replica term 
{is taken to be vanishing in the $R\rightarrow1$ replica limit}.
{Because of the vanishing equal replica term, $f(R)=c(R)-Rc$
for this fine-tuned replica action has a zero at $R=1$.}
Then assuming that $f(R)$ has zeroes only at $R=0$ and $R=1$ in the interval $R\in [0,1]$ ~\footnote{Recall that the zero of $f(R)$ at $R=0$ results trivially from the definition of the replica trick.}, we can make the above argument to obtain Eq. \eqref{EqReverseCeffThm}, which says that the effective central charge $c_{\text{eff}}^{(R\rightarrow0)}$ in the $R\rightarrow0$ replica limit for the fine-tuned replica action is greater than the central charge $c$ of the unperturbed CFT $S_*$.
A case in point for the latter is the work by Shimada on the quenched disordered $O(n)$ model with $n<1$~\cite{SHIMADA2009707}, where the equal replica terms
in the replica action are not irrelevant in the RG sense.
From the one-loop beta functions
for the disordered $O(n)$ model~\cite{SHIMADA2009707}, one can see that {for $n_*<n<1$ with some $n_*>0$ (also see \cite{ShimadaJacobsenKamiya}),} there is a non-trivial  {attractive} fixed point 
{which in the $R\rightarrow1$ replica limit} occurs at a non-vanishing strength of the \textit{unequal} replica term and at the vanishing strength of the \textit{equal} replica term.
Away from $R\rightarrow1$ replica limit, we can tune the coupling constant of the equal replica term such that the replica field theory sits
on the
critical manifold of the
fixed point which in the $R\rightarrow1$ replica limit deforms into the above non-trivial {attractive} fixed point.
Then, from our above argument for the replica limit $R\rightarrow0$ effective central charge {$c_{\text{eff}}^{(R\rightarrow0)}$, $c_{\text{eff}}^{(R\rightarrow0)}$} for this fine-tuned replica action must be greater than the central charge $c$ of the clean $O(n)$ ($n<1$) model. From the epsilon-expansion results in Ref. \cite{SHIMADA2009707}, it is clear that the
effective central charge in the $R\rightarrow0$ replica limit for the fine-tuned replica action, at least for small values of epsilon, is indeed greater than the central charge of the clean, disorder-free critical $O(n)$ ($n<1$) model.
}}
\par
{{An analogous argument,} under the analogous assumptions, also holds true for the effective Affleck-Ludwig boundary entropy $s_{\text{eff}}^{(R\rightarrow0)}=ds(R)/{dR}|_{R=0}$ of the defect replica action in Eq. \eqref{EqDefectReplicaFieldTheoryIntro} in the $R\rightarrow0$ replica limit. 
In particular, for exactly analogous reasons as those discussed above for 
{$f(R)=c(R)-R.c$,}
$s(R)=\ln g(R)$ for defect replica field theories in Eq. \eqref{EqDefectReplicaFieldTheoryIntro} also has two `natural' zeroes at $R=0$ and $R=1$.
Therefore, following the above discussion for the effective central charge in the $R\rightarrow0$ replica limit,
if we assume that the 
{boundary entropy} $s(R)=\ln g(R)$ for the defect replica field theories in Eq. \eqref{EqDefectReplicaFieldTheoryIntro} is nonzero in the interval $R\in [0,1]$ except for the natural zeroes at the endpoints {of} this interval {and that it is a differentiable function of $R$ in this interval}~\footnote{{This is intimately related to the validity of analogous assumptions as those stated at the end of Sec. \ref{Sec:Intro}, {in the \textit{entire} interval $R\in [0,1]$,} now for the defect replica field theory.}}, our $g_{\text{eff}}$ theorem in the $R\rightarrow1$ replica limit
implies that the 
effective 
boundary entropy
$s_{\text{eff}}^{(R\rightarrow0)}=ds(R)/{dR}|_{R=0}$ 
for the same defect replica field theory, but in the different $R\rightarrow0$ replica limit, must be greater than zero (i.e. greater than it's UV value).
A case in point
for this are the defect replica field theories for a random, uncorrelated defect line of the leading energy field in the tricritical $Q-$state Potts model \cite{DengBloteNienhuis2004PRE,Nienhuis1982ExactTricrit,NienhuisBerkerRiedelSchick,PatilLudwig2024} ($2\leq Q\leq 4$) which are studied in an epsilon expansion
in Ref. \cite{JengLudwig}.
It is clear from 
{the results of Ref. \cite{JengLudwig} that,}
at least for small enough values of epsilon, the replica limit $R\rightarrow0$ effective
boundary entropy $s_{\text{eff}}^{(R\rightarrow0)}$ for these defect replica field theories, which are of the form in Eq. \eqref{EqDefectReplicaFieldTheoryIntro}, is greater than zero (i.e., 
it {increases}
in going from UV to IR).
}
{Analogous to the above discussion for the effective central charge $c_{\text{eff}}^{(R\rightarrow0)}$ 
for
the replica action in the $R\rightarrow0$ replica limit
{with a relevant equal replica term}, if the corresponding equal replica term 
in the defect replica action 
is relevant, we cannot make the above argument for the effective 
boundary entropy $s_{\text{eff}}^{(R\rightarrow0)}$  {in the $R\rightarrow0$ replica limit}.
However, again 
one can consider the fine-tuned defect replica action,
{which for each $R<1$}
sits on the RG trajectory (critical manifold) of the infrared {\textit{defect}} fixed point
{that}
smoothly deforms in the $R\rightarrow1$ replica limit to the {\textit{defect}} fixed point we get by setting the coupling constant of the equal replica term to zero. 
For these fine-tuned defect replica actions, $s(R)=\ln g(R)$ has a zero at $R=1$, our $g_{\text{eff}}$ theorem in the $R\rightarrow1$ replica limit holds true~\footnote{{provided of course that no other additive operator renormalizations are required.}}, and $s(R)=\ln g(R)$ is differentiable in the interval $R\in[0,1]$ by construction.
Then assuming that the function $s(R)=\ln g(R)$ has zeroes only at $R=0$ and $R=1$ in the interval $R\in [0,1]$, 
we can conclude that the effective
boundary entropy $s_{\text{eff}}^{(R\rightarrow0)}$ for this fine-tuned defect replica action in the replica limit $R\rightarrow0$ is {also} greater than zero, i.e. its UV value.
}
\vskip 0.8cm
\section{Discussion and Outlook\label{LabelDiscuss}}
{In this work, {we have presented two theorems. 
{The first}
of them is the {``$c_{\text{eff}}$ theorem''}, where}
we 
have demonstrated non-perturbatively the decrease of the effective central charge (Eq. \eqref{EqDefinitionOfEffC}),
for 
replica field theories {in the limit of $R\to 1$ replicas,}}
{under the RG flow that arises from}
perturbing a CFT $S_*$ by a RG relevant replica interaction of the form shown in Eq. \eqref{EqDisorderReplicaAction}.
Under the fairly standard assumptions stated at the end of Sec. \ref{Sec:Intro}, we have shown that  
{the $R\rightarrow1$  effective central charge (Eq. \eqref{EqDefinitionOfEffC})}
is always 
{\it less}
than the central charge of the unperturbed 
CFT $S_{*}$ {(i.e.
{\it less}
than the $R\rightarrow1$ effective {central} charge
at the ultraviolet fixed point).}
{As already mentioned, {the} replica field theories we considered in Eq. \eqref{EqDisorderReplicaAction} have been shown in Ref. \cite{NahumJacobsen} to describe  {$2D$} weakly monitored classical critical systems, 
 where measurements introduce a form of quenched randomness via Bayes' theorem. 
 These are classical {statistical mechanics}
 problems
 where measurements update our knowledge of the {unmeasured} statistical ensemble according to Bayes' theorem,
and we are interested in the properties of the resulting conditioned ensemble. 
{Following Ref. \cite{PutzGarrattNishimoriTrebstZhu}, and as we discuss in 
{general terms in}
our App. \ref{AppRKWavefunctionsAndMeasurements}, these 
{monitored}
classical 
problems are equivalent to
corresponding problems of performing {suitable} \textit{quantum} Born-rule measurements 
on 
the
quantum Rokhsar-Kivelson (RK) 
wavefunction~\cite{CLHenley_2004,ArdonneFedleyFradkin,IsakovFendleyLudwigTrebstTroyer} 
corresponding to the unmeasured classical stat-mech system.
As mentioned already, the
effective central charge $c_{\text{eff}}$ defined in Eq. \eqref{EqDefinitionOfEffC}, together with the central charge $c$ of the unmeasured critical system, has been shown~\cite{NahumJacobsen} to characterize the universal finite-size scaling of the 
{Shannon entropy}
of the measurement record in the $2D$ 
monitored classical problem,
where the latter is 
{also}
identical 
to the Shannon entropy of the measurement record calculated with the Born-rule probability distribution in the corresponding problem of quantum measurements on $2D$ quantum RK wavefunctions (see App. \ref{AppRKWavefunctionsAndMeasurements}).
Therefore, our $c_{\text{eff}}$ theorem 
{readily finds applications in}
the problem of 
{two-dimensional}
weakly monitored classical critical systems and {also} the corresponding, equivalent problem of performing weak 
{Born-rule}
measurements on $2D$ critical RK wavefunctions.}
{{It}
would be{, of course,} interesting to see applications of our $c$-effective theorem in the $R\rightarrow1$ replica limit
{to}
other problems where $R\rightarrow1$ replica field theories, although of a different form than Eq. \eqref{EqDisorderReplicaAction}, are known to provide an effective long-wavelength description {including, e.g.,} systems with monitored quantum dynamics {(see e.g. \cite{JianYouVasseurLudwig2019,BaoChoiAltman2019})}, 
classical disordered systems with analogues of the Nishimori line \cite{Nishimori_1980} which are also relevant for problems in quantum error correction {(see e.g. \cite{DennisKitaevLandahlPreskill,ZhuTantivasadakarnVishwanathTrebstVerresen,LeeJiBiFisher})}.
Therefore, the main challenge in applying our theorem
{directly}
to these problems lies in checking if the replica action for any such problem can be shown,
{by a suitable choice of ultraviolet formulation,} to be equivalent to 
{an action of}
the form of Eq. \eqref{EqDisorderReplicaAction}.}
\par
{The second theorem 
that we
presented is the
{``$g_{\text{eff}}$ theorem''}, 
which demonstrates
non-perturbatively the
decrease under the RG flow of the effective 
Affleck-Ludwig~\cite{AffleckLudwig1991PRL}
boundary entropy $s_{\text{eff}}=\ln g_{\text{eff}}$ (Eq. \eqref{EqSeffIntro}) for $2D$ defect replica field theories in the $R\rightarrow1$ replica limit with a relevant defect replica 
{interaction of the form shown in Eq. \eqref{EqDefectReplicaFieldTheoryIntro}.}
{Under fairly standard assumptions which are entirely analogous to those stated at the end of Sec. \ref{Sec:Intro} in the context of
$c_{\rm eff}$,
we have proved this
``$g_{\text{eff}}$ theorem''}
in App. \ref{AppGEffTheorem}.
The 
defect replica field theories Eq. \eqref{EqDefectReplicaFieldTheoryIntro} in the replica limit $R\rightarrow1$ govern the long-distance physics 
of the 
problem of
performing weak measurements on one-dimensional quantum critical ground states (see e.g. \cite{GarrattWeinsteinAltman2022,WeinsteinSajithAltmanGaratt,YangMaoJian,PatilLudwig2024}). 
In particular, as discussed in Ref. \cite{PatilLudwig2024}, the 
boundary entropy $s_{\text{eff}}$ of {the defect replica field theory in the $R\rightarrow1$ replica limit} characterizes the universal finite-size scaling behavior of the Shannon entropy of the measurement record on 
the
one-dimensional quantum critical ground state.}
{Interestingly, our $g_{\text{eff}}$ theorem is equivalent to the statement that the Shannon entropy of the measurement record in the corresponding measurement problem on the one-dimensional critical ground state is a concave function of the length $L$ of the critical ground state (periodic boundary conditions),
{Eq.~\eqref{EqInequalityOnShannon},}
{for large enough values of $L$ where the defect replica field theory description is valid}.
{In particular, we note that both our theorems -- the $c_{\text{eff}}$ theorem and the $g_{\text{eff}}$ theorem -- 
{demonstrate the
decrease
under the RG flow of quantities characterizing}
the
universal information
content in the Shannon entropy of
the measurement record in the respective measurement problems corresponding to the replica
field theories Eq. \eqref{EqDisorderReplicaAction} and \eqref{EqDefectReplicaFieldTheoryIntro}, respectively.}
{It would} be interesting to see if our theorems, can be proved using information-theoretic
tools or ideas.
}
\par
{Finally, we discussed
a possible 
{consequence}
of our $c_{\text{eff}}$ theorem in the $R\rightarrow1$ replica limit, 
{for}
the 
effective central charge $c_{\text{eff}}^{(R\rightarrow0)}$ [Eq. \eqref{EqDisorderReplicaZeroEffC}]
{of}
the same replica field theory in Eq. \eqref{EqDisorderReplicaAction} but in the $R\rightarrow0$ replica limit. 
{As 
discussed, the function {$f(R)=c(R)-R.c$}, which represents
the difference in the central charge of the $R$-copy replica action in Eq. \eqref{EqDisorderReplicaAction} and of the (unperturbed) action $\sum_aS_*^{(a)}$, has two `natural' 
zeroes {at $R=0$ and $R=1$, respectively.}}
If we assume that $f(R)$ does not have any other zeroes in the interval $R\in[0,1]$ (an assumption substantiated by epsilon expansion results in Ref. \cite{LUDWIGCardy,SHIMADA2009707}),
{and that it is 
{a differentiable function}
in this interval}, our $c_{\text{eff}}$ theorem in the replica limit $R\rightarrow1$ then implies that the 
effective central charge $c_{\text{eff}}^{(R\rightarrow0)}$ for the same replica field theory in Eq. \eqref{EqDisorderReplicaAction} but in the $R\rightarrow0$ replica limit is \textit{greater} than the central charge of the unperturbed CFT $S_*$. 
We 
note that to describe classical systems with generic, uncorrelated quenched impurity disorder, replica field theories in Eq. \eqref{EqDisorderReplicaAction} in the replica limit $R\rightarrow0$ 
require 
{in addition}
an additional ``equal replica term".
{However,} if the latter equal replica term is irrelevant in RG sense, one
{can}
apply the argument in Sec. \ref{SecReplicaLimitZeroCEff} 
{to}
the effective central charge $c_{\text{eff}}^{(R\rightarrow0)}$ 
of
the 
replica action {in {the} $R\rightarrow0$ replica limit,}
{which governs}
the classical critical system with generic, uncorrelated impurity-type quenched disorder.
{Then} by implying that the 
effective central charge $c_{\text{eff}}^{(R\rightarrow0)}$ is greater than the already 
{non-vanishing}
central charge of the clean, disorder-free classical critical system, 
{the argument}
guarantees the criticality of the system subjected to generic, uncorrelated quenched impurity-type disorder.
{Despite being based on 
the 
above assumption regarding zeroes of the function 
{$f(R)=c(R)-R.c$,}
we think that this is a surprising 
result.}
{In the case where the equal replica term is relevant, 
{an analogous} argument for the increase of the
effective central charge $c_{\text{eff}}^{(R\rightarrow0)}$ in the replica limit $R\rightarrow0$ from {the  UV to the IR} can be made
{by placing the replica action onto a RG crossover trajectory to a different fixed point, which is}
obtained by tuning the coupling constant of the equal replica term
{so that the system resides on the crossover trajectory (the `critical manifold').}
{We note that} analogous arguments, under analogous assumptions, can also be made for the {increase under the RG flow of the} effective
boundary entropy $s_{\text{eff}}^{(R\rightarrow0)}$ for defect replica field theories in Eq. \eqref{EqDefectReplicaFieldTheoryIntro} in the $R\rightarrow0$ replica limit.
}}
 \vskip 1.5cm
\begin{acknowledgments}
One of us (R.A.P.) thanks Jake Hauser for stimulating discussions on related topics.
\end{acknowledgments}
\vskip .8cm
\appendix
\section{(Re-)Introducing Randomness\label{AppReIntroOfRandom}}
Replica field theories, one example of which is given in Eq. \eqref{EqDisorderReplicaAction}, are usually obtained by using the replica trick and averaging over some form of randomness that can be introduced by, e.g., {generic {(uncorrelated)} quenched ``impurity-type} disorder" \cite{EdwardsAnderson_1975} or {from measurements in (quantum or classical) monitored systems} \cite{JianYouVasseurLudwig2019,BaoChoiAltman2019,NahumJacobsen}.
In particular, it is always possible to trace back the steps of this averaging and to reintroduce {a} ``randomness" field. 
We will now introduce such a  ``randomness" field $h(x)$ to rewrite the {$R\rightarrow1$} replica theory in Eq. \eqref{EqDisorderReplicaAction} as a theory of a single copy 
{of CFT} 
$S_{*}$ interacting with the randomness field $h(x)$.\par 
The path integral for the replica field theory 
{with a number $R$}
of replicas in Eq. \eqref{EqDisorderReplicaAction} is given by
\begin{align}\label{EqReversingToRamdomnessFirstStep}
    Z_{R}&=\int 
    {\left [\prod_{a=1}^{R} D\phi^{(a)}\right ]}\;e^{-\sum_{a}S_{*}^{(a)}+\Delta\int d^2x\sum_{a\neq b}\varphi^{(a)}\varphi^{(b)}}
\end{align}
where the path integral field $\phi$ is a placeholder field for the definition of the path integral \footnote{{The field $\varphi$ appearing in the replica action Eq. \eqref{EqDisorderReplicaAction} is some function of the path integral field $\phi$}}. 
Using 
{the Gaussian integral formula,}
the partition function $Z_R$ can be 
{written, with an implicit 
normalization}~\footnote{{The normalization of the path integral measure $\int Dh(x)$ for field $h(x)$ is determined by requiring that 
$\int Dh(x)\,e^{-\int\,d^2x\frac{h(x)^2}{4\Delta}}=1$.}}
{of the path integral measure
$ Dh(x)$, as}
\begin{align}
    Z_R&=\int Dh(x)e^{-\int d^2x \frac{h^2(x)}{4\Delta}}\times\nonumber\\ &\hspace{-0.75cm}\times\int 
     \prod_{a=1}^{R} D\phi^{(a)}e^{-\sum_{a=1}^{R}\big[S_{*}^{(a)}+\Delta\int d^2x\,(\varphi^{(a)}(x)-\frac{h(x)}{2\Delta})^2-\int d^2x \frac{h^2(x)}{4\Delta}\big]}\\
    &\hspace{-0.75cm}=\int Dh(x)e^{-\int d^2x \frac{h^2(x)}{4\Delta}}\int \prod_{a=1}^{R} D\phi^{(a)}e^{-\sum_{a=1}^{R}\mathcal{S}^{(a)}[\{h(x)\}]},
    \label{LabelEqZR}
\end{align}
{where we have defined} the action $\mathcal{S}[\{h(x)\}]$ for a fixed realization $h(x)$ of randomness
{by}~\footnote{{The action in Eq. \eqref{EqInteractingTheoryWithRandomness} differs by the extra term ``$-\frac{1}{4\Delta}\int d^2x\,{h^2(x)}$" from the action for a fixed realization of measurement outcomes ``$h(x)$" considered in Ref. \cite{NahumJacobsen}. (Also see our App. \ref{AppRKWavefunctionsAndMeasurements} and Eq.~\eqref{EqBornRuleProbRKWFEnergy}.) Due to this extra term, the equal replica term $(R-1)\sum_{a}\varphi^{(a)}\varphi^{(a)}$ that is present in the corresponding replica action in Ref.~\cite{NahumJacobsen} is not present in the corresponding replica action for
the fixed randomness $h(x)$ action $\mathcal{S}[\{h(x)\}]$
in Eq.~\eqref{EqInteractingTheoryWithRandomness}.
However, this difference is inconsequential for our discussion because, as demonstrated in App. \ref{App:VanishingSingleReplicaPerturbation}, the presence of the latter equal replica term $(R-1)\sum_{a}\varphi^{(a)}\varphi^{(a)}$, which clearly vanishes from the replica action in the $R\rightarrow1$ replica limit, does not affect the conclusion of our $c_{\text{eff}}$ theorem in Eq. \eqref{EqCEffectiveTheorem}.}}
\begin{eqnarray}
\label{EqInteractingTheoryWithRandomness}
&&\mathcal{S}[\{h(x)\}]:=
\\ \nonumber
&&
=S_{*}+\Delta\int d^2x \, \bigg(\varphi(x)-
\frac{h(x)}{2\Delta}\bigg)^2-\frac{1}{4\Delta}\int d^2x \, h^2(x).
\end{eqnarray}
{The} actions $\mathcal{S}^{(a)}[\{h(x)\}]$ denote replica copies of the action $\mathcal{S}[\{h(x)\}]$ for replica indices $a=1,\cdots, R$.
Then the replica partition function 
{$Z_{R}$, 
 in the form of the last line of Eq.~\eqref{LabelEqZR},}
{can be written as}
\begin{align}\nonumber
   &Z_R= \\ \nonumber
   =&\int Dh(x)\,e^{-\int d^2x \frac{h^2(x)}{4\Delta}}\int \prod_{a=1}^{R} D\phi^{(a)}e^{-\sum_{a=1}^{R}\mathcal{S}^{(a)}[\{h(x)\}]}\\ \nonumber
    =&\int Dhe^{-\int_x  \frac{h^2}{4\Delta}}\bigg(\ \int D\phi e^{-\mathcal{S}[\{h\}]}\bigg)^R\\ 
    =&\int Dh\int D\phi\,e^{-\int_{x}  \frac{h^2}{4\Delta}-\mathcal{S}[\{h\}]}\bigg(\ \int D\phi e^{-\mathcal{S}[\{h\}]}\bigg)^{R-1},
\label{AnIntermediateEquationforZR}
\end{align}
where in the above equations, for compactness of notation, we have denoted the integrals $\int d^2x$ in the exponentials by just $\int_{x}$, and the fields $\varphi(x)$ and $h(x)$ by just $\varphi$ and $h$, respectively. 
We will now define the following average denoted by the 
{overbar $\overline{(\cdots)}$} symbol
\footnote{{Since $S_{*}$, the scalar field $\varphi(x)$ and the randomness field $h(x)$ are all real valued, the action $\mathcal{S}[\{h(x)\}]$ in Eq. \eqref{EqInteractingTheoryWithRandomness} is also real valued, and hence the averaging weight defined in Eq. \eqref{EqWhatDoesanOverlineMean} is positive.}}
,
\begin{equation}\label{EqWhatDoesanOverlineMean}
    \overline{(\cdots)}:=\frac{1}{Z_*}\int Dh\int D\phi\,e^{-\int_x  \frac{h^2}{4\Delta}-\mathcal{S}[\{h\}]}(\cdots),
\end{equation}
where $Z_*=\int D\phi\, e^{-S_{*}}$ is the partition function of the unperturbed CFT $S_{*}$. 
As discussed in Sec. \ref{Sec:Intro}, analogues of this average denoted by 
{an  overbar $\overline{(\cdots)}$},
where 
{in addition to}
averaging over the randomness $h(x)$ (with some weight of its own) we also have an additional copy of the
{theory $\mathcal{S}[\{h(x)\}]$
 in the averaging}
{weight}, can naturally arise in monitored systems.
In particular, 
\begin{align}
    \overline{\,1\,}&=\frac{1}{Z_*}\int Dh D\phi e^{-\int_x  \frac{h^2}{4\Delta}-\mathcal{S}[\{h\}]}\nonumber\\
    &=\frac{1}{Z_*}\int Dh D\phi e^{-\int_x  \frac{h^2}{4\Delta}
    -\left(S_{*}+\Delta\int_x \,\varphi \varphi-\int_x  h\varphi \right)}\nonumber
    \\
    &=\frac{1}{Z_*}\int  D\phi e^{-\left(S_{*}+\Delta\int_x \,\varphi \varphi\right)+\Delta\int_x \varphi\varphi}=1,
\end{align}
and the replica partition function $Z_{R}$ in Eq. \ref{AnIntermediateEquationforZR} is given by,
\begin{equation}\label{EqReversingToRamdomnessLastStep}
    Z_R=Z_*\times\Bigg(\overline{\Big(\ \int D\phi\, e^{-\mathcal{S}[\{h\}]}\Big)^{R-1}}\Bigg).\end{equation}
Therefore, the $R\rightarrow1$ limit of the replica partition function $Z_{R}$ is
\begin{equation}\label{EqZRisZ}
    \lim_{R\rightarrow1}Z_R=Z_*,
\end{equation}
where $Z_*=\int D\phi\, e^{-S_{*}}$ is the partition function of the unperturbed CFT $S_{*}$.
\par
{Let us now compare correlation functions in the $R\rightarrow1$ replica theory of Eq. \eqref{EqDisorderReplicaAction}, with the correlation functions and products of correlation functions in the theory $\mathcal{S}[\{h(x)\}]$ of Eq. \eqref{EqInteractingTheoryWithRandomness} that are averaged over all realizations of randomness $h(x)$ with the average 
{$\overline{(\cdots)}$}
defined in Eq. \eqref{EqWhatDoesanOverlineMean}.}
For example, consider the following product of correlation functions in the theory $\mathcal{S}[\{h(x)\}]$ of Eq. \eqref{EqInteractingTheoryWithRandomness} for a \textit{fixed} realization of
{the} randomness field $h(x)$,
\begin{align}
    \langle O_1\rangle_{\{h\}}^N&\langle O_2\rangle_{\{h\}}^M=\nonumber\\
    &\bigg(\frac{\int D\phi\,O_1 e^{-\mathcal{S}[\{h(x)\}]}}{\int D\phi\,e^{-\mathcal{S}[\{h(x)\}]}}\bigg)^N \bigg(\frac{\int D\phi\,O_2e^{-\mathcal{S}[\{h(x)\}]}}{\int D\phi\,e^{-\mathcal{S}[\{h(x)\}]}}\bigg)^M,
\end{align}
{where each of $O_1$ and $O_2$ is a field (e.g. $\varphi(r)$) or a product of fields (e.g. $\varphi(r)\varphi(0)$) in the CFT $S_*$.}
Then averaging over the randomness with an extra copy of the theory $\mathcal{S}[\{h(x)\}]$ as given in Eq. \eqref{EqWhatDoesanOverlineMean}, we obtain
\begin{align}
&\overline{\langle O_1\rangle_{\{h\}}^N\langle O_2\rangle_{\{h\}}^M}=\nonumber\\
=&\overline{\bigg(\frac{\int D\phi\,O_1 e^{-\mathcal{S}[\{h(x)\}]}}{\int D\phi\,e^{-\mathcal{S}[\{h(x)\}]}}\bigg)^N \bigg(\frac{\int D\phi\,O_2e^{-\mathcal{S}[\{h(x)\}]}}{\int D\phi\,e^{-\mathcal{S}[\{h(x)\}]}}\bigg)^M}\\
    =&\frac{1}{Z_*}\int Dh \int D\phi e^{-\int_x \frac{h^2}{4\Delta}-\mathcal{S}[\{h(x)\}]}\bigg[\bigg(\frac{\int D\phi\,O_1 e^{-\mathcal{S}[\{h(x)\}]}}{\int D\phi\,e^{-\mathcal{S}[\{h(x)\}]}}\bigg)^N\nonumber\\
    &\;\;\;\;\;\;\;\;\;\;\;\;\;\;\;\;\;\;\;\;\;\;\;\;\;\;\;\;\;\;\;\;\;\;\;\;\;\;\;\;\;\;\;\times\bigg(\frac{\int D\phi\,O_2 e^{-\mathcal{S}[\{h(x)\}]}}{\int D\phi\,e^{-\mathcal{S}[\{h(x)\}]}}\bigg)^M\bigg].    
\end{align}
Then using the replica trick, which involves essentially tracing back the steps from Eq. \eqref{EqReversingToRamdomnessLastStep} to \eqref{EqReversingToRamdomnessFirstStep}, one can see that the correlation function in the above equation can be written as
\begin{align}
    &\overline{\langle O_1\rangle_{\{h\}}^N\langle O_2\rangle_{\{h\}}^M}=
    \nonumber
    \\
    =&\frac{1}{Z_*}\int Dh e^{-\int_x \frac{h^2}{4\Delta}}\bigg[\Big(\int D\phi e^{-\mathcal{S}[\{h(x)\}]}\Big)^{1-N-M}\times\nonumber\\
    &\;\;\;\;\;\;\;\times\Big({\int D\phi\,O_1 e^{-\mathcal{S}[\{h(x)\}]}}\Big)^N\Big({\int D\phi\,O_2 e^{-\mathcal{S}[\{h(x)\}]}}\Big)^M\bigg]  \\
   =&\frac{1}{Z_*}\lim_{R\rightarrow1}\int Dh e^{-\int_x \frac{h^2}{4\Delta}}\bigg[\Big(\int D\phi e^{-\mathcal{S}[\{h(x)\}]}\Big)^{R-N-M}\times\nonumber\\
    &\;\;\;\;\;\;\;\times\Big({\int D\phi\,O_1 e^{-\mathcal{S}[\{h(x)\}]}}\Big)^N\Big({\int D\phi\,O_2 e^{-\mathcal{S}[\{h(x)\}]}}\Big)^M\bigg]  \\
    =&\frac{1}{Z_*}\lim_{R\rightarrow1}\int \prod_{a=1}^{R} D\phi^{(a)} O_1^{(1)}\dots O_1^{(N)}O_2^{(1)}\dots O_2^{(M)}e^{-\mathbb{S}}\label{EqCorrelationInReplicaTheory},
\end{align}
where $\mathbb{S}$ is precisely the replica action given in Eq. \eqref{EqDisorderReplicaAction}, and 
{the fields $O_{1}^{(a)}$ and $O_2^{(a)}$ denote the copies of fields $O_1$ and $O_2$, respectively, in replica copy $a$.}\par
For the purpose of our discussion, we will write Eq. \eqref{EqCorrelationInReplicaTheory} in the following form
\begin{align}
    &\overline{\langle O_1\rangle_{\{h\}}^N\langle O_2\rangle_{\{h\}}^M}=\nonumber\\&\lim_{R\rightarrow1}\frac{Z_{R}}{Z_*}\times \frac{\int \prod_{a=1}^{R} D\phi^{(a)} O_1^{(1)}\dots O_1^{(N)}O_2^{(1)}\dots O_2^{(M)}e^{-\mathbb{S}}}{\int \prod_{a=1}^{R} D\phi^{(a)} e^{-\mathbb{S}}},
\end{align}
where $Z_{R}=\int \prod_{a=1}^{R} D\phi^{(a)} e^{-\mathbb{S}}$ is the partition function of the replica theory
{(Eq.~\eqref{EqReversingToRamdomnessFirstStep}, 
where $\mathbb{S}$ is defined in Eq. \eqref{EqDisorderReplicaAction})}
and $Z_*=\int D\phi \,e^{-S_{*}}$ is the partition function of the unperturbed CFT $S_{*}$.
Therefore,
\begin{equation}
    \overline{\langle O_1\rangle_{\{h\}}^N\langle O_2\rangle_{\{h\}}^M}=\lim_{R\rightarrow1}\frac{Z_{R}}{Z_*}\langle O_1^{(1)}\dots O_1^{(N)}O_2^{(1)}\dots O_2^{(M)}\rangle_R
\end{equation}
where the subscript $R$ on $\langle \dots\rangle_{R}$ on the right hand side of the above equation indicates that the correlation function is evaluated in the $R$-replica theory of Eq. \eqref{EqDisorderReplicaAction}. 
Assuming that the partition function $Z_R$ and the correlation functions of interest are analytic functions of the number of replicas $R$ 
{(at least for small enough values of $R$),}
we obtain
{from Eq. \eqref{EqZRisZ}} that
\begin{equation}\label{EqReplicaTrickCorrelationFunction}
    \langle O_1^{(1)}\dots O_1^{(N)}O_2^{(1)}\dots O_2^{(M)}\rangle_{R=1}=\overline{\langle O_1\rangle_{\{h\}}^N\langle O_2\rangle_{\{h\}}^M}.
\end{equation}
{The above equation then allows us to obtain the equalities in Eq. \eqref{EqReplicaCorrelationFunctionsToRandomnessOnes} of the main text of this paper.}

\vskip .8cm
\section{Additive Renormalization of the Operator \texorpdfstring{$\sum_{a\neq b}\varphi^{(a)}\varphi^{(b)}$}{Lg} and Perturbing the Replica Action with \texorpdfstring{$(R-1)\sum_{a}\chi^{(a)}$}{Lg}\label{App:VanishingSingleReplicaPerturbation}}
The multiple OPE of the 
field $\sum_{a\neq b}\varphi^{(a)}\varphi^{(b)}$ with itself, given by
\begin{equation}\label{EqPhiTimesN}
\big(\underbrace{\sum_{a_1\neq b_1}\varphi^{(a_1)}\varphi^{(b_1)}\big)\times\big(\sum_{a_2\neq b_2}\varphi^{(a_2)}\varphi^{(b_2)}\big)\dots\big(\sum_{a_n\neq b_n}\varphi^{(a_n)}\varphi^{(b_n)}\big)}_{n\;=\;\text{the}\; \# \;\text{of} \;\sum_{a\neq b}\varphi^{(a)}\varphi^{(b)}},
\end{equation}
This then gives us the terms of the form in Eq. \eqref{EqOperatorsInMultipleOPE}, where $\chi_j$s are fields (which 
also includes 
the identity field) in the 
{closed OPE subalgebra generated by $\varphi$.}
Since some of the {fields} $\chi_j$ could be equal to the identity (trivial) field, we conclude that the most general term in the multiple OPE of Eq. \eqref{EqPhiTimesN} is given by Eq. \eqref{EqOperatorsInMultipleOPE} where $\chi_j$ are only the 
{non-identity}
fields in the closed OPE subalgebra of $\varphi$, and we allow $p$ to be any integer between $1$ 
{to}
$2n$.
\begin{equation}\label{EqOperatorsInMultipleOPE}
\sum_{\substack{a_1,a_2,\dots,a_{p}}}^{\substack{\text{all indices are}\\\text{pairwise distinct}}}\chi_1^{(a_1)}\chi_2^{(a_2)}\dots\chi_p^{(a_p)},
\end{equation} 
where 
{$p\in\{1,2,\dots,2n\}$},
$a_{j}\in\{1,2,\dots,R\}$ and $\chi_j$ 
are 
{non-identity}
scaling fields in the closed OPE subalgebra generated by the field $\varphi$
\footnote{I.e. the closed OPE subalgebra {formed by the field $\varphi$,} fields in the multiple OPE of {the field} $\varphi$, 
and {also} the fields in the OPEs of the fields in the multiple OPE of {the field} $\varphi$. }.
The operator $\sum_{a\neq b}\varphi^{(a)}\varphi^{(b)}$ will require additive renormalization if there exists a field  with scaling dimension $X$ in Eq. \eqref{EqOperatorsInMultipleOPE} that appears in the multiple OPE of Eq. \eqref{EqPhiTimesN} such that \cite{CardyLesHouche} {(also see the discussion in Ref. \cite{ZamolodchikovIntegrable})},
\begin{equation}\label{EqRenormalizability}
    X\leq 2X_{\varphi}-(n-1)(2-2X_{\varphi})<2X_{\varphi},
\end{equation}
where $X_{\varphi}$ is the scaling dimension of the field $\varphi$,
{and $n$ is the number of $\sum_{a\neq b}\varphi^{(a)}\varphi^{(b)}$ operators in the OPE of Eq. \eqref{EqPhiTimesN}.}
{Since $n\geq2$, we have used Eq. \eqref{EqPerturbationIsRelevant} to obtain the last inequality of Eq. \eqref{EqRenormalizability}.}
If the field $\varphi$ in a single copy of CFT $S_{*}$ is the most relevant operator in the closed OPE subalgebra generated by itself, it is clear that the two and higher replica index operators of the form in Eq. \eqref{EqOperatorsInMultipleOPE} with $p\geq2$ \textit{cannot} satisfy Eq. \eqref{EqRenormalizability}, 
 hence do not cause any new ultraviolet {(UV)} divergences that require {additive} renormalization. 
The only operators that can possibly satisfy Eq. \eqref{EqRenormalizability} are single replica operators of {the} form $\sum_{a}\chi^{(a)}$ for some scaling field $\chi$ in the closed OPE subalgebra of the field $\varphi$. 
However, it can be verified that all single replica fields of form $\sum_{a}\chi^{(a)}$ {occur in the multiple OPE of Eq. \eqref{EqPhiTimesN}} with a 
{prefactor of
$(R-1)$} or {of} a combinatorial factor $ \binom{R-1}{k}$ for {some} integer $k$ \cite{PatilLudwig20252}.  
{The argument for this is analogous to the one in App. E of Ref. \cite{PatilLudwig2024}, which makes this argument for the case where $S_{*}$ is the Ising CFT and the field $\varphi$ is the spin field of the Ising CFT.
Therefore, in the replica limit $R\rightarrow1$ the terms of form $\sum_{a}\chi^{(a)}$ disappear from the OPE of Eq. \eqref{EqPhiTimesN} and are not required as counter terms to the replica theory of Eq. \eqref{EqDisorderReplicaAction} in the replica limit $R\rightarrow1$.}
\par 
{However, one can wonder if the presence of a counter term of the following form in the replica action of Eq. \eqref{EqDisorderReplicaAction} could impact the conclusion of our theorem in Eq. \eqref{EqCEffectiveTheorem}
\begin{align}\label{EqSingleReplicaPerturbation}
    &(R-1)g\int d^2x\sum_{a}\chi^{(a)}(x),
\end{align}
here $\chi$ is a scaling field in the closed OPE subalgebra of the field $\varphi$, 
and $g$ is a finite, non-vanishing coupling constant in the replica limit $R\rightarrow1$}.
Apart from the technicalities of additive renormalization and {the UV divergences in the replica field theory}, the question about the presence of the term in Eq. \eqref{EqSingleReplicaPerturbation} in the replica action of Eq. \eqref{EqDisorderReplicaAction} is also of importance for applications to monitored classical critical systems.
For certain measurement protocols, the replica actions for monitored classical systems {studied in} Ref. \cite{NahumJacobsen} can contain 
{the term}
$(R-1)\sum_{a}\varphi^{(a)}(x)\varphi^{(a)}(x)$ (see footnote \cite{Note2}), which is of the form in Eq. \eqref{EqSingleReplicaPerturbation} after using point splitting and the OPE of {the} field $\varphi$ with itself. 
We will {now} show
that the theorem in Eq. \eqref{EqCEffectiveTheorem} is agnostic to the presence of {the} term in Eq. \eqref{EqSingleReplicaPerturbation} in the replica action.
In fact, we will show that the infrared $c_{\text{eff}}$ (Eq. \eqref{EqDefinitionOfEffC}) for the replica action in Eq. \eqref{EqDisorderReplicaAction} is the same as that of the replica action in Eq. \eqref{EqDisorderReplicaAction} perturbed by the term in Eq. \eqref{EqSingleReplicaPerturbation}.
\par
The replica action in Eq. \eqref{EqDisorderReplicaAction} with the term in Eq. \eqref{EqSingleReplicaPerturbation} is given by,
\begin{align}\label{EqModifiedReplicaAction}
    -\mathbb{S}=-\sum_{a=1}^{R}S_{*}^{(a)}+\Delta\int d^2 x \sum_{\substack{a,b=1\\ a\neq b}}^{R}\varphi^{(a)}(x)\varphi^{(b)}(x)\nonumber\\+(R-1)g \int d^2x\sum_{a}\chi^{(a)}(x),
\end{align}
and {assuming the action (Eq. \eqref{EqModifiedReplicaAction}) now satisfies the first assumption in 
the list 
at the end of Sec. \ref{Sec:Intro},} 
the trace of the stress-energy tensor in Eq. \eqref{EqTraceOfTheStressEnergyTensor} is then given by,
\begin{align}
    \Theta(x)=-2\pi\Delta(2-2X_{\varphi})\sum_{\substack{a,b=1\\ a\neq b}}^{R}\varphi^{(a)}(x)\varphi^{(b)}(x)+\nonumber\\-2\pi(R-1)g(2-X_{\chi})\sum_{a}\chi^{(a)}(x),
\end{align}
where $X_{\chi}$ is the scaling dimension of the field $\chi$ in the unperturbed CFT $S_{*}$.
Then, we can consider applying the 
{differential equation}
in Eq. \eqref{EqCtheorem} to the theory in Eq. \eqref{EqModifiedReplicaAction}.
In addition to the terms already present in Eq. \eqref{EqSumRuleAppliedToReplicaTheory}, 
we have to consider the following terms which will contribute to the sum rule in Eq. \eqref{EqSumRuleAppliedToReplicaTheory} for the new replica theory in Eq. \eqref{EqModifiedReplicaAction},
\begin{subequations}
\begin{align}
&(R-1)\times\langle \sum_{\substack{a,b=1\\ a\neq b}}^{R}\varphi^{(a)}(0)\varphi^{(b)}(0)\rangle_R\langle\sum_{c=1}^{R}\chi^{(c)}(0)\rangle_{R}\label{EqB1}\\
&(R-1)^2\times\langle\sum_{c=1}^{R}\chi^{(c)}(0)\rangle_{R}\langle\sum_{c=1}^{R}\chi^{(c)}(0)\rangle_{R}\label{EqB2}\\
&(R-1)^2\times\langle \sum_{c=1}^{R}\chi^{(c)}(r)\sum_{c=1}^{R}\chi^{(c)}(0)\rangle_{R}\label{EqB3}\\
&(R-1)\times\langle \sum_{\substack{a,b=1\\ a\neq b}}^{R}\varphi^{(a)}(r)\varphi^{(b)}(r)\sum_{c=1}^{R}\chi^{(c)}(0)\rangle_{R} \label{EqB4}\\
&(R-1)\times\langle \sum_{\substack{a,b=1\\ a\neq b}}^{R}\varphi^{(a)}(0)\varphi^{(b)}(0)\sum_{c=1}^{R}\chi^{(c)}(r)\rangle_{R}\label{EqB5}
\end{align}
\end{subequations}
The terms in the above equation are listed with the pre-factors of $(R-1)$ that will appear with them in Eq. \eqref{EqSumRuleAppliedToReplicaTheory}, and other prefactors 
{accompanying}
these terms
which are non-vanishing and finite in {the} replica limit $R\rightarrow1$ will not be important for our discussion. 
{Using permutation symmetry of the replica indices, we obtain that} 
\begin{align}
   \langle \sum_{\substack{a,b=1\\ a\neq b}}^{R}\varphi^{(a)}(0)\varphi^{(b)}(0)\rangle_R=R(R-1)\langle\varphi^{(1)}(0)\varphi^{(2)}(0)\rangle_{R},
\end{align}
{and thus}
in the replica limit $R=1$ the expressions in Eq. \eqref{EqB1}, \eqref{EqB2} and \eqref{EqB3} are $\mathcal{O}((R-1)^2)$.
The term in Eq. \eqref{EqB4} can be also {simplified using permutation symmetry of the replica indices} as 
follows
\begin{align}\label{EqNontrivialR=1}
    &(R-1)\times\langle \sum_{\substack{a,b=1\\ a\neq b}}^{R}\varphi^{(a)}(r)\varphi^{(b)}(r)\sum_{c=1}^{R}\chi^{(c)}(0)\rangle_{R}=\nonumber\\
    &=(R-1)(R(R-1)(R-2)\langle\varphi^{(1)}(r)\varphi^{(2)}(r)\chi^{(3)}(0)\rangle_R+\nonumber\\
    &\;\;\;\;\;\;\;\;\;\;\;\;\;\;\;\;\;\;\;\;\;\;\;\;+2R(R-1)\langle\varphi^{(1)}(r)\varphi^{(2)}(r)\chi^{(2)}(0)\rangle_R)\nonumber\\
    &=(R-1)^2(R(R-2)\langle\varphi^{(1)}(r)\varphi^{(2)}(r)\chi^{(3)}(0)\rangle_R+\nonumber\\
    &\;\;\;\;\;\;\;\;\;\;\;\;\;\;\;\;\;\;\;\;\;\;\;\;+2R\langle\varphi^{(1)}(r)\varphi^{(2)}(r)\chi^{(2)}(0)\rangle_R).  
\end{align}
Since the correlation functions in Eq. \eqref{EqB4} and \eqref{EqB5} are related by translational invariance {in the replica field theory}, the above equation implies that both the expressions in Eq. \eqref{EqB4} and \eqref{EqB5} are also $\mathcal{O}((R-1)^2)$ in the replica limit $R\rightarrow1$.
{Therefore, all the ``extra" contributions (Eq. \eqref{EqB1}-\eqref{EqB5}) to the 
{differential equation}
in Eq. \eqref{EqSumRuleAppliedToReplicaTheory} that one gets due to the extra term $(R-1)\sum_{a}\chi^{(a)}$ in the replica action of Eq. \eqref{EqModifiedReplicaAction} are $\mathcal{O}((R-1)^2)$, and hence they can be absorbed in the $\mathcal{O}((R-1)^2)$ corrections in Eq. \eqref{EqReplicaSumRuleSimplified}.
Therefore,
the correlation functions that contribute to the calculation of 
{the `effective' 
$C$-function
$\frac{dC_R(r)}{dR}|_{R=1}$
}
for the replica action in Eq. \eqref{EqModifiedReplicaAction} are 
 the same as 
 those
 for the replica action of Eq. \eqref{EqDisorderReplicaAction}, i.e. those in Eq. \eqref{EqReplicaSumRuleSimplified}.
 In particular, since the replica 
 {field theories}
 in Eq. \eqref{EqDisorderReplicaAction} and Eq. \eqref{EqModifiedReplicaAction} are identical in the replica limit $R\rightarrow1$, the three correlation functions in Eq. \eqref{EqReplicaSumRuleSimplified}  $\langle \varphi^{(1)}(r)\varphi^{(1)}(0)\varphi^{(2)}(r)\varphi^{(2)}(0)\rangle_{R=1}$, $\langle\varphi^{(1)}(r)\varphi^{(2)}(r)\varphi^{(3)}(0)\varphi^{(4)}(0)\rangle_{R=1}$ and $\langle \varphi^{1}(r)\varphi^{(1)}(0)\varphi^{(2)}(r)\varphi^{(3)}(0)\rangle_{R=1}$ are also identical in the two replica field theories, therefore 
{the `effective' $C$-function 
$\frac{dC_R(r)}{dR}|_{R=1}$
}
which is obtained by taking the first derivative of the
{differential equation}
in Eq. \eqref{EqReplicaSumRuleSimplified} w.r.t. $R$ at $R=1$, is also the same for the two replica theories in Eq. \eqref{EqDisorderReplicaAction} and Eq. \eqref{EqModifiedReplicaAction}. 
 Finally, Eq. 
 {\eqref{EqLimitsOfCEfffunction} together with Eq. \eqref{EqCEffectiveTheorem}}
 then implies that the $c_{\text{eff}}$ for the replica action of Eq. \eqref{EqModifiedReplicaAction} is also less than the central charge $c$ of the unperturbed CFT $S_{*}$.}

 \vskip .8cm

\section{Conformal Quantum Criticality and ``Rokhsar-Kivelson'' formulation\label{AppRKWavefunctionsAndMeasurements}}

{In this Appendix, we present a general formulation of performing {\it quantum} {Born rule} measurements on the Rokhsar-Kivelson wave
function~\cite{CLHenley_2004,ArdonneFedleyFradkin,IsakovFendleyLudwigTrebstTroyer}
obtained from a classical partition function.
This yields the same 
{replica theory in the $R\rightarrow1$ limit}{, and the same expression}
for measurement-averaged {moments of}
observables, Eq.~\eqref{EqSimplifiedMeasurementAveragedMomentsFinal},
as 
{that}
presented in Ref.~\onlinecite{NahumJacobsen} using the Baysian formulation.}
\subsection{Setup}

Consider a general Landau{-Ginzburg}
lattice Boltzmann weight 
\begin{eqnarray}
\label{LabelEqBoltzmannWeightLatt}
w[\{s_x\}]= \exp\left ( - {\cal H}[\{s_x\}]\right )
\end{eqnarray}
of classical statistical mechanics, 
where $x$ labels positions on a $d$-dimensional lattice, and $s_x$ are the classical degrees of freedom at lattice site $x$ (e.g., Ising-, Potts-, Heisenberg model spins, ...). 
For every configuration 
$\{s_x\}$ of the classical variables we associate an element 
\begin{eqnarray}
\label{LabelEqONbasis}
|\{s_x\}\rangle
\end{eqnarray}
of an orthonormal basis of a Hilbert space 
$$
\langle \{s_x\}|\{{s'}_x\}\rangle = \prod_x \delta_{s_x, {s'}_x}.
$$
We furthermore define mutually commuting~\footnote{Hermitian, or normal {(if eigenvalues are complex)}} operators
${\hat s}_x$ at all lattice sites, of which the states in Eq.~\eqref{LabelEqONbasis} are simultaneous eigenstates,
$$
{\hat s}_{x'} |\{s_x\}\rangle
=
s_{x'} |\{s_x\}\rangle.
$$
The (unnormalized) quantum state associated with the Boltzmann weight in Eq.~\eqref{LabelEqBoltzmannWeightLatt} is defined to be~\cite{CLHenley_2004,ArdonneFedleyFradkin,IsakovFendleyLudwigTrebstTroyer}
\begin{eqnarray}
\nonumber
|{\tilde \Psi}\rangle :=
\sum_{\{s_x\}} \  \sqrt{w[\{s_x\}]} \ 
|\{s_x\}\rangle
\end{eqnarray}
whose squared norm is the classical partition function
\begin{eqnarray}
\label{LabelEqClassicalPartitionFunctionLatt}
Z = 
\langle{\tilde \Psi}|{\tilde \Psi}\rangle
=\sum_{\{s_x\}}
\exp\left ( - {\cal H}[\{s_x\}]\right ).
\end{eqnarray}
The normalized Rohksar-Kivelson (RK)  quantum state is thus
\begin{eqnarray}
\label{LabelEqNormalizedRKstateLatt}
|\Psi\rangle=
\sum_{\{s_x\}}
\Psi[\{s_x\}] \ \ |\{s_x\}\rangle
\end{eqnarray}
where
\begin{eqnarray}
\label{LabelEqRKWavefunctionLatt}
\Psi[\{s_x\}] =
\frac{1}{\sqrt{Z}}
\exp\left ( - \frac{1}{2} {\cal H}[\{s_x\}]\right )
\end{eqnarray}
is the (normalized) RK wave function.

\subsection{Weak measurements on RK wave function\label{Labelsubsubsectionweaklattice}}

\noindent{\it (i): Order Parameter Measurements}
\vskip .1cm

\noindent We start by first considering {weak} measurement of the  operator ${\hat s}_x$.  At position $x$ we define the following Kraus operator
\begin{eqnarray}
\label{LabelEqDefKrausOp}
{\hat K}_{m_x} := 
\exp\left ( -{\frac{\Delta}{2}} \left({\hat s}_x - m_x\right )^2\right ).
\end{eqnarray}
{Here we discuss the real case, where $-\infty < m_x < \infty$. [The 
case 
 of complex measurement outcomes
$m_x$ of normal (as opposed to Hermitian) operators ${\hat s}_x$
follows analogously.]}
Upon a suitable choice of the normalization constant ${\cal N}$, 
the so-defined Kraus operator satisfies the
POVM~\footnote{{Positive Operator Valued Measure}}
condition at
{any}
site $x'$, as we can make shift of the integration variable $m_{x'}$ by the eigenvalue
$s_{x'}$ obtained by action {of ${\hat s}_{x'}$} on a basis state,
\begin{eqnarray}
\nonumber
&&
\int_{-\infty}^{+\infty}
\frac{d m_{x'}}{{\cal N}}
\ {\hat K}^\dagger_{m_{x'}} {\hat K}_{m_{x'}} \ \ |\{s_x\}\rangle = \\
\nonumber
&&
=
\int_{-\infty}^{+\infty}
 \frac{d m_{x'}}{{\cal N}}
 \exp\left(-
 {\Delta} (s_{x'} - m_{x'})^2
 \right)
   |\{s_x\}\rangle = 
   \\  \nonumber
   &&
   =
\int_{-\infty}^{+\infty}
\frac{d m_{x'}}{{\cal N}}
 \exp\left(-{\Delta} m_{x'}^2 
 \right)
   |\{s_x\}\rangle =
   \\ 
\label{LabelEqOnsitePOVMlattice}
&&=|\{s_x\}\rangle.
\end{eqnarray}
{That is, at each site $x'$
\begin{eqnarray}
    \Rightarrow \;
\int_{-\infty}^{+\infty}
\frac{d m_{x'}}{{\cal N}}
\ {\hat K}^\dagger_{m_{x'}}{\hat K}_{m_{x'}}={\hat {\bf 1}},
\end{eqnarray}
where ${\hat {\bf 1}}$ is the identity operator.}
The  Kraus operator for the entire system (i.e. for measurements at all positions $x$), defined by
\begin{eqnarray}
{\hat K}_{\vec m}
:=
\exp\left ( -{\frac{\Delta}{2}}\sum_x \left({\hat s}_x - m_x\right )^2\right ),\label{EqLabelFullKrusOperatorOP}
\end{eqnarray}
where we introduced the abbreviation ${\vec m}=$ $\{m_x\}$,
satisfies the POVM condition
\begin{eqnarray}
\label{LabelEqPOVMspinx}
\int_{\vec m}
{\hat K}_{\vec m}^\dagger{\hat K}_{\vec m}
 = {\hat {\bf 1}},
\end{eqnarray}
in view of Eq.~\eqref{LabelEqOnsitePOVMlattice}, with the integration measure defined to be
\begin{eqnarray}
\label{LabelEqMeasureDm}
\int_{\vec m}
\ := 
\prod_{x} \left [\int_{-\infty}^\infty 
\frac{d m_{x'}}{{\cal N}}
\right ].
\end{eqnarray}
We can now define the Born-rule probability
\begin{eqnarray}
\nonumber
&& P[{\vec m}]
:={\bra{\Psi}{\hat K}^\dagger_{\vec m}{\hat K}_{\vec m}\ket{\Psi}}=\text{Tr}_{\{s_x\}} \left (
{\hat K}_{\vec m}\  |\Psi\rangle \langle\Psi| \  
{\hat K}^\dagger_{\vec m}
\right)
\\ 
\label{EqBornRuleProbRKWF}
&&
\qquad\;\;\;=\frac{1}{Z}
\sum_{\{s_x\}}
 \exp - {\cal H}[\{s_x\}; \{m_x\}]
 \\ \nonumber
&& {\rm where}
 \qquad \qquad 
 \\ \nonumber
 &&
 {\cal H}[\{s_x\}; \{m_x\}] :=
  {\cal H}[\{s_x\}]
  +{\Delta}\sum_x \left(s_x - m_x\right )^2,
  \qquad
\end{eqnarray}
which is normalized
in view of Eq.~\eqref{LabelEqPOVMspinx}.
{The Born-rule probability 
in Eq. \eqref{EqBornRuleProbRKWF} 
{for}
the measurement outcomes $\vec{m}$
{of}
quantum 
measurements 
on the RK wavefunction is equal to the probability 
(up to trivial factors) of obtaining the \textit{same} measurement outcomes $\vec{m}$ in the corresponding classical monitored problem~\cite{NahumJacobsen}
with classical measurements on the corresponding classical stat-mech model in Eq. \ref{LabelEqBoltzmannWeightLatt}. 
{[As mentioned already, in the latter problem}
classical measurements update our knowledge of the classical stat-mech ensemble via Bayes' theorem.] 
As an immediate consequence of this,
the Shannon entropy of the measurement record
\begin{equation}
    S=-\int_{\vec{m}}P[\vec{m}]\ln P[\vec{m}]
\end{equation}
in the two problems is also identical~\cite{NahumJacobsen} (up to trivial factors which are non-universal).
}
{Coming back to Eq. \eqref{EqBornRuleProbRKWF}, the state obtained upon measurements and corresponding to outcomes $\vec{m}$
is given by
\begin{equation}\label{EqStateObtainedUponMeasurement}
\ket{\Psi_{\vec m}}
=
{{\hat {K}}_{\vec m} {\ket 0}
\over
\sqrt{{\bra 0}
{\hat { K}}^\dagger_{\vec m}
 {\hat { K}}_{\vec m}
{\ket 0}
}
}={{\hat {K}}_{\vec m} {\ket 0}
\over
\sqrt{P[m]}
}{.}
\end{equation}
}

\vskip .4cm
\noindent{\it (ii): Measurements of general operators}
\vskip .1cm

\noindent 
{We will be interested in the case of}
operators 
{which}
are built from the operators ${\hat s}_x$ and thus all mutually commute, and  are diagonal in the basis of Eq.~\eqref{LabelEqONbasis}.
We can measure any of 
{these operators}
using analogous Kraus operators {as that in Eq. \eqref{LabelEqDefKrausOp}}. As an example, consider the local  ``energy operator'' located on a 
{link}
{$\ell$}
joining two nearest neighbor lattice sites 
{$\ell
=\langle x', x'' \rangle$}
\begin{eqnarray}
\label{LabelEqDEFenergyOperator}
&&{\hat {\mathfrak{e}}}_{\ell}
:= {\hat s}_{x'} {\hat s}_{x''}, 
\qquad  \qquad   ({\ell}=\langle x', x'' \rangle = {\rm link}) \   \ \ 
\\ \nonumber
&&
{\hat {\mathfrak{e}}}_{\ell} \  |\{s_x\}\rangle=
 {\mathfrak{e}}_{\ell}   \   |\{s_x\}\rangle
 \qquad
\qquad \qquad
\end{eqnarray}
The  Kraus operator for the entire system defined analogously {(compare Eq. \eqref{EqLabelFullKrusOperatorOP})}, while now denoting 
${\vec {\mathfrak m}}=$ $\{ {\mathfrak{m}}_{\ell}\}$, by
\begin{eqnarray}
\label{EqKrausOperatorEnergy}
{\hat K}_{\vec {\mathfrak{m}}}
:=
\exp\left ( -{\frac{\Delta}{2}}\sum_{\ell}
\left({\hat {\mathfrak{e}}}_{\ell} 
-
{\mathfrak{m}}_{\ell}\right )^2\right ),
\end{eqnarray}
satisfies again the POVM condition
\begin{eqnarray}
\label{LabelEqPOVMenergyb}
\int_{{\vec {\mathfrak m}}}{\hat K}_{\vec {\mathfrak{m}}}^\dagger
{\hat K}_{\vec {\mathfrak{m}}}
 = {\hat {\bf 1}}.
\end{eqnarray}
We can now  analogously define the Born-rule probability
\begin{eqnarray}
\nonumber
&& P[{\vec m}]
:={\bra{\Psi}{\hat K}^\dagger_{\vec m}{\hat K}_{\vec m}\ket{\Psi}}=\text{Tr}_{\{s_x\}} \left (
{\hat K}_{\vec m}\  |\Psi\rangle \langle\Psi| \  
{\hat K}^\dagger_{\vec m}
\right)
\\ 
\label{EqBornRuleProbRKWFEnergy}
&&
\qquad\;\;=
{1\over Z}
\sum_{\{s_x\}}
 \exp - {\cal H}[\{s_x\};
 \{\mathfrak{m}_{\ell}\}]
 \\ \nonumber
 &&{\rm where}
 \\ \nonumber
 &&
 {\cal H}[\{s_x\};
 \{{\mathfrak{m}_{\ell}}\}] :=
  {\cal H}[\{s_x\}]
  + {\Delta}\sum_{\ell} 
  \left({\mathfrak{e}}_{\ell} - 
  {\mathfrak{m}}_{\ell}\right )^2,
\end{eqnarray}
which is normalized
in view of Eq.~\eqref{LabelEqPOVMenergyb}. 
{The state obtained upon measurements and corresponding to measurement outcomes $\vec{\mathfrak{m}}$ is again given by Eq. \eqref{EqStateObtainedUponMeasurement} with Kraus operator in Eq. \eqref{EqKrausOperatorEnergy}.
Again, the Born-rule probability in Eq. \eqref{EqBornRuleProbRKWFEnergy} is the same as the probability of obtaining the same measurement outcomes $\vec{\mathfrak{m}}$ in the corresponding classical monitored system~\cite{NahumJacobsen} (up to trivial factors); and, up to trivial, non-universal constants, the Shannon entropy of the measurement record in the two problems is also identical.
}

\subsection{Expectation values of observables}

The following discussion proceeds in complete analogy to that in Ref.~\cite{PatilLudwig2024}. Let  
${\hat {\cal O}}_1$, ${\hat {\cal O}}_2$, ...,
${\hat {\cal O}}_N$  be a set of operators or products of operators of the RK quantum mechanics. 
{We will be interested in the case where each operator
${\hat {\cal O}}_1$, ${\hat {\cal O}}_2$, ...,
${\hat {\cal O}}_N$
is}
built from products of the commuting operators ${\hat s}_x$, and thus all commute with any  Kraus operators.

We now pick a particular measurement operator, for example 
the 
${\hat {\mathfrak{e}}}_\ell$ 
in Eq.~\eqref{LabelEqDEFenergyOperator}
and denote the set of measurement {outcomes}
{on all links $\ell$}
simply by $\vec{\mathfrak{m}}$.
The
{Born rule} average of a product of
ground state expectation values over measurement outcomes, denoted by an overbar, then reads
\begin{eqnarray}
\label{LabelEqOperatorExpectationValues}
&&
\overline{
\langle {\hat {\cal O}}_1\rangle_{\vec{\mathfrak{m}}}
\langle {\hat {\cal O}}_2\rangle_{\vec{\mathfrak{m}}}
...
\langle {\hat {\cal O}}_N\rangle_{\vec{\mathfrak{m}}}
}=
\\ \nonumber
&&
=
\int_{{\vec {\mathfrak m}}}
P[\vec{\mathfrak{m}}] \ 
{
\langle \Psi|
{\hat K}^\dagger_{\vec{\mathfrak{m}}} {\hat {\cal O}}_1 {\hat K}_{\vec{\mathfrak{m}}}
|\Psi\rangle ...
\langle \Psi|
{\hat K}^\dagger_{\vec{\mathfrak{m}}} {\hat {\cal O}}_N  {\hat K}_{\vec{\mathfrak{m}}}
|\Psi\rangle
\over (P[\vec{\mathfrak{m}}])^N
}\nonumber\\
&&=\nonumber
\int_{{\vec {\mathfrak m}}}
P[\vec{\mathfrak{m}}]^{1-N} \ 
\langle \Psi|
{\hat K}^\dagger_{\vec{\mathfrak{m}}} {\hat {\cal O}}_1 {\hat K}_{\vec{\mathfrak{m}}}
|\Psi\rangle ...
\langle \Psi|
{\hat K}^\dagger_{\vec{\mathfrak{m}}} {\hat {\cal O}}_N  {\hat K}_{\vec{\mathfrak{m}}}
|\Psi\rangle
\\
&&=\nonumber
\int_{{\vec {\mathfrak m}}}
\langle \Psi|
{\hat K}^\dagger_{\vec{\mathfrak{m}}}{\hat K}_{\vec{\mathfrak{m}}}
|\Psi\rangle^{1-N} \ 
\langle \Psi|
{\hat K}^\dagger_{\vec{\mathfrak{m}}}  {\hat K}_{\vec{\mathfrak{m}}}{\hat {\cal O}}_1
|\Psi\rangle\times\\
\nonumber
&&\;\;\;\;\;\;\;\;\;\;\;\;\;\;\;\times\langle \Psi|
 {\hat K}^\dagger_{\vec{\mathfrak{m}}} {\hat K}_{\vec{\mathfrak{m}}}{\hat {\cal O}}_2
|\Psi\rangle\cdots
\langle \Psi|
 {\hat K}^\dagger_{\vec{\mathfrak{m}}}  {\hat K}_{\vec{\mathfrak{m}}} {\hat {\cal O}}_N 
|\Psi\rangle\\
\label{EqMeasurementAveragedCorrFnPenultimateStep}\\
&&\nonumber=
\int_{{\vec {\mathfrak m}}}
\bigg(\frac{\sum_{\{s_x\}}\exp{\{-{\cal H}[\{s_x\}]-{\Delta}\sum_{\ell}\left({\mathfrak{e}}_{\ell} -{\mathfrak{m}}_{\ell}\right )^2\}}}{Z}\bigg)^{1-N}\nonumber\\
&&\times \bigg(\frac{\sum_{\{s_x\}}\mathcal{O}_1\exp{\{-{\cal H}[\{s_x\}]-{\Delta}\sum_{\ell}\left({\mathfrak{e}}_{\ell} -{\mathfrak{m}}_{\ell}\right )^2\}}}{Z}\bigg)\times\nonumber\\
&&\cdots \bigg(\frac{\sum_{\{s_x\}}\mathcal{O}_N\exp{\{-{\cal H}[\{s_x\}]-{\Delta}\sum_{\ell}\left({\mathfrak{e}}_{\ell} -{\mathfrak{m}}_{\ell}\right )^2\}}}{Z}\bigg)\nonumber
\\
\end{eqnarray}
where in the above equation we have used Eq. \eqref{EqBornRuleProbRKWFEnergy} and eigenvalues ${\mathcal{O}}_{k}$  of operators $\hat{\mathcal{O}}_{k}$ in the diagonal basis $\ket{\{s_x\}}$. Simplifying the above equation, we obtain the following
\begin{eqnarray}
&&\overline{
\langle {\hat {\cal O}}_1\rangle_{\vec{\mathfrak{m}}}
\langle {\hat {\cal O}}_2\rangle_{\vec{\mathfrak{m}}}
...
\langle {\hat {\cal O}}_N\rangle_{\vec{\mathfrak{m}}}
}=\\
&&=\frac{1}{Z}
\int_{{\vec {\mathfrak m}}}e^{-\Delta\sum_{\ell}\mathfrak{m}_\ell^2}\times\nonumber\\
&&\times \bigg({\sum_{\{s_x\}}\exp{\{-{\cal H}[\{s_x\}]-{\Delta}\sum_{\ell}{(\mathfrak{e}}_{\ell}^2 -2{\mathfrak{e}}_{\ell}{\mathfrak{m}}_{\ell})\}}}\bigg)^{1-N}\times\nonumber\\
&&\times \bigg({\sum_{\{s_x\}}\mathcal{O}_1\exp{\{-{\cal H}[\{s_x\}]-{\Delta}\sum_{\ell}{(\mathfrak{e}}_{\ell}^2 -2{\mathfrak{e}}_{\ell}{\mathfrak{m}}_{\ell})\}}}\bigg)\times\nonumber\\
&&\cdots \bigg({\sum_{\{s_x\}}\mathcal{O}_N\exp{\{-{\cal H}[\{s_x\}]-{\Delta}\sum_{\ell}{(\mathfrak{e}}_{\ell}^2 -2{\mathfrak{e}}_{\ell}{\mathfrak{m}}_{\ell})\}}}\bigg)\nonumber
\\
&&=\lim_{R\rightarrow1}\frac{1}{Z}
\int_{{\vec {\mathfrak m}}}e^{-\Delta\sum_{\ell}\mathfrak{m}_\ell^2}\times\nonumber\\
&&\times \bigg({\sum_{\{s_x\}}\exp{\{-{\cal H}[\{s_x\}]-{\Delta}\sum_{\ell}{(\mathfrak{e}}_{\ell}^2 -2{\mathfrak{e}}_{\ell}{\mathfrak{m}}_{\ell})\}}}\bigg)^{R-N}\times\nonumber\\
&&\times \bigg({\sum_{\{s_x\}}\mathcal{O}_1\exp{\{-{\cal H}[\{s_x\}]-{\Delta}\sum_{\ell}{(\mathfrak{e}}_{\ell}^2 -2{\mathfrak{e}}_{\ell}{\mathfrak{m}}_{\ell})\}}}\bigg)\times\nonumber\\
&&\cdots \bigg({\sum_{\{s_x\}}\mathcal{O}_N\exp{\{-{\cal H}[\{s_x\}]-{\Delta}\sum_{\ell}{(\mathfrak{e}}_{\ell}^2 -2{\mathfrak{e}}_{\ell}{\mathfrak{m}}_{\ell})\}}}\bigg)\nonumber
\\
&&=\lim_{R\rightarrow1}\frac{1}{Z}
\int_{{\vec {\mathfrak m}}}e^{-\Delta\sum_{\ell}\mathfrak{m}_\ell^2}\times\nonumber\\
&&\times \bigg(\sum_{\{s_x^{(a)}\}_{a=1}^{R}}\mathcal{O}^{(1)}_1\mathcal{O}^{(2)}_2\cdots \mathcal{O}^{(N)}_N\exp{\bigg\{-\sum_{a=1}^{R}{\cal H}[\{s_x^{(a)}\}]}+\nonumber\\
&&\qquad\qquad\qquad\qquad\qquad-{\Delta}\sum_{\ell}\sum_{a=1}^{R}{((\mathfrak{e}}^{(a)}_{\ell})^2 -2\mathfrak{e}^{(a)}_{\ell}{\mathfrak{m}}_{\ell})\bigg\}\bigg)\nonumber\\
\label{EqMeasurementAveragedMomentsSimplified}
\end{eqnarray}
where in the 
{second-to-last}
step above, we have used the replica trick.
It is easy to do the Gaussian integration over $\mathfrak{m}_{\ell}$ in the above equation by noting that
\begin{equation}
    \int_{{\vec {\mathfrak m}}} e^{-\Delta\sum_{\ell}\mathfrak{m}_\ell^2}e^{+2\Delta\sum_{\ell}\mathfrak{m}_\ell(\sum_{a=1}^R\mathfrak{e}_\ell^{(a)})}=e^{\Delta\sum_{\ell}(\sum_{a=1}^R\mathfrak{e}_\ell^{(a)})^2}.\label{EqGaussianIntegrationIsDone}
\end{equation}
Then using Eq.~\eqref{EqMeasurementAveragedMomentsSimplified} and Eq.~\eqref{EqGaussianIntegrationIsDone}, we obtain
\begin{align}
&&\overline{\langle {\hat {\cal O}}_1\rangle_{\vec{\mathfrak{m}}}
\langle {\hat {\cal O}}_2\rangle_{\vec{\mathfrak{m}}}
...
\langle {\hat {\cal O}}_N\rangle_{\vec{\mathfrak{m}}}}=\lim_{R\rightarrow1}\frac{1}{Z}\sum_{\{s_x^{(a)}\}_{a=1}^{R}}\mathcal{O}^{(1)}_1\cdots \mathcal{O}^{(N)}_N\nonumber\\&&\times\exp{\bigg\{-\sum_{a=1}^{R}{\cal H}[\{s_x^{(a)}\}]}+{\Delta}\sum_{\ell}\sum_{\substack{a\neq b\\a,b=1}}^{R}{\mathfrak{e}}^{(a)}_{\ell}{\mathfrak{e}}^{(b)}_{\ell}\bigg\}.\label{EqSimplifiedMeasurementAveragedMomentsFinal}
\end{align}
Clearly, the weight in the 
{exponential} of 
Eq. \eqref{EqSimplifiedMeasurementAveragedMomentsFinal} is just the lattice version of the replica action in Eq. \eqref{EqDisorderReplicaAction}
{for the case under consideration where the field $\varphi(x)$ is the energy field ${\mathfrak{e}}_{\ell}$ at link $\ell$
in the classical model we started out with}. 
Therefore, the replica action in Eq. \eqref{EqDisorderReplicaAction} in the replica limit $R\rightarrow1$ governs the long-distance physics of the problem of performing weak measurements on the {$2D$} Rokhsar-Kivelson wavefunction in Eq. \ref{LabelEqRKWavefunctionLatt}.
\vskip 0.8cm 
\section{The Proof of \texorpdfstring{$g_{\text{eff}}$}{Lg} Theorem\label{AppGEffTheorem}}
\subsection{The Proof\label{SubSecProofofGEffThm}}
{In this appendix, we prove our $g_{\text{eff}}$ theorem 
presented in 
Sec. \ref{SecGEffThm} and demonstrate
the 
decrease of {$s_{\text{eff}}$ ($=\ln g_{\text{eff}}$, Eq. \eqref{EqDefOFGeffInTermsOfSeff}) in Eq.
\eqref{EqSeffIntro}}
under the RG flow for
the defect replica field theory 
Eq. \eqref{EqDefectReplicaFieldTheoryIntro}. 
{In particular, we prove our Eq. \eqref{EqGeffTheorem}.}
To prove this theorem, we will consider the equivalent boundary field theory that is obtained by folding the 
defect field 
{theory {in Eq.~\eqref{EqDefectReplicaFieldTheoryIntro}}. 
As mentioned {in Sec. \ref{SecGEffThm}}, 
{we consider}
the {\it defect theory} {Eq.~\eqref{EqDefectReplicaFieldTheoryIntro}}
defined on an {\it infinite} cylinder of 
circumference $L$ with defect at axial coordinate 
``$\tau=0$".}
{After folding, the {\it defect} turns
(see the general formulation of this folding procedure presented
in Ref. \cite{LeclairLudwig}) into a {\it boundary}, located at the open end of the 
{semi-infinite}
cylinder, on a CFT in the bulk of the 
{semi-infinite}
cylinder which is the tensor product of the original CFT (described by 
$\sum_{a}S_*^{(a)}$
above) with itself, and thus has twice the central charge. At the same time, the
defect perturbation in Eq.~\eqref{EqDefectReplicaFieldTheoryIntro} becomes a boundary perturbation on this folded bulk theory
$CFT \otimes CFT$. (See Sec. \ref{SubSecDefectFolding} of this appendix for details.)}
The {quantity}
$s_{\text{eff}}$
for the defect field theory can be obtained by studying this corresponding boundary field theory.}
\par
{For unitary boundary CFTs, the 
{decrease}
of 
$s=\ln g$ 
under 
RG 
{flow, the ``g-theorem'',}
was proposed by Affleck and Ludwig in Ref. \cite{AffleckLudwig1991PRL} and it was
{later} 
proved by Friedan and Konechny \cite{FriedanKonechny}.}
The proof of the $g$-theorem for unitary boundary CFTs \cite{AffleckLudwig1991PRL} by Friedan and Konechny \cite{FriedanKonechny,FriedanKonechny2006} involves ``a gradient formula"{, that}
forms the analogue of Zamolodchikov's differential equation for
{the so-called}
$C$-function in his proof of {the} $c$-theorem \cite{Zamolodchikovctheorem}, 
{\footnote{Note that we used 
Zamolodchikov's differential equation for the $C$-function (Eq. \eqref{EqCtheorem}) in our proof of the $c_{\text{eff}}$-theorem in Sec. \ref{Sec:Proof}.}, 
and it reads~\footnote{{See in particular, e.g., Eqs. (3), (12), (13), (16) of \cite{FriedanKonechny2006}}}}
\begin{eqnarray}
    &&\frac{d\mathfrak{s}(L)}{d \ln L}=\nonumber\\
    &&
    =
    -\frac{2\pi^2}{L}\int_{0}^{L}dx \Big(\frac{L}{\pi}\sin\big(\frac{\pi x}{L}\big)\Big)^2 \bigg(\langle \tilde{\theta}(x)\tilde{\theta}(0)\rangle-\langle\tilde{\theta}(0)\rangle^2\bigg),
    \nonumber\\ \label{EqFKGradientFormulaVersion0}
\end{eqnarray}
{where $\tilde{\theta}(x)$ is the trace of the \textit{boundary} stress-energy tensor that is supported on the 
one-dimensional boundary 
at the open-end of
a semi-infinite 
cylinder of circumference $L$,
{and $\mathfrak{s}(L)$ is the subleading piece in the
boundary 
{free energy $:= -\ln z(L)$,}
defined as,
\begin{equation}\label{DefGothicS}
    \mathfrak{s}(L)=\Big(1-L\frac{\partial}{\partial L}\Big)\ln z(L).
\end{equation}
In particular, $\mathfrak{s}(L)$ then interpolates between the Affleck-Ludwig boundary entropy $s_{UV}=\ln g_{UV}$ of the ultraviolet boundary fixed point,
\begin{equation}
   \mathfrak{s}(L\rightarrow0)=s_{UV}=\ln g_{UV}
\end{equation}
and the Affleck-Ludwig boundary entropy $s_{IR}=\ln g_{IR}$ of the infrared boundary fixed point,
\begin{equation}
   \mathfrak{s}(L\rightarrow\infty)=s_{IR}=\ln g_{IR}.
\end{equation}}}
\par
{In 
{the present}
paper, we use a different convention for the stress-energy tensor (that is more standard in CFT, see e.g. \cite{BelavinPolyakovZamolodchikov,CardyLesHouche}), which differs from the convention of Friedan and Konechny in Ref. \cite{FriedanKonechny,FriedanKonechny2006} by a factor of $2\pi$. 
In particular, the trace of the boundary stress energy tensor $\theta(x)$ in our convention is equal to 
$2\pi\times \tilde{\theta}(x)$, where $\tilde{\theta}(x)$, appearing in Eq. \eqref{EqFKGradientFormulaVersion0},
is the trace of the \textit{boundary} stress-energy tensor in the convention used by Friedan and Konechny in Ref. \cite{FriedanKonechny,FriedanKonechny2006}\footnote{{The components of the bulk stress-energy tensor in the convention of Friedan and Konechny are 
given by multiplying the components of bulk stress-energy 
in our convention, given in Eqs.~\eqref{LabelEqStressTensorComponentszz}, \eqref{LabelEqStressTensorComponentsz*z*}, \eqref{LabelEqStressTensorComponentszz*}, by $\frac{1}{2\pi}$.}}. 
Then in our convention the gradient formula in Eq. \eqref{EqFKGradientFormulaVersion0} reads}
\begin{eqnarray}
    &&\frac{d\mathfrak{s}(L)}{d \ln L}=\nonumber\\
    &&
    =
    -\frac{1}{2L}\int_{0}^{L}dx \Big(\frac{L}{\pi}\sin\big(\frac{\pi x}{L}\big)\Big)^2 \bigg(\langle \theta(x)\theta(0)\rangle-\langle\theta(0)\rangle^2\bigg).
    \nonumber\\ \label{EqFKGradientFormula}
\end{eqnarray}
In the case of the
{defect}
replica field theory in Eq. \eqref{EqDefectReplicaFieldTheoryIntro} with a relevant coupling constant $\Delta$, the trace of the 
{boundary}
stress-energy tensor 
{of the boundary 
{theory}
obtained by folding the defect 
{theory}}
is, {in our convention}, {given} by (see Sec. \ref{SubSecOnDerivationOftheta(x)} and \ref{SubSecDefectFolding} for a derivation)
\begin{equation}\label{EqTraceofBoundaryStressETensor}
    \theta(x)=-2\pi \Delta(1-2X_{\varphi})\sum_{\substack{a,b=1\\ a\neq b}}^{R}\varphi^{(a)}(x,0)\varphi^{(b)}(x,0).
\end{equation}
{This equation is exact
\cite{CardyLesHouche,ZamolodchikovIntegrable,GhoshalZamolodchikov},
provided that}
no additive operator renormalizations (``counter-terms") are required for the defect replica field theory in Eq. \eqref{EqDefectReplicaFieldTheoryIntro}. ({This is the precise ``boundary-analogue''} of our {``Assumption 1.''} in the list at the end of Sec. \ref{Sec:Intro}). {
That is, in} {
this case}
there are no higher order corrections in $\Delta$ to Eq. \eqref{EqTraceofBoundaryStressETensor}. [Moreover, analogous to App. \ref{App:VanishingSingleReplicaPerturbation}, we can show that no additive operator renormalizations are required in the replica limit $R\rightarrow1$ whenever $\varphi$ is the most relevant field in the closed OPE subalgebra generated by itself.]
\\
Then we can consider applying the gradient formula of
{Friedan and Konechny, Eq.~\eqref{EqFKGradientFormula} above,}
to the defect replica field theory in Eq. \eqref{EqDefectReplicaFieldTheoryIntro}. 
Using Eq. \eqref{EqTraceofBoundaryStressETensor}, we obtain that
\begin{align}
   &\frac{d \mathfrak{s}_R(L)}{d\ln L}=
   \nonumber-\frac{2\pi^2}{L}\Delta^2(1-2X_{\varphi})^2\int_{0}^{L}dx\bigg[ \Big(\frac{L}{\pi}\sin\big(\frac{\pi x}{L}\big)\Big)^2 \times\nonumber\\&\;\;\;\;\;\times
   \big(\langle \sum_{\substack{a,b=1\\ a\neq b}}^{R}\varphi^{(a)}(x,0)\varphi^{(b)}(x,0)\sum_{\substack{c,d=1\\ c\neq d}}^{R}\varphi^{(c)}(0,0)\varphi^{(d)}(0,0)\rangle_{R}\nonumber\\
   &\;\;\;\;\;\;\;\;\;\;\;\;\;\;\;\;\;\;\;\;\;\;\;\;\;\;\;\;\;\;\;\;\;\;-\big(\langle \sum_{\substack{a,b=1\\ a\neq b}}^{R}\varphi^{(a)}(0,0)\varphi^{(b)}(0,0)\rangle_R\big)^2
   \big)\bigg],
\label{EqSumRuleAppliedToDefectReplicaTheory} 
\end{align}
where {both} the correlation functions $\langle\cdots\rangle_{R}$ on the RHS {and the function 
$\mathfrak{s}_R(L)$
on the LHS} of the above equation are evaluated in the $R$-copy defect replica field theory of Eq. \eqref{EqDefectReplicaFieldTheoryIntro}.
Exactly analogous to Eq. \eqref{EqTraceSquare} and \eqref{EqCorrelatorSimplified2},  the correlation functions on the RHS of the above formula can be simplified using permutation symmetry of replica indices on the fields. In particular, analogous to Eq. \eqref{EqTraceSquare},
\begin{align}
   &\big(\langle \sum_{\substack{a,b=1\\ a\neq b}}^{R}\varphi^{(a)}(0,0)\varphi^{(b)}(0,0)\rangle_R\big)^2\nonumber\\&\;\;\;\;\;\;\;\;\;\;\;\;\;\;=R^2(R-1)^2(\langle\varphi^{(1)}(0,0)\varphi^{(2)}(0,0)\rangle_{R})^2\label{EqDefectTraceSquare}
\end{align}
and analogous to Eq. \eqref{EqCorrelatorSimplified2}
\begin{align}
     &\langle \sum_{\substack{a,b=1\\ a\neq b}}^{R}\varphi^{(a)}(x,0)\varphi^{(b)}(x,0)\sum_{\substack{c,d=1\\ c\neq d}}^{R}\varphi^{(c)}(0,0)\varphi^{(d)}(0,0)\rangle_{R}\nonumber\\
     &=2(R-1) \bigg(\langle \varphi^{(1)}(x,0)\varphi^{(1)}(0,0)\varphi^{(2)}(x,0)\varphi^{(2)}(x,0)\rangle_{R=1}\nonumber\\&\;\;\;\;\;\;\;\;\;\;\;\;\;+\langle\varphi^{(1)}(x,0)\varphi^{(2)}(x,0)\varphi^{(3)}(0,0)\varphi^{(4)}(0,0)\rangle_{R=1}\nonumber\\&\;\;\;\;\;\;\;\;\;\;\;\;\;-2\langle \varphi^{1}(x,0)\varphi^{(1)}(0,0)\varphi^{(2)}(x,0)\varphi^{(3)}(0,0)\rangle_{R=1}\bigg)+\nonumber\\&\;\;\;\;\;\;\;\;\;\;\;\;\;\;\;\;\;\;\;\;\;\;\;\;\;\;\;\;\;\;\;\;\;\;\;\;\;\;\;\;\;\;\;\;\;\;\;\;\;\;\;\;\;+ \mathcal{O}((R-1)^2).\label{EqDefectCorrelatorSimplified}
\end{align}
Using Eq. \eqref{EqSumRuleAppliedToDefectReplicaTheory}, \eqref{EqDefectTraceSquare} and \eqref{EqDefectCorrelatorSimplified}, we obtain
\begin{align}
   &\frac{d\mathfrak{s}_R(L)}{d\ln L}=\nonumber\\
   & \nonumber-\frac{(2\pi)^2}{L}(R-1)\Delta^2(1-2X_{\varphi})^2\int_{0}^{L}dx\bigg[ \Big(\frac{L}{\pi}\sin\big(\frac{\pi x}{L}\big)\Big)^2 \times\nonumber\\&\;\;\;\;\;\;\;\;\times\bigg(\langle \varphi^{(1)}(x,0)\varphi^{(1)}(0,0)\varphi^{(2)}(x,0)\varphi^{(2)}(x,0)\rangle_{R=1}\nonumber\\&\;\;\;\;\;\;\;\;\;\;\;\;\;\;\;\;+\langle\varphi^{(1)}(x,0)\varphi^{(2)}(x,0)\varphi^{(3)}(0,0)\varphi^{(4)}(0,0)\rangle_{R=1}\nonumber\\&\;\;\;\;\;\;\;\;\;\;\;\;\;\;\;\;-2\langle \varphi^{1}(x,0)\varphi^{(1)}(0,0)\varphi^{(2)}(x,0)\varphi^{(3)}(0,0)\rangle_{R=1}\bigg)\bigg]\nonumber\\&\;\;\;\;\;\;\;\;\;\;\;\;\;\;\;\;\;\;\;\;\;\;\;\;\;\;\;\;\;\;\;\;\;\;\;\;\;\;\;\;\;\;\;\;\;\;\;\;\;\;\;\;\;+ \mathcal{O}((R-1)^2).
\label{EqSimplifiedSumRuleAppliedToDefectReplicaTheory} 
\end{align}
where we have absorbed the contribution from Eq. \eqref{EqDefectTraceSquare} in the corrections of order $\mathcal{O}((R-1)^2)$ in the above equation.
Differentiating the above equation w.r.t. $R$ at $R=1$ 
{and}
 {following {Eq. \eqref{EqSeffDefInTermsOfS(L))}},}
we obtain,
\begin{align}
   &\frac{d\mathfrak{s}_{\text{eff}}(L)}{d\ln L}=-\frac{(2\pi)^2}{L}\Delta^2(1-2X_{\varphi})^2\int_{0}^{L}dx\bigg[ \Big(\frac{L}{\pi}\sin\big(\frac{\pi x}{L}\big)\Big)^2 \times\nonumber\\&\;\;\;\;\;\;\;\;\;\;\;\;\times\bigg(\langle \varphi^{(1)}(x,0)\varphi^{(1)}(0,0)\varphi^{(2)}(x,0)\varphi^{(2)}(x,0)\rangle_{R=1}\nonumber\\&\;\;\;\;\;\;\;\;\;\;\;\;\;\;\;\;+\langle\varphi^{(1)}(x,0)\varphi^{(2)}(x,0)\varphi^{(3)}(0,0)\varphi^{(4)}(0,0)\rangle_{R=1}\nonumber\\&\;\;\;\;\;\;\;\;\;\;\;\;\;\;\;\;-2\langle \varphi^{1}(x,0)\varphi^{(1)}(0,0)\varphi^{(2)}(x,0)\varphi^{(3)}(0,0)\rangle_{R=1}\bigg)\bigg].
\label{EqPenultimateSumRuleDefectReplicaTheory} 
\end{align}
where on the left-hand side of the above equation we have used {Eq. \eqref{EqSeffDefInTermsOfS(L))}} for $\mathfrak{s}_{\text{eff}}(L)$.
{Now by introducing a randomness field $h(x)$
{supported solely on the defect, and by proceeding in complete analogy with App. \ref{AppReIntroOfRandom},}}
we can write the above correlation functions $\langle\cdots\rangle_{R=1}$ of replicated copies of fields in the defect replica field theory of Eq. \eqref{EqDefectReplicaFieldTheoryIntro} as randomness averaged products of correlation functions in a single copy theory {$\tilde{\mathcal{S}}[\{h(x)\}]$} where the randomness field $h(x)$ is coupled to the field $\varphi(x,0)$ \textit{on the defect line} situated at $\tau=0$.
In particular, analogous to Eq. \eqref{EqReplicaCorrelationFunctionsToRandomnessOnes}, we can write the correlation functions on the right-hand side of 
{Eq. \eqref{EqPenultimateSumRuleDefectReplicaTheory}}
as
\begin{subequations}\label{EqDefectCorrelationFunctionsToRandomnessOnes}
\begin{eqnarray}
      &&\langle \varphi^{(1)}(x,0)\varphi^{(1)}(0,0)\varphi^{(2)}(x,0)\varphi^{(2)}(0,0)\rangle_{R=1}\nonumber\\&&\;\;\;\;\;\;\;\;\;\;\;\;\;\;\;\;\;\;\;\;\;\;\;\;\;\;\;\;\;\;\;\;\;\;\;\;\;\;\;=\overline{\langle\varphi(x,0)\varphi(0,0)\rangle_{\{h\}}^2}\\
    &&\langle\varphi^{(1)}(x,0)\varphi^{(2)}(x,0)\varphi^{(3)}(0,0)\varphi^{(4)}(0,0)\rangle_{R=1}\nonumber\\&&\;\;\;\;\;\;\;\;\;\;\;\;\;\;\;\;\;\;\;\;\;\;\;\;\;\;\;\;\;\;\;\;\;\;\;\;=\overline{\langle\varphi(x,0)\rangle_{\{h\}}^2\langle\varphi(0,0)\rangle_{\{h\}}^2}\\
    &&\langle \varphi^{1}(x,0)\varphi^{(1)}(0,0)\varphi^{(2)}(x,0)\varphi^{(3)}(0,0)\rangle_{R=1}\nonumber\\&&\;\;\;\;\;\;\;\;\;\;\;\;\;\;\;\;\;\;=\overline{\langle \varphi(x,0)\varphi(0,0)\rangle_{\{h\}}\langle\varphi(x,0)\rangle_{\{h\}}\langle\varphi(0,0)\rangle_{\{h\}}}.\nonumber\\
\end{eqnarray}
\end{subequations}
where the correlation functions $\langle\dots\rangle_{\{h\}}$ on the right-hand side of the above equation are evaluated in the single copy theory 
{$\tilde{\mathcal{S}}[\{h(x)\}]$}
given by 
{(analogous to Eq. \eqref{EqInteractingTheoryWithRandomness})}
\begin{eqnarray}
\label{EqDefectInteractingTheoryWithRandomness}
&&
{\tilde{\mathcal{S}}[\{h(x)\}]}:=
\\ \nonumber
&&
=S_{*}+\Delta\int_0^{L} dx \, \bigg(\varphi(x,0)-
\frac{h(x)}{2\Delta}\bigg)^2-\frac{1}{4\Delta}\int_0^{L} dx \, h^2(x),
\end{eqnarray}
{which is interacting with a \textit{fixed} realization of
{the randomness field}
$h(x)$ \text{on the $\tau=0$ defect line.}}
{The}
average denoted by 
an overbar
$\overline{(\dots)}$ in Eq. \eqref{EqDefectCorrelationFunctionsToRandomnessOnes} is 
{defined as}
(analogous to Eq. \eqref{EqWhatDoesanOverlineMean})
\begin{equation}
     \overline{(\cdots)}:=\frac{1}{Z_*}\int Dh\int D\phi\,e^{-\int_0^L dx  \frac{h(x)^2}{4\Delta}-
     {\tilde{\mathcal{S}}[\{h(x)\}]}
     }(\cdots).
\end{equation}
{Now, again completely analogous}
to our discussion in Sec. \ref{Sec:Proof}, if the defect replica field theory in Eq. \eqref{EqDefectReplicaFieldTheoryIntro} is obtained from a given problem with
weak measurements on a one-dimensional quantum critical {ground} state {corresponding to $(1+1)D$ CFT $S_*$}, the above procedure of introducing the ``artificial" randomness field $h(x)$ is not necessary. In particular, in such a case, we can directly write the replica correlation functions {in the $R\rightarrow1$ replica limit} on the left-hand side of Eq. \eqref{EqDefectCorrelationFunctionsToRandomnessOnes} as
suitable measurement-averaged products of correlation functions, where the  precise
form for the latter is the same as that of the corresponding correlation functions on the right-hand side of Eq. \eqref{EqDefectCorrelationFunctionsToRandomnessOnes}, except that $\langle \cdots \rangle_{\{h\}}$ now represents the correlation
functions evaluated in a given  “measurement
trajectory $\{h(x)\}$” and the overbar $\overline{(\cdots)}$ denotes the average over measurement outcomes {$\{h(x)\}$} with Born-rule probability. (See e.g. Refs. \cite{GarrattWeinsteinAltman2022,YangMaoJian,WeinsteinSajithAltmanGaratt,PatilLudwig2024}.)\par
Then using Eq. \eqref{EqPenultimateSumRuleDefectReplicaTheory} and Eq. \eqref{EqDefectCorrelationFunctionsToRandomnessOnes}, we obtain
\begin{align}
   &\frac{d\mathfrak{s}_{\text{eff}}(L)}{d\ln L}=-\frac{(2\pi)^2}{L}\Delta^2(1-2X_{\varphi})^2\int_{0}^{L}dx\bigg[ \Big(\frac{L}{\pi}\sin\big(\frac{\pi x}{L}\big)\Big)^2 \times\nonumber\\&\;\;\;\;\;\;\;\;\;\;\;\times\overline{(\langle\varphi(x,0)\varphi(0,0)\rangle_{\{h\}}-\langle\varphi(x,0)\rangle_{\{h\}}\langle\varphi(0,0)\rangle_{\{h\}})^2}\bigg].
\label{EqFinalSumRuleDefectReplicaTheory} 
\end{align}
From the above equation it is clear that, since the integrand in the above equation is a positive number~\footnote{{In complete analogy to the discussion in footnote 
\cite{Note23}, any}
correlation functions 
for the
scalar field 
$\varphi(x,0)$ in the theory $\tilde{S}[\{h(x)\}]$ (Eq. \eqref{EqDefectInteractingTheoryWithRandomness}), in which the real randomness field $h(x)$ is
coupled to $\varphi(x,0)$ on the defect, can be expanded (in a functional Taylor expansion {in $h(x)$}) in terms of 
multipoint correlation functions of the field $\varphi(x,0)$ in the unperturbed CFT $S_{*}$, which are \textit{real} {in a unitary CFT $S_*$}, and the \textit{real} randomness field $h(x)$. Therefore the correlation function $\langle\varphi(x,0)\varphi(0,0)\rangle_{\{h\}}-\langle\varphi(x,0)\rangle_{\{h\}}\langle\varphi(0,0)\rangle_{\{h\}}$ is real and the integrand in Eq. \eqref{EqFinalSumRuleDefectReplicaTheory} is a positive number.},
{we obtain}
due to the overall negative sign,
\begin{equation}
    \frac{d\mathfrak{s}_{\text{eff}}(L)}{d\ln L}\leq0,
\end{equation}
{i.e. we obtain Eq. \eqref{EqGeffTheorem} in Sec. \ref{SecGEffThm} of the main text of this paper
that
demonstrates the 
decrease of $s_{\text{eff}}=\ln g_{\text{eff}}$ under the RG flow. (See Sec. \ref{SecGEffThm} for discussion.)}\par
Finally, we {note that, one} can integrate the sum-rule in Eq. \eqref{EqFinalSumRuleDefectReplicaTheory} from $L=0$ to $L=\infty$ to obtain an expression for the infrared value of $s_{\text{eff}}=\ln g_{\text{eff}}$ in terms of integrals of the randomness-averaged square of the connected
correlation function of the field $\varphi$. {(See Eq.~\eqref{EQFinalSumRuleIntegrated} for analogous expression for $c_{\text{eff}}$ in `bulk' replica field theories Eq.~\eqref{EqDisorderReplicaAction} in the $R\rightarrow1$ replica limit.)}
{The latter}
relation {might be} useful
{in its own right} in the problem of weak {Born-rule} measurements {with a local operator corresponding to field $\varphi$} on one-dimensional quantum critical ground 
{state corresponding to $(1+1)D$ CFT $S_*$}, where as mentioned above the correlation functions $\langle\cdots\rangle_{\{h\}}$ on the right-hand side of Eq. \eqref{EqFinalSumRuleDefectReplicaTheory} have a direct natural interpretation in the measurement problem as correlation functions evaluated in a given measurement trajectory $\{h(x)\}$ and the average $\overline{(\cdots)}$ corresponds to the average over measurement outcomes with Born-rule probability. \par
\subsection{The trace of the boundary stress-energy tensor: \texorpdfstring{$\theta(x)$}{Lg}
\label{SubSecOnDerivationOftheta(x)}}
In this subsection, we will derive the expression for the trace of the {boundary} stress-energy tensor $\theta(x)$ for a 
{bulk CFT with action}
$S_{*}$ defined on the upper half 
{plane, $z=x+iy$, $\bar{z}=x-iy$ with
$y\geq0$, in the presence of an RG relevant boundary perturbation localized on the real axis.}

{Specifically, we consider a particular conformally invariant boundary condition on the real axis $y=\Im z=0$. We use the complex components
of the  bulk stress-energy tensor having Cartesian components  $T_{\mu\nu}$
(where $\mu, \nu = x, y$),
 \begin{eqnarray}
 \label{LabelEqStressTensorComponentszz}
     &&T \equiv T_{zz} = \frac{1}{4} (T_{xx} - T_{yy} - 2i T_{xy}), \\ 
     \label{LabelEqStressTensorComponentsz*z*}
     &&\overline{T} \equiv T_{\bar{z} \bar{z}} = \frac{1}{4} (T_{xx} - T_{yy} + 2i T_{xy}),\\ 
\label{LabelEqStressTensorComponentszz*}
     &&
     \Theta \equiv 4 T_{z\bar{z}} = 4 T_{z z^\star} = T_{xx} + T_{yy}.
 \end{eqnarray}
Due to conformal invariance of the bulk, $T$ and $\overline{T}$ are analytic functions of $z$ and ${\bar z}$, respectively, which 
implies~\footnote{{Due to the conservation law
$\partial_{\bar z}T = -\frac{1}{4}\partial_{z}\Theta$ and
$\partial{z} \overline{T}=-\frac{1}{4}\partial_{\bar{z}} \Theta$}} 
that 
$\Theta=0$ in the bulk. It is known~\cite{CARDY1984surface}
that
conformal invariance of the boundary condition is characterized}
{by the crucial condition
\begin{eqnarray}
\label{LabelEqConformalInvarianceOfBoundary}
\lim_{y \to 0}
\left ( T(x+iy)
-\overline{T}(x-iy)
\right)=
(-i) T_{xy}|_{y=0}=0.
\quad
\end{eqnarray}
This means that $\overline{T}$ evaluated in the upper half complex plane, is nothing else but the analytic continuation of $T$ (defined initially also in the upper half complex plane) into the lower half complex plane.}

{
We will be interested in adding a 
boundary perturbation
$\int_{-\infty}^{+\infty} dx \,\chi(x)$ 
to the action 
which is relevant in the RG sense, 
where the field 
$\chi(x)$
is a boundary field (supported solely on the boundary) that exists at the given boundary condition we are considering. In the presence of the perturbation, the action thus reads}
\begin{equation}\label{EqBoundaryTheoryPerturbed}
     S=S_*-\lambda\int_{-\infty}^{+\infty} dx \,\chi(x).
 \end{equation}
{A key quantity of interest to us will be the
{\it trace $\theta(x)$ of the boundary stress-energy tensor}, defined 
by
\begin{eqnarray}
\label{LabelEqDefBoundaryStressEnergyTrace}
\Theta(x,y)
=
\delta(y)  \ \theta(x),
\end{eqnarray}
which vanishes in the absence of the boundary perturbation, i.e. $\theta(x)=0$
 when $\lambda=0$. However, as we will now review following Ghoshal and Zamolodchikov \cite{GhoshalZamolodchikov},
 the right hand side of
 Eq.~\eqref{LabelEqConformalInvarianceOfBoundary} no longer vanishes
 when $\lambda \not =0$, but instead satisfies
 \begin{align}
&\lim_{y\rightarrow0}\,i(T(x+iy)-\overline{T}(x-iy))=
T_{xy}|_{y=0} = -\frac{d}{dx}\theta(x),
\label{EqSecondDefinitionForSmallTheta(x)}
\end{align}
where $\theta(x)$ is related as follows
to the boundary perturbation in Eq.~\eqref{EqBoundaryTheoryPerturbed}:}
We will consider the following correlation function
{evaluated}
with {the boundary} action {in} Eq. \eqref{EqBoundaryTheoryPerturbed}
where 
{the ellipsis} ``$\cdots$'' represents arbitrary fields 
from the CFT,
\begin{align}
    &\Big\langle i(T(x+iy)-\overline{T}(x-iy))\cdots\Big\rangle_{S}=
    \nonumber\\=& 
    \Big\langle i(T(x+iy)-\overline{T}(x-iy))
    \cdots
e^{\lambda\int_{-\infty}^{+\infty} dx' \,\chi(x')}\Big\rangle_{S_*}=
 \nonumber\\=& 
    \Big\langle i(T(x+iy)-\overline{T}(x-iy))\cdots\Big\rangle_{S_*}+\nonumber\\
&+\lambda\Big\langle i(T(x+iy)-\overline{T}(x-iy))
\int_{-\infty}^{+\infty} dx'\chi(x')
\cdots\Big\rangle_{S_*}+\nonumber\\&+\mathcal{O}(\lambda^2).\label{EqTaylorExpansion}
\end{align}
{The correlation functions 
after the first and second equality sign
are evaluated, as indicated, with the unperturbed action $S_*$, i.e. at $\lambda=0$. A particular consequence of this is that $\overline{T}(x-iy)$
is evaluated in the presence of the conformally invariant boundary condition, at which {Eq.} \eqref{LabelEqConformalInvarianceOfBoundary} holds.
Thus, as mentioned in the paragraph below {Eq.} \eqref{LabelEqConformalInvarianceOfBoundary}, $\overline{T}(x-iy)$ is replaced by the analytic continuation of $T(x+iy)$ into the lower half complex plane, that is $\overline{T}(x-iy)\to $ $T(x-iy)$.
The first correlation function
after the second equality sign
vanishes identically in any case in the $y\rightarrow0$ limit
{by virtue of {Eq.} \eqref{LabelEqConformalInvarianceOfBoundary}.}
In order to evaluate the first order contribution in $\lambda$, we use the OPE of $T(z)$ with the  operator $\chi(x)$ which, being a boundary operator, is a holomorphic field {of conformal scaling dimension $h$ (assumed here to be primary)} supported entirely on the boundary.
Using the standard  OPE of the stress-{energy} tensor with a holomorphic primary field,}
\begin{subequations}\label{EqConformalWardIdentity}
 \begin{eqnarray}
 &&T(z)\chi(x')\sim
 \frac{h \chi(x')}{(z-x')^2}+\frac{
 {d\over dx'} \chi(x')
 }{(z-x')}\\
 &&T(z^*)\chi(x')\sim
 \frac{h\chi(x')}{(z^*-x')^2}+\frac{\frac{d }{d x'}\chi(x')}{(z^*-x')}
 \end{eqnarray}
 \end{subequations}
we can simplify the second term 
{after the second equality sign of
Eq. \eqref{EqTaylorExpansion} to read}
 \begin{align}
    &\lambda\Big\langle i(T(x+iy)-T(x-iy))\int_{-\infty}^{+\infty} dx'\chi(x')\cdots\Big\rangle_{S_*}\nonumber\\&=i\lambda \Big\langle \int_{-\infty}^{+\infty} dx'\bigg(\frac{h\chi(x')}{(x+iy-x')^2}+\frac{\frac{d}{d x'}\chi(x')}{(x+iy-x')}\nonumber\\&\;\;\;\;\;\;\;\;\;\;\;\;\;\;\;\;\;\;\;\;\;\;\;-\frac{h\chi(x')}{(x-iy-x')^2}-\frac{\frac{d }{d x'}\chi(x')}{(x-iy-x')}
    \bigg)\dots \Big\rangle_{S_{*}}\nonumber\\
    &=i\lambda \Big\langle \int_{-\infty}^{+\infty} dx'\bigg(\frac{-4iy(x-x')h}{((x-x')^2+y^2)^2}\chi(x')\nonumber\\&\qquad\qquad\qquad\qquad+\frac{-2iy}{(x-x')^2+y^2}{\frac{d \chi(x')}{dx'}}
    \bigg)\dots \Big\rangle_{S_{*}}\nonumber\\
    &=i\lambda \Big\langle \int_{-\infty}^{+\infty} dx'\bigg(2\pi ih\frac{d \ \delta(x-x')}{dx}\chi(x')+\nonumber\\&\qquad\qquad\qquad\qquad-2\pi i\delta(x-x') {\frac{d \chi(x')}{d x'}}
    \bigg)\dots \Big\rangle_{S_{*}}
 \end{align}
where in the last line we have used the following representation of the delta function and its derivative:
\begin{equation}
    2\pi\delta(x)=\lim_{y\rightarrow0} \frac{2y}{x^2+y^2},\;\;\;2\pi\frac{d\delta(x)}{dx}=\lim_{y\rightarrow0} \frac{-4xy}{(x^2+y^2)^2}.
\end{equation}
{Hence} we obtain
\begin{align}
    \lambda\Big\langle i(T(x+iy)&-
    T(x-iy)\int_{-\infty}^{+\infty} dx'\chi(x')\cdots\Big\rangle_{S_*}\nonumber=\\
    =&2\pi\lambda \Big\langle\big(-h\frac{d\chi(x)}{dx}+ \frac{d\chi(x)}{dx}
   \bigg)\dots \Big\rangle_{S_{*}}\nonumber\\
=&2\pi\lambda (1-h)\Big\langle\frac{d\chi(x)}{dx}\dots \Big\rangle_{S_{*}}.\label{EqFirstOrderTermInTaylorExp}
\end{align}
{Therefore, using Eq. \eqref{EqSecondDefinitionForSmallTheta(x)}, the expression for the trace of the boundary stress-energy tensor is given by}
\begin{equation}
    \theta(x)=-2\pi\lambda(1-h)\chi(x)+\mathcal{O}(\lambda^2).
\end{equation}
{
{
Corrections of higher than first order in $\lambda$ in the above equation,
which relates the field $\theta(x)$ to the field
$\chi(x)$,
vanish}
identically provided the field theory in 
Eq.~\eqref{EqBoundaryTheoryPerturbed} is well defined in the ultraviolet without the need for counter terms, i.e.  {without} additive operator renormalizations} 
{of the field 
$\chi(x)$} (see e.g. \cite{CardyLesHouche,ZamolodchikovIntegrable}), and we {obtain}
 {the exact relation~\footnote{{Of course, say, the connected 2-point function of the field $\theta(x)$ and that of the field 
 $\chi(x)$
 evaluated with the action Eq.~\eqref{EqBoundaryTheoryPerturbed} will in general contain contributions to all orders in $\lambda$, but 
those are all (exactly) related to each other by the
 exact equation  Eq.~\eqref{LabelEqExactRelationthetachi} 
 below.}}}
\begin{equation}
\label{LabelEqExactRelationthetachi}
    \theta(x)=-2\pi\lambda(1-h)\chi(x).
\end{equation}
{This concludes our derivation of the trace of the boundary stress-energy tensor $\theta(x)$ for boundary perturbations 
{of a boundary of a bulk CFT.}~\footnote{{Note that the only way the particular boundary condition on the bulk CFT makes its appearance
in the relationship between $\theta(x)$ and $\chi(x)$ is through the requirement that
$\chi(x)$
must be an allowed boundary field at the particular boundary condition. In other words, this field must be contained in the spectrum of all possible boundary operators at the particular boundary condition considered {(see e.g. Sec. 4 of Ref. \cite{AffleckLudwig1991NPB}).}
Thus, while the particular linear combination of the conformal blocks of any multipoint correlation function of the boundary field $\chi(x)$, being a holomorphic field, depends in general on the particular boundary condition 
\cite{LEWELLEN1992654,RUNKEL1999563},
the multipoint correlation function of the trace of the boundary stress tensor $\theta(x)$ will thus follow {the same}
linear combination of conformal blocks due to the exact relationship Eq.~\eqref{LabelEqExactRelationthetachi}}} In the next subsection, we will describe the `folding' procedure~\cite{LeclairLudwig}
to obtain a boundary CFT perturbed by a boundary perturbation from a problem of
{a defect perturbation of a bulk} CFT.}
\subsection{Folding the defect CFT into a boundary CFT\label{SubSecDefectFolding}}
As mentioned in Sec. \ref{SubSecProofofGEffThm}, given a $2D$ CFT $\mathcal{A}_{*}$ with a defect along the $x$-axis ($y=0$) we can obtain a boundary theory by folding the theory on the defect.
The defect perturbation {then} turns into {a} boundary perturbation on the boundary at $y=0$
{of}
the folded bulk theory with
 a tensor product of two copies of the CFT $\mathcal{A}_{*}$ (i.e. $\mathcal{A}_{*}\otimes \mathcal{A}_*$)~\cite{LeclairLudwig}.
To illustrate this, we will consider a CFT {with action} $\mathcal{A}_
*$ perturbed with a defect perturbation $\int dx \,\chi(x,0)$ supported on the $x$-axis
\begin{equation}\label{EqDefectTheoryGeneral}
    \mathcal{A}=\mathcal{A}_*-\lambda\int dx \,\chi(x,0),
\end{equation}
where $\chi$ is a bulk scaling field from the CFT $S_*$. Our defect replica field theory of interest in Eq. \eqref{EqDefectReplicaFieldTheoryIntro} is of this form, where the action $\mathcal{A}_*=\sum_{a}S_*^{(a)}$ and the field $\chi=\sum_{\substack{a,b=1\\a\neq b}}^R\varphi^{(a)}\varphi^{(b)}$.
Following Ref. \cite{LeclairLudwig}, we can turn 
{the}
general problem of defect perturbation of a CFT $\mathcal{A}_*$ into a problem of {a} 
boundary perturbation 
{of}
the tensor product of CFT $\mathcal{A}_*$ with itself, i.e. $\mathcal{A}_*\otimes\mathcal{A}_*$.
Using the components of stress-energy tensor $T$ and $\overline{T}$ for the 
unperturbed CFT $\mathcal{A}_*$,
we can define the components of stress-energy tensor $T_f$ and $\overline{T}_f$ for the folded boundary CFT $\mathcal{A}_*\otimes\mathcal{A}_*$ as
\begin{equation}\label{EqSETensorForTheFoldedTheory}
\left.\begin{aligned}
&T_f(z,\bar{z})=T(z)+\overline{T}({z})\\
&\overline{T}_f(z,\bar{z})=T(z^{*})+\overline{T}({z}^*)         
\end{aligned} \right\}\;\; \cdots\;\;\text{Im}(z)\geq0
\end{equation}
where the folded theory $\mathcal{A}_*\otimes\mathcal{A}_*$ is, of course,
defined only in the upper-half plane $y\geq 0$, i.e. $y=\text{Im}(z)\geq 0$. From the above two equations it is clear that,
\begin{equation}
    (T_{f}-\overline{T}_{f})|_{y=0}=0,
\end{equation}
which defines a conformal boundary condition on the folded CFT $\mathcal{A}_*\otimes\mathcal{A}_*$ defined in the upper-half plane.\par
We will now discuss boundary scaling fields in the folded CFT $\mathcal{A}_*\otimes\mathcal{A}_*$. 
Any bulk scalar scaling field $\Psi(x,y)$  with bulk scaling dimension $=X$ (i.e. conformal dimensions $h=X/2$ and $\bar{h}=X/2$) from the unfolded CFT $\mathcal{A}_*$, is a \textit{boundary} scaling field when situated at $y=0$ and 
has \textit{boundary} scaling dimension $X$ in the folded CFT $\mathcal{A}_*\otimes\mathcal{A}_*$. To see this consider the OPE of such a scalar field 
$\Psi(x,y=0)$ situated on $y=0$ with the holomorphic stress-energy tensor of the folded boundary CFT in Eq. \eqref{EqSETensorForTheFoldedTheory}, 
\begin{align}
&T_f(z)\Psi(x,0)= T(z)\Psi(x,0)+\overline{T}(z)\Psi(x,0)\sim\nonumber\\
&\sim \frac{\frac{X}{2}\Psi(x,0)}{(z-x)^2}+\frac{\partial \Psi(x,0)}{(z-x)}+\frac{\frac{X}{2}\Psi(x,0)}{(z-x)^2}+\frac{\bar{\partial} \Psi(x,0)}{(z-x)}, 
\end{align}
where in the above equation we have used the `bulk' conformal Ward identity/OPE of the field $\Psi$ and the `bulk' stress-energy tensor components $T$ and $\overline{T}$ of the unfolded CFT $\mathcal{A}_*$. Then using $\partial+\bar{\partial}=\frac{d}{dx}$, we obtain
\begin{align}
T_f(z)\Psi(x,0)
\sim \frac{{X}\Psi(x,0)}{(z-x)^2}+\frac{ \frac{d}{dx}\Psi(x,0)}{(z-x)},
\end{align}
and similarly
\begin{align}
\overline{T}_f(\overline{z})\Psi(x,0)
\sim \frac{{X}\Psi(x,0)}{(z^*-x)^2}+\frac{ \frac{d}{dx}\Psi(x,0)}{(z^*-x)}.
\end{align}
Therefore, the OPEs of the field $\Psi(x,0)$ with the components of the stress-energy tensor $T_{f}$ and $\overline{T}_{f}$ of the folded boundary CFT $\mathcal{A}_*\otimes\mathcal{A}_*$ precisely satisfy the general form of {the} boundary conformal Ward identity/OPE in Eq. \eqref{EqConformalWardIdentity}, and 
the \textit{boundary} scaling dimension of the field $\Psi(x,0)$ in the folded CFT $\mathcal{A}_*\otimes\mathcal{A}_*$ is equal to $X$, i.e. its bulk scaling dimension in the unfolded CFT $\mathcal{A}_*$.\par
In particular, following the previous section Sec. \ref{SubSecOnDerivationOftheta(x)} 
and assuming that the defect theory in Eq. \eqref{EqDefectTheoryGeneral} is well defined in the UV without the need for counter-terms  ({i.e. without} additive operator renormalization),
the expression for the trace of the boundary stress-energy tensor for the folded {boundary} theory corresponding to the defect theory in Eq. \eqref{EqDefectTheoryGeneral} is given by
\begin{equation}\label{Eqsmallthetageneraldefecttheory}
    \theta(x)=-2\pi \lambda(1-X)\chi(x,0),
\end{equation}
where $X$ is the bulk scaling dimension of the field $\chi$ in the unperturbed, unfolded CFT $\mathcal{A}_*$. 
The expression for $\theta(x)$ in Eq. \eqref{EqTraceofBoundaryStressETensor} for the defect replica field theory in Eq. \eqref{EqDefectReplicaFieldTheoryIntro} then follows, using the bulk scaling dimension $=2X_{\varphi}$ of the field $\sum_{\substack{a,b=1\\a\neq b}}^R\varphi^{(a)}\varphi^{(b)}$ in the unperturbed CFT $\sum_aS_*^{(a)}$.

\bibliography{ceffthm}
\end{document}